\documentclass[smallcondensed]{svjour3}

\usepackage{lineno,hyperref,xcolor,graphicx,amsmath,natbib,amssymb,xcolor}
\modulolinenumbers[5]

\RequirePackage{fix-cm}
\smartqed  
\usepackage{graphicx}
\newcommand{\aap}{{Astron. Astrophys.}}
\newcommand{\apj}{{Astrophys. J.}}

\newcommand{\apjl}{{Astrophys. J. Lett.}}
\newcommand{\apjs}{{\it Astrophys. J. Supp. Series}}
\newcommand{\apss}{{\it Astrophys. Space Sci.}}

\newcommand{\prl}{{\it Phys. Rev. Lett.}}
\newcommand{\solphys}{{Solar Phys.}}
\newcommand{\ssr}{{Space Sci. Rev.}}

\newcommand{\jgr}{{J. Geophys. Res.}}
\newcommand{\mnras}{MNRAS}
\newcommand{\pasj}{{\it Pub. Astron. Soc. Japan}}
\newcommand{\ion}[2]{#1{\sc #2}}
\newcommand{\kms}{km\,s$^{-1}$}

\begin{document}

\title{Kink oscillations of coronal loops}

\titlerunning{Kink oscillations of coronal loops}        
\authorrunning{V. M. Nakariakov et al.}
\author{V.~M. Nakariakov        \and
             S.~A.  Anfinogentov  \and
             P.~Antolin \and
             R.~Jain \and 
             D.~Y.~Kolotkov \and
             E.~G.~Kupriyanova \and
             D. Li \and
             N.~Magyar \and
             G.~Nistic\`o \and
             D.~J.~Pascoe \and
             A.~K.~Srivastava \and
             J.~Terradas \and
             S.~Vashegani Farahani \and
        G.~Verth \and 
         D.~Yuan \and 
         I.~V.~Zimovets
}

\authorrunning{V. M. Nakariakov et al.}

\institute{
V.~M. Nakariakov \at
	Centre for Fusion, Space and Astrophysics, Department of Physics, University of Warwick, Coventry CV4 7AL, UK,\\
	School of Space Research, Kyung Hee University, Yongin, 446-701, Gyeonggi, Korea,\\
	St Petersburg Branch, Special Astrophysical Observatory, Russian Academy of Sciences, St Petersburg 196140, Russia\\
	\email{V.Nakariakov@warwick.ac.uk}           
	\and
S.~A.  Anfinogentov \at 
	Institute of Solar-Terrestrial Physics SB RAS, Lermontov St. 126, Irkutsk 664033, Russia	
           \and
P.~Antolin \at
	Department of Mathematics, Physics \& Electrical Engineering, Northumbria University, Newcastle Upon Tyne, NE18ST, UK
	\and
R.~Jain \at
	School of Mathematics and Statistics, University of Sheffield, Sheffield S3 7RH, UK
	\and
D.~Y.~Kolotkov \at
	Centre for Fusion, Space and Astrophysics, Department of Physics, University of Warwick, Coventry CV4 7AL, UK,\\
	Institute of Solar-Terrestrial Physics, Lermontov St., 126a, Irkutsk 664033, Russia
	\and
E.G.~Kupriyanova \at
	Central Astronomical Observatory at Pulkovo of RAS, Saint Petersburg, Russia
	\and
D.~Li \at
	Key Laboratory of Dark Matter and Space Astronomy, Purple Mountain Observatory, CAS, Nanjing, 210033, China\\
	State Key Laboratory of Space Weather, Chinese Academy of Sciences, Beijing, 100190, China\\
	CAS Key Laboratory of Solar Activity, National Astronomical Observatories, Beijing, 100101, China
	\and
N. Magyar \at 
	Centre for Fusion, Space and Astrophysics, Physics Department, University of Warwick, Coventry CV4 7AL, UK
	\and
G. Nistic\`o \at
              Dipartimento di Fisica, Universit\`a della Calabria, via P. Bucci, Cubo 31C, 87036, Arcavacata di Rende (CS), Italy
	\and
D.J. Pascoe \at 
	Centre for mathematical Plasma Astrophysics, Department of Mathematics, KU~Leuven, Celestijnenlaan 200B bus 2400, B-3001 Leuven, Belgium
	\and
A.~K. Srivastava \at 
	Department of Physics, Indian Institute of Technology (BHU), Varanasi-221005, India
	\and
J. Terradas \at
	Departamet de F\'isica, Universitat de les Illes Balears, 07122, Palma de Mallorca, Spain,\\ 
	Institut d'Aplicacions Computacionals de Codi Comunitari (IAC), Universitat de les Illes Balears, 07122 Palma de Mallorca, Spain
	\and
S. Vasheghani Farahani \at 
	Department of Physics, Tafresh University, Tafresh 39518 79611, Iran
	\and
G. Verth \at
	Plasma Dynamics Group, School of Mathematics and Statistics, University of Sheffield, Sheffield, UK
	\and
D.~Yuan \at
	Institute of Space Science and Applied Technology, Harbin Institute of Technology, Shenzhen, Guangdong 518055, China
	\and
I.~V.~Zimovets \at
	Space Research Institute of the Russian Academy of Sciences (IKI RAS), 84/32 Profsoyuznaya St., Moscow 117997, Russia
}

\date{Received: date / Accepted: date}

\maketitle

\begin{abstract}
Kink oscillations of coronal loops,{ i.e., standing kink waves,} is one of the most studied dynamic phenomena in the solar corona. The oscillations are excited by impulsive energy releases, such as low coronal eruptions. Typical periods of the oscillations are from a few to several minutes, and are found to increase linearly with the increase in the major radius of the oscillating loops. It clearly demonstrates that kink oscillations are natural modes of the loops, and can be described as standing fast magnetoacoustic waves with the wavelength determined by the length of the loop. Kink oscillations are observed in two different regimes. In the rapidly decaying regime, the apparent displacement amplitude reaches several minor radii of the loop. The damping time which is about several oscillation periods decreases with the increase in the oscillation amplitude, suggesting a nonlinear nature of the damping. In the decayless regime, the amplitudes are smaller than a minor radius, and the driver is still debated. The review summarises major findings obtained during the last decade, and covers both observational and theoretical results. Observational results include creation and analysis of comprehensive catalogues of the oscillation events, and detection of kink oscillations with imaging and spectral instruments in the EUV and microwave bands. Theoretical results include various approaches to modelling in terms of the magnetohydrodynamic wave theory.  Properties of kink oscillations are found to depend on parameters of the oscillating loop, such as the magnetic twist, stratification, steady flows, temperature variations and so on, which make kink oscillations a natural probe of these parameters by the method of magnetohydrodynamic seismology. 
\end{abstract}

\tableofcontents

\section{Introduction}
\label{intro}

One of the key features which makes the plasma of the solar corona different from other natural plasma environments, for example, the Earth's magnetosphere, is a pronounced field-aligned filamentation of the macroscopic parameters, such as the density and temperature. In particular, coronal active regions consist of dense hot plasma loops which are clearly observed as bright structures in the EUV and soft X-ray, and sometimes in microwaves \citep[see][for a comprehensive review]{2014LRSP...11....4R}. The loops are anchored at the chromosphere at locations known as footpoints. Usually both footpoints of a loop are located in the same active region, while sometimes loops link different active regions. The typical minor radii of coronal loops are a few Mm, while major radii reach several hundred Mm. The longest, transequatorial loops link active regions situated in different hemispheres. Individual loops are separated by apparently more rarefied plasma regions. The density contrast inside and outside a loop is from a few tens percent up to one hundred times or more in flaring active regions. Coronal loops, being the main building block of active regions, have been remaining a puzzling, intensively debated plasma structure for several decades. The key open questions are the mechanisms responsible for their appearance, typical spatial scales and their lifetimes of several hours or longer; why the loop's minor radius is usually constant along the loop; whether the loop has a fine structure in density and/or temperature, and if yes, which one (a bundle of threads, a set of co-axial shells, or something else). As dense and hot plasma objects, coronal loops are directly linked with the coronal heating problem. {In addition, equilibrium solutions describing coronal loops remain unknown.} 

Coronal loops are observed to be dynamic objects. The brightness determined by the temperature and density, and geometry, may vary in time. In addition, there could be field-aligned plasma flows of various nature, and wave motions. Repetitive transverse displacements of the loop axis are called kink oscillations. Kink oscillations of coronal loops, {i.e., standing kink waves or oscillatory bouncing of the loops,} predicted theoretically in 1970s \citep{1975IGAFS..37....3Z, 1976JETP...43..491R}, and discovered observationally in the high-resolution EUV data {obtained with the Transition Region And Coronal Explorer (TRACE ) \citep{1999SoPh..187..229H}} in late 1990s \citep{1999ApJ...520..880A, 1999Sci...285..862N}, have become one of the most intensively studied magnetohydrodynamic (MHD) wave phenomena in the solar corona. The major breakthrough in the observational study of kink oscillations is associated with their detection with the Atmospheric Imaging Assembly (AIA) on the Solar Dynamics Observatory (SDO) spacecraft. The first detection of kink oscillations of coronal loops with AIA was reported by \citep{2011ApJ...736..102A}. Since that time, several hundred kink oscillation events have been found in AIA data, usually in the 171~\AA\ passband \citep{2019ApJS..241...31N}. 

Since the discovery, our understanding of kink oscillations has evolved through several observational and theoretical advances. Almost immediately after the first detection of decaying oscillations, the decay was linked with resonant absorption \citep{2002ApJ...577..475R, 2002A&A...394L..39G}, i.e., a linear transformation of the observed collective transverse movements of the loop as a whole to unresolved torsional movements localised at certain magnetic surfaces. That seminal result allowed the research community to use the wealth of the theoretical knowledge on resonant absorption for the interpretation of observations. \cite{2002ApJ...576L.153O} used the empirical scaling of the observed damping times and periods to demonstrate its consistency with the resonant absorption theory. \cite{2004A&A...424.1065V} demonstrated that the effect of the curvature of the loop on kink oscillations and their resonant absorption is weak, and hence justified the applicability of the straight cylinder model. On the other hand, alternative interpretations {were proposed too, which link kink motions with either eigenmodes of coronal arcades \citep{2014ApJ...784..103H}, or with a propagating fast magnetoacoustic wave train \citep{2003SoPh..218...17U, 2005A&A...441..371T}. }

\cite{2005SoPh..229...79D} developed a mathematical formalism for taking into account a non-uniformity of the equilibrium plasma parameters along the loop. \cite{2008ApJ...687L.115T} demonstrated numerically that the sheared layer between the loop and the external medium formed by the transverse movements of the loop is subject to {Kelvin-Helmholtz instability (KHI)}. \cite{2019FrP.....7...85A} established that resonant absorption further enhances the instability.
Resonant absorption was shown to work effectively in a cylinder with an irregular cross-section \citep{2011ApJ...731...73P}. It was established that the oscillation damping could have either an exponential or Gaussian profile \citep{2012A&A...539A..37P}. Analysing a catalogue of decaying kink oscillations, \cite{2015A&A...577A...4Z} established empirically that the oscillations are preferentially excited by a displacement of the loop by a low coronal eruption. In addition, the linear correlation of oscillation periods and lengths of the oscillating loops unequivocally demonstrated that kink oscillations are eigenmodes of the loops. Moreover, it was found that the oscillation quality-factor is proportional to the initial amplitude to the power of minus two thirds \citep{2016A&A...590L...5G}. 

In addition to the decaying kink oscillations, a decayless regime of kink oscillations was observationally found. The first observational evidence is reported by \cite{2012ApJ...751L..27W} as an event of transverse oscillations of EUV loops, growing in amplitude. Further observations showed that a loop could oscillate in either decaying or decayless regime in different time intervals, while the oscillation period remains the same \citep{2013A&A...552A..57N}. \cite{2015A&A...583A.136A} found that decayless kink oscillations are ubiquitous in coronal active regions, and occur even in the absence of solar flares, eruptions or other impulsive energy releases. Oscillation periods of decayless oscillations correlate linearly with the lengths of the oscillating loops, and hence the oscillations are natural modes of the loop. \citet{2016A&A...591L...5N} demonstrated that in the statistical distribution of decayless oscillation amplitudes there are no peaks associated with a certain period, which disproved their excitation by 5-min or 3-min oscillations in the lower layers of the solar atmosphere. It was suggested that the oscillation damping by resonant absorption could be compensated by either steady or random flows near footpoints of the loop. In the former case, the oscillations are actually self-oscillations with the amplitudes determined by the balance between energy losses and gains. This list of major new results is by all means incomplete and not exclusive. In parallel, there are intensive studies of propagating kink waves in loops and streamers, see, for example, \citep{2009ApJ...697.1384T} and \citep{2014SoPh..289.1663C}, respectively, which are out of scope of this review. An exception was made for kink oscillations of hot plasma jets, as it is often  difficult to disentangle the phase speed and the speed of the equilibrium plasma flow, and hence the kink oscillatory movements could be standing.  

The confident and frequent detection of kink oscillations makes them a promising probe of physical conditions in the coronal loops hosting them, i.e., kink oscillations are a tool for coronal seismology. Kink oscillations are used for estimating the absolute value of the magnetic field \citep[e.g.,][]{2001A&A...372L..53N} and its variation along the loop \citep{2008A&A...486.1015V}, the spatial scale of the density stratification \citep[e.g.,][]{2005ApJ...624L..57A}, and give information about transverse profiles of the Alfv\'en speed and mass density \citep{2003ApJ...598.1375A}, including its steepness \citep{2016A&A...589A.136P}. 
More recently, it was demonstrated that the ubiquitous kink oscillations observed in coronal active regions during the quiet time periods, allow for the seismological mapping of the Alfv\'en speed and magnetic field \citep[e.g.,][]{2019ApJ...884L..40A}. Kink oscillations in loops of the sigmoid shape have been shown to carry the information about the free magnetic energy associated with non-potential field geometry \citep[e.g.,][]{2020ApJ...894L..23M}. 

In this review we concentrate mainly on the results obtained in the last ten years, i.e., since 2010.  Previous comprehensive reviews which address kink oscillations of coronal loops include \cite{2009SSRv..149...31A, 2009SSRv..149....3A, 2009SSRv..149..199R, 2009SSRv..149..255T, 2009SSRv..149..299V, 2012RSPTA.370.3193D, 2014SoPh..289.3233L}. A possible role of kink oscillations in the coronal heating problem, which has been intensively studied since their first detection, is reviewed in \citet{2020SSRv..216..140V}.  

The review is organised as follows. In Section~\ref{sec:scal} we describe empirical properties of decaying kink oscillations, including relationships between their parameters and also parameters of the host loops.
In Section~\ref{sec:zser} we review the standard model of kink oscillations of a plasma cylinder.
Section~\ref{sec:ra} is dedicated to linear mechanisms for the damping of kink oscillations. 
Section~\ref{sec:twisted} addresses the effect of the magnetic twist.
Section~\ref{sec:mirr} presents kink oscillations of a current-carrying loop, caused by the magnetic mirror force.
In Section~\ref{sec:flow} we discuss kink oscillations of in the presence of parallel shear plasma flows, including negative energy wave instabilities. 
In Section~\ref{sec:cool}, kink oscillations in loops undergoing plasma cooling are considered. 
In Section~\ref{sec:p1_2p2} we review period ratios of different parallel harmonics, their detection in observations, and seismological inferences.
In Section~\ref{sec:nonlin} nonlinear effects are discussed. 
Section~\ref{sec:excit} describes excitation mechanisms. 
In Section~\ref{sec:decless} we consider observations and theoretical models of decayless kink oscillations. 
Section~\ref{sec:radio} presents possible detections of kink oscillations in the radio band including microwaves. 
In Section~\ref{sec:conc} we summarise outstanding problems and draw conclusions.  

\section{Empirical properties of decaying kink oscillations}
\label{sec:scal}

\begin{figure*}
\centering
\includegraphics[width=0.37\textwidth]{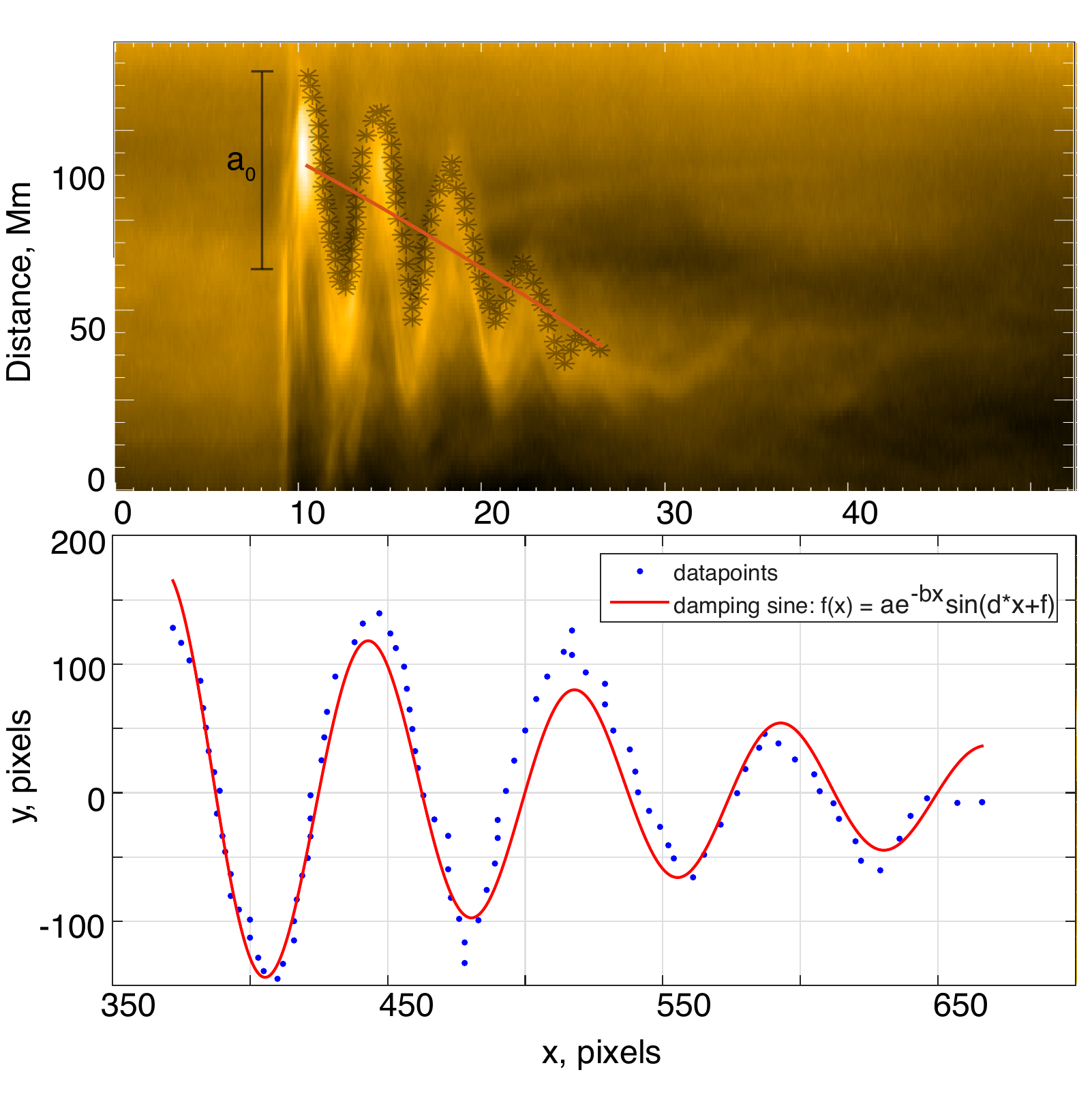}
\includegraphics[width=0.52\textwidth]{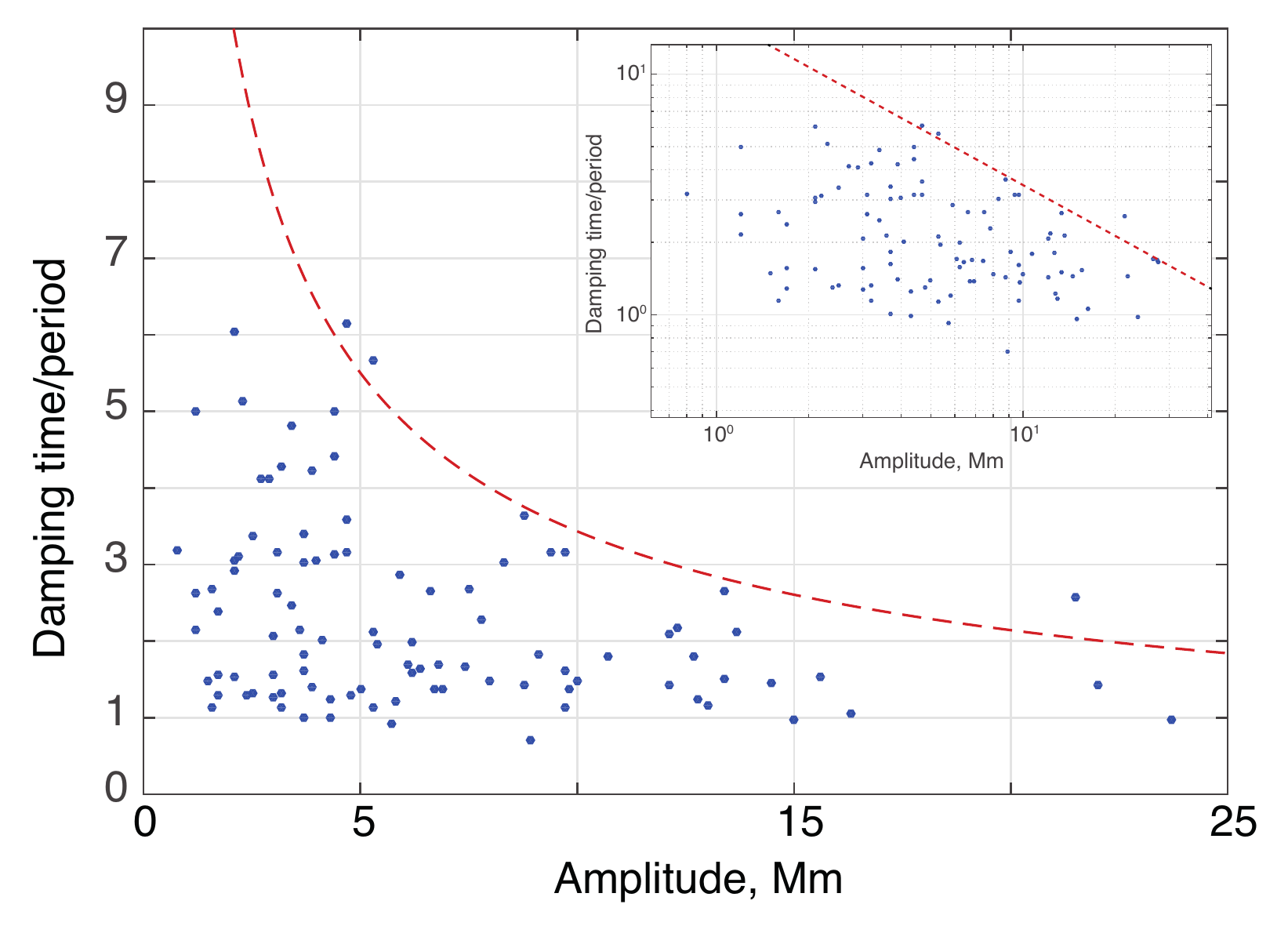}
 \caption{Left: Example of the determination of the oscillation parameters. (Top panel): the semi-transparent black stars show the instantaneous positions of the boundary of the oscillating loop, picked by hand for tracking the oscillation.
The power-law background trend is shown by the red line. The initial
displacement $a_0$ is also shown. (Bottom panel): the blue dots show the
detrended signal. The red curve shows the best-fitting exponentially decaying
harmonic oscillation. 
Right: Quality factor of 101 kink oscillations of coronal loops plotted against the projected oscillation amplitude.
The inset shows the same dependence in the log--log plot. The dotted and dashed curves show the approximation of the upper-right boundary of the data clouds with a linear function in the log--log plot and the corresponding power-law function in the linear plot, respectively.
Figures are taken from \cite{2019ApJS..241...31N}.  }
\label{fig:nech1}
\end{figure*}

The abundant observational detection of kink oscillations of coronal loops with AIA allows for the search for correlations between different parameters of the oscillations and properties of the oscillating loops. 
To the date, the most comprehensive catalogue of impulsively excited \textit{decaying} kink oscillations\footnote{We use this term to distinguish them from decayless kink oscillations which are not seen to result from an impulsive event. The decayless regime is discussed in Section~\ref{sec:decless}.}, which  includes information about 223 oscillating loops in 96 oscillation events has been compiled by \cite{2019ApJS..241...31N}. Statistical properties of kink oscillations were determined by approximating the evolution of detrended transverse displacements of the loop's segments by an exponentially decaying harmonic function (see Figure~\ref{fig:nech1}). The oscillation periods range from 1 to 28 minutes, with 74\% of the detections in the range of 2--10 minutes. About 90\% of the oscillations have the apparent displacement  amplitude in the plane of the sky, i.e., the amplitude reduced by the projection effect, in the range of 1--10~Mm. The lengths of oscillating loops are 70--600~Mm. Kink oscillations in shorter loops may be missing because of insufficient time  resolution. The typical apparent displacement amplitude is about 1\% of the loop length, although it is higher in terms of the loop minor radius.

\begin{figure*}
\centering
\includegraphics[width=0.45\textwidth]{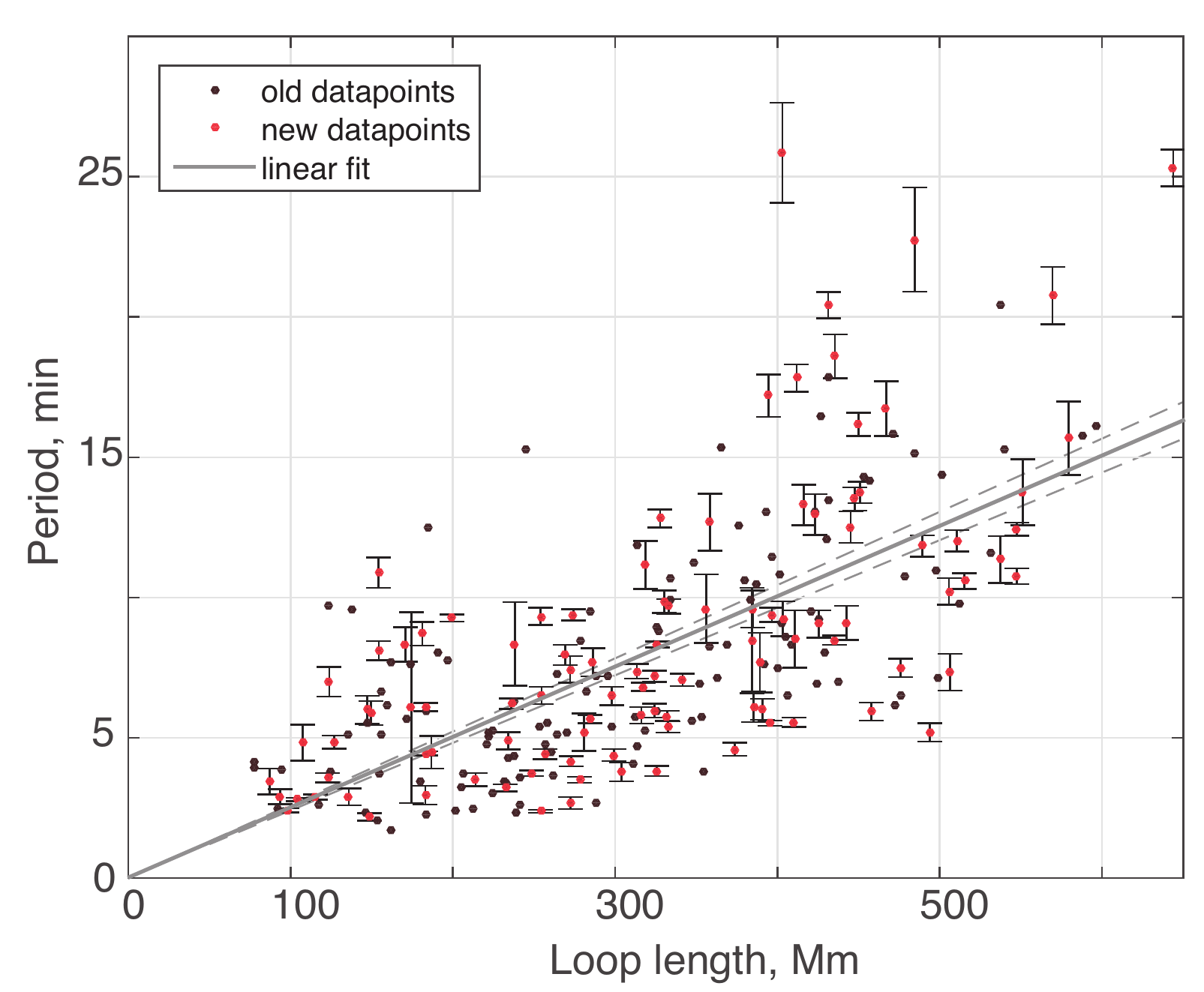}
\includegraphics[width=0.45\textwidth]{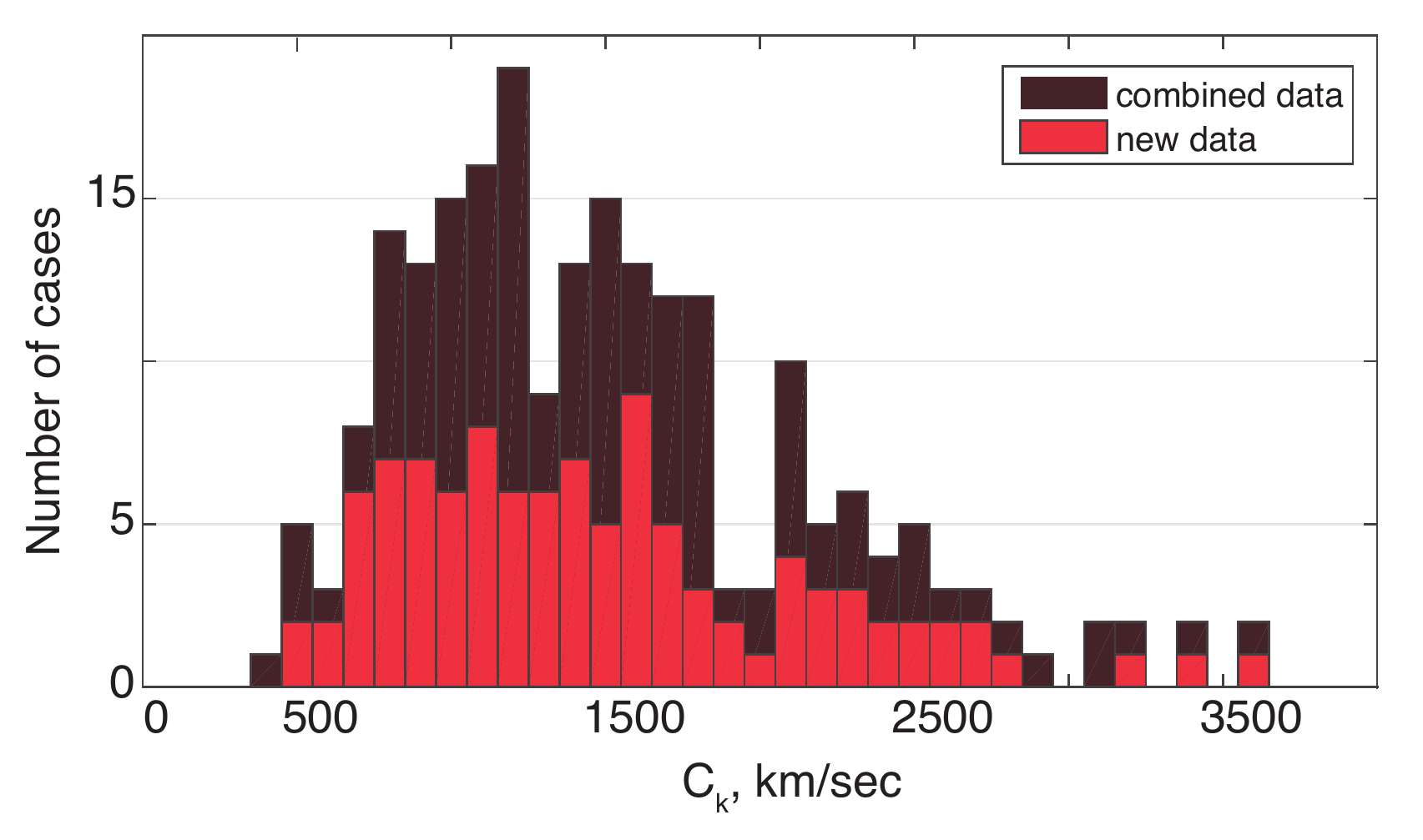}
\caption{Left: Empirical scaling of kink oscillation periods in the decaying regime with the length of the oscillating loop. The \lq\lq old\rq\rq\ and \lq\lq new\rq\rq\ data denote the data summarised in the old \citep{2015A&A...577A...4Z, 2016A&A...585A.137G} and updated \citep{2019ApJS..241...31N} catalogues. The solid line indicates the best-fitting linear function, with the error bars shown by the dashed curves. Right: The histogram of kink speeds estimated by their periods and lengths of the oscillating loops. Figures are taken from \cite{2019ApJS..241...31N}. }
\label{fig:nech2}
\end{figure*}

An important empirical result is the increase in the kink oscillation period with the length of the oscillating loop, demonstrated by Figure~\ref{fig:nech2}. Such a scaling, together with the synchronous displacement of different segments of the oscillating loop, seen in movies made by time sequences of 2D EUV images, indicates that kink oscillations are normal modes of the loop. In a vast majority of observed cases, the displacement has a single maximum, i.e., an antinode, at the loop top, and two nodes, at footpoints. Hence, kink oscillations are fundamental normal modes in those cases. In some rare cases, second harmonics are detected too, with the third node at the loop top, and two antinodes in the legs \citep[see, e.g.,][]{2005ApJ...624L..57A, 2013ApJ...765L..23A, 2016A&A...593A..53P, 2017ApJ...842...99L}. The third harmonic has been detected too \citep{2019A&A...632A..64D}.  The harmonics discussed here are harmonics in the axial direction, along the field. There could also be radial harmonics. The linear dependence of the oscillation period $P_\mathrm{k}$ on the loop length $L$ allows one to estimate the kink wave speed which for the fundamental harmonic is $C_\mathrm{k} = 2L/P_\mathrm{kink}$. The histogram of the empirically determined values of $C_\mathrm{k}$ is shown in the right panel of Fig.~\ref{fig:nech2}. The average value of  the kink speed is $C_\mathrm{k} = 1328\pm 53$~km\,s$^{-1}$. The value of $C_\mathrm{k}$ is determined by the densities and magnetic fields in the loop and its environment (see Section~\ref{longwave}). The broadness of the distribution reflects the broadness of the physical conditions in the oscillating loops.

Analysis of the kink oscillation damping performed under the assumption of an exponential decay confirmed the linear scaling of the damping time with the oscillation period, established by \cite{2002ApJ...576L.153O} on a much shorter dataset. The linear correlation coefficient was estimated as 0.640. On average, the exponential damping time is $1.79\pm 0.14$ the oscillation period. A surprising finding was the dependence of the oscillation quality factor $Q$ defined as the ratio of the exponential damping time over the oscillation period on the {initial} amplitude $A$  \citep{2016A&A...590L...5G} (see their Figure~\ref{fig:nech1}~right). The amplitude is defined as the projected initial displacement. As the angle between the oscillation plane and the line-of-sight is usually unknown, the apparent amplitude is lower than its actual value. Thus, the actual scaling should be determined by the outer boundary of the data cloud in the figure, i.e., by several points only. It was found that $\log Q  =  - (0.68 \pm 0.13)\log (A/\mathrm{Mm}) + (2.80 \pm 0.37)$. Thus, the quality factor scales as $A^{-2/3}$.


\subsection{Spectroscopic observations of kink oscillations of coronal loops} 
\label{sec:spec}

A promising new trend in the observational study of kink oscillations of coronal loops is their detection in coronal emission spectra. Periodic movements of a plasma loop along the line-of-sight appear as Doppler shift oscillations of the corresponding spectral lines. The lack of simultaneous modulation of the emission intensity could be taken as the evidence of a kink oscillation which is weakly compressive in the long-wavelength regime. However, in the optically thin regime, kink oscillations could modulate the spectral line intensity by the modulation of the column depths along the line-of-sight \citep[see, e.g.,][]{2003A&A...397..765C, 2016ApJS..223...23Y, 2017ApJ...836..219A}. {The amplitude of the intensity modulation can readily reach several percent \citep{2009ApJ...698..397V}. }

\begin{figure}
\centering
\includegraphics[width=\textwidth]{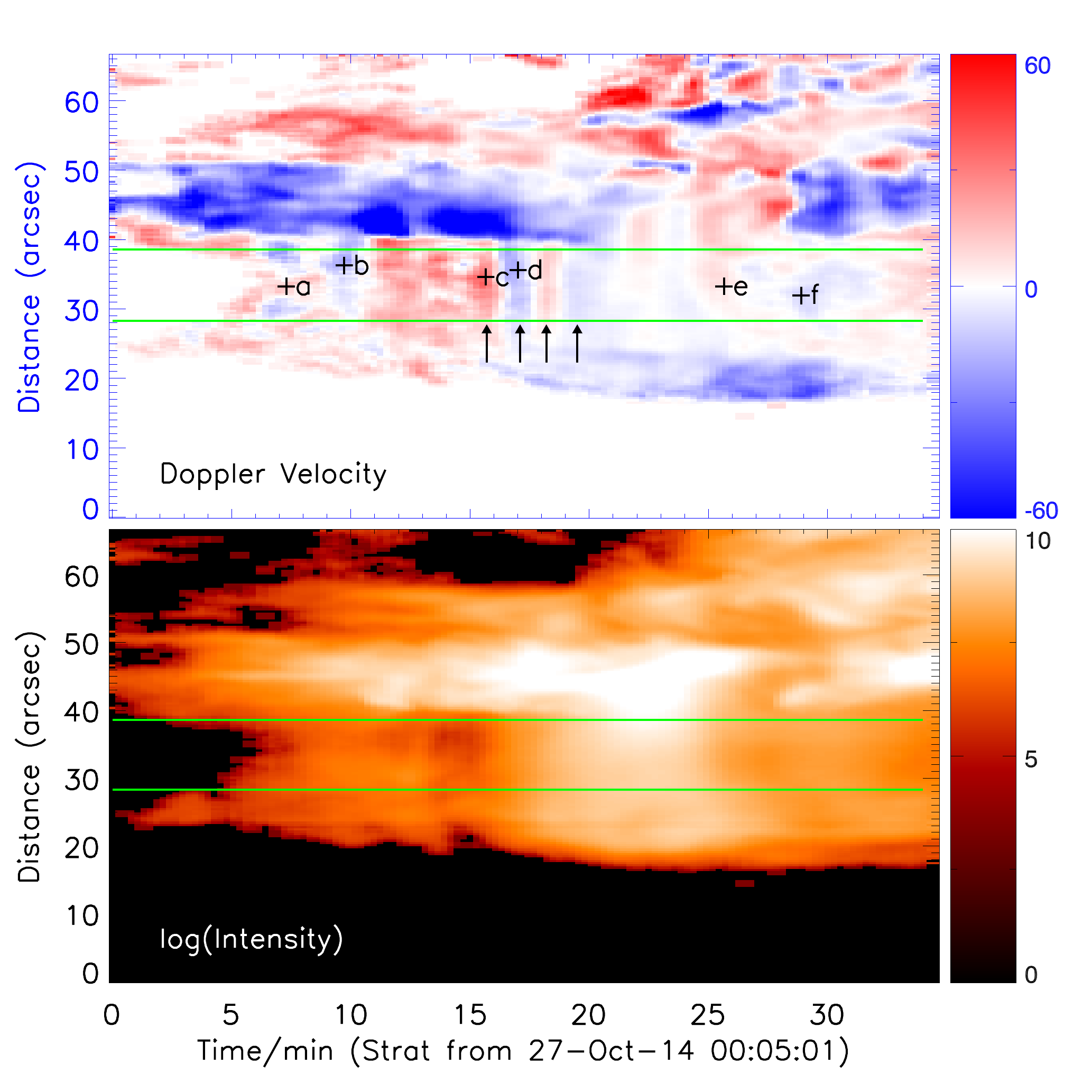}
\caption{Upper: Spacetime image of Doppler velocities at \ion{Fe}{XXI} line. Lower: Spacetime image of line-integrated intensities with a
logarithmic brightness scale. Two green lines are the bounds of the selected loop-top region. The colour bars show the line-of-sight speed in \kms. Adaptation of Figure 3 in \cite{2017ApJ...849..113L} 
\label{fig:li2017}}
\end{figure}

In contrast to the ubiquitous observations of kink oscillations with EUV imaging telescopes, detections of kink oscillations in spectral data remain sporadic. This is probably due to the line-of-sight  superposition, which strongly affects the Doppler velocity measurements as demonstrated by forward modelling of numerical simulations of kink oscillations \citep{2012ApJ...746...31D, 2017ApJ...836..219A}. \cite{2017ApJ...849..113L} presented a detailed study of the decay oscillation in a hot flare loop on 27 October 2014. Periodic changes from red to blue shifts were clearly seen in Doppler velocities of \ion{Fe}{XXI} (Figure~\ref{fig:li2017}, top panel), with a dominant period of about 3.1~minutes. Such an oscillation was not seen in the line-integrated intensity, as seen in Figure~\ref{fig:li2017}, bottom panel. The authors concluded that the hot flare loop was most likely oscillating in a weakly-compressive MHD mode, such as the standing kink oscillation. The observational results were consistent with those from MHD modelling which simulated the manifestation of a standing kink oscillation of coronal loops in Doppler velocities \citep[e.g.,][]{2015A&A...581A.137C, 2016ApJS..223...23Y}. A magnetic field strength of around 68~G was estimated in the loop. In another event, on 10 September 2014 \citet{2016ApJ...823L..19Z} found a standing kink oscillation during the precursor phase of a flare. The precursor oscillation with alternating red and blueshifts was seen in Doppler velocities of \ion{Fe}{XXI}. The oscillation period was $\sim$280~s, and the oscillation lasted for about 13 minutes. The authors stated that the precursor oscillation was triggered by a periodic energy release via magnetic reconnection. 

\section{The Zaitsev--Stepanov--Edwin--Roberts model}
\label{sec:zser}

The model that has proven to be standard in the kink oscillation modelling is based on linear MHD perturbations of a plasma cylinder  surrounded by a plasma with different properties, with both the internal and external plasmas being penetrated by a straight magnetic field. Such a cylinder could also be considered as a magnetic flux tube. Because of the frozen-in condition, its displacement is accompanied by perturbation of the magnetic field. The cylinder is a one-dimensional non-uniformity of MHD parameters of the plasma, such as the temperature, mass density, the magnetic field, and, possibly, stationary flows along the boundary of the cylinder. Equilibrium quantities must satisfy the magnetostatic force balance condition. 
Kink perturbations of such a cylinder are transverse, and hence can be of linear, elliptical or circular polarisation. In the cylindrical frame of reference with the $z$ axis coinciding with the cylinder's axis, the kink perturbation has the azimuthal wave number $|m|=1$, with different signs of $m$ corresponding to the opposite senses of the circular polarisation. The key difference of the kink mode from MHD modes with $m\neq 1$ is the displacement of the axis of the cylinder. 

Dispersion relations for coronal kink oscillations in the important case of the radial non-uniformity given by a step function, and a straight and uniform magnetic field parallel to the axis of the cylinder, and in the absence of equilibrium mass flows, were independently derived by \cite{1975IGAFS..37....3Z} and \cite{1983SoPh...88..179E}. We shall refer to this model as the ZSER model. The ZSER dispersion relation is a transcendental algebraic equation 
\begin{equation}
\label{zser}
\frac{\rho_{0\mathrm{i}}}{\rho_{0\mathrm{e}}}\,\frac{ (C_{\mathrm{Ai}}^{2}-V_\mathrm{ph}^{2})}{( C_{\mathrm{Ae}}^{2}-V_\mathrm{ph}^2)} \frac{\kappa_\mathrm{e}}{n_\mathrm{i}}\frac{K^{\prime}_{1}\,(\kappa_\mathrm{e}\,a)}{K_{1}\,(\kappa_\mathrm{e}\,a)} = \frac{J^{\prime}_{1}\,(n_\mathrm{i}\,a)}{J_{1}\,(n_\mathrm{i}\,a)},
\end{equation}
where $\rho_{0\mathrm{i}}$ and $\rho_{0\mathrm{e}}$ respectively represent the equilibrium internal and external mass densities; $a$ is the radius of the cylinder in equilibrium, the indices $\mathrm{i}$ and $\mathrm{e}$ indicate the internal or external media; and the functions $J_1$ and $K_1$ represent the Bessel function of the first kind and the modified Bessel of the second kind, both being of the first order, the prime denotes the derivative with respect to the argument. The Bessel functions describe the radial structure of the perturbations. 
The effective radial wave number is 
\begin{eqnarray}
\kappa_\alpha^2=\frac{(C_{\mathrm{A}\alpha}^2-V_\mathrm{ph}^2) (C_{\mathrm{s}\alpha}^2-V_\mathrm{ph}^2)}{(C_{\mathrm{A}\alpha}^2+C_{\mathrm{s}\alpha}^2)(C_{\mathrm{t}\alpha}^2-V_\mathrm{ph}^2)}k_z^2,
\label{mexp}
\end{eqnarray}
where $k_z$ is the axial wave number, $V_\mathrm{ph} = \omega/{k_z}$ is the phase speed along the axis of the cylinder, note that for the internal medium $n_\mathrm{i}^{2}= - \kappa_\mathrm{i}^2>0$; $C_{\mathrm{A}\alpha}$ and $C_{\mathrm{s}\alpha}$ respectively represent the Alfv\'en and sound speeds; and the index $\alpha$ stands for either $\mathrm{i}$ or $\mathrm{e}$. In typical coronal loops, $C_\mathrm{Ai} < C_\mathrm{Ae}$ and $\max\{C_\mathrm{si}, C_\mathrm{se}\} < \min\{C_\mathrm{Ai}, C_\mathrm{Ae}\}$ \citep[e.g.][]{2014LRSP...11....4R}.
 The tube speed is $C_{\mathrm{t}\alpha} = C_{\mathrm{s}\alpha}C_{\mathrm{A}\alpha}/(C_{\mathrm{s}\alpha}^{2}+C_{\mathrm{A}\alpha}^{2})^{1/2}$. As typically in coronal loops the plasma parameter $\beta$ is much less than unity, the cold plasma approximation, $\beta = 0$, with $C_{\mathrm{s}\alpha} = C_{\mathrm{t}\alpha} = 0$, is often used. In standing kink oscillations, the axial wave numbers are discrete.  
 
The ZSER dispersion model describes an infinite number of radial harmonics of kink oscillations. In the long wavelength regime, $k_za \ll 1$, all of them except the lowest one are leaky, i.e., their radial structure outside the cylinder is given by a Henkel function which appears because of the negative sign of $m_\mathrm{e}^2$. In contrast, the fundamental radial harmonic of the kink mode is trapped, i.e., its radial structure outside the cylinder is evanescent for all values of $k_z$. 
The long wavelength limit is often referred to as the thin tube limit in the consideration of the lowest radial harmonics. 
In this limit, its phase speed can be approximated as
\begin{equation}
\label{longwave}
V_\mathrm{ph} \approx C_\mathrm{k} \left[ 1 - {\cal Q}  K_0({\cal T} |k_z|a) (k_za)^2 \right],
\end{equation}
where
$$
{\cal Q} = \frac{\rho_\mathrm{0i} \rho_\mathrm{0e}{\cal T}^2 (C_\mathrm{Ae}^2 - C_\mathrm{Ai}^2)}{2 C_\mathrm{k}^2 (\rho_\mathrm{0i} + \rho_\mathrm{0e})^2}, \mbox{\ \ \ \ } {\cal T} = \left( 1 -  \frac{C_\mathrm{k}^2}{C_\mathrm{Ae}^2}\right)^{1/2},
$$
and
$$
C_\mathrm{k} = \left( \frac{\rho_\mathrm{0i}C_\mathrm{Ai}^2 +  \rho_\mathrm{0e}C_\mathrm{Ae}^2}{\rho_\mathrm{0i} + \rho_\mathrm{0e}}\right)^{1/2},
$$ 
which is a density-weighted average of the internal and external Alfv\'en speeds, is the kink speed first defined by \cite{1976JETP...43..491R}. 
The kink speed is always lower than $C_\mathrm{Ae}$. The phase speed decreases with the decrease in the axial wavelength. 
In the limit of a zero-$\beta$ plasma, for which the internal and external magnetic fields in the cylinder are equal,  the phase speed of the kink mode in the long wavelength limit reduces to
\begin{equation}
C_\mathrm{k} = C_\mathrm{Ai} \sqrt{\frac{2 \zeta}{\zeta + 1}},
\label{eq:ck}
\end{equation}
where $C_\mathrm{Ai}$ is now the Alfv\'en speed at the axis of the cylinder, and $\zeta = \rho_\mathrm{0i} / \rho_\mathrm{0e}$ is the ratio of the internal and external plasma densities in the equilibrium.

\begin{figure*}
\centering
\includegraphics[width=0.9\textwidth]{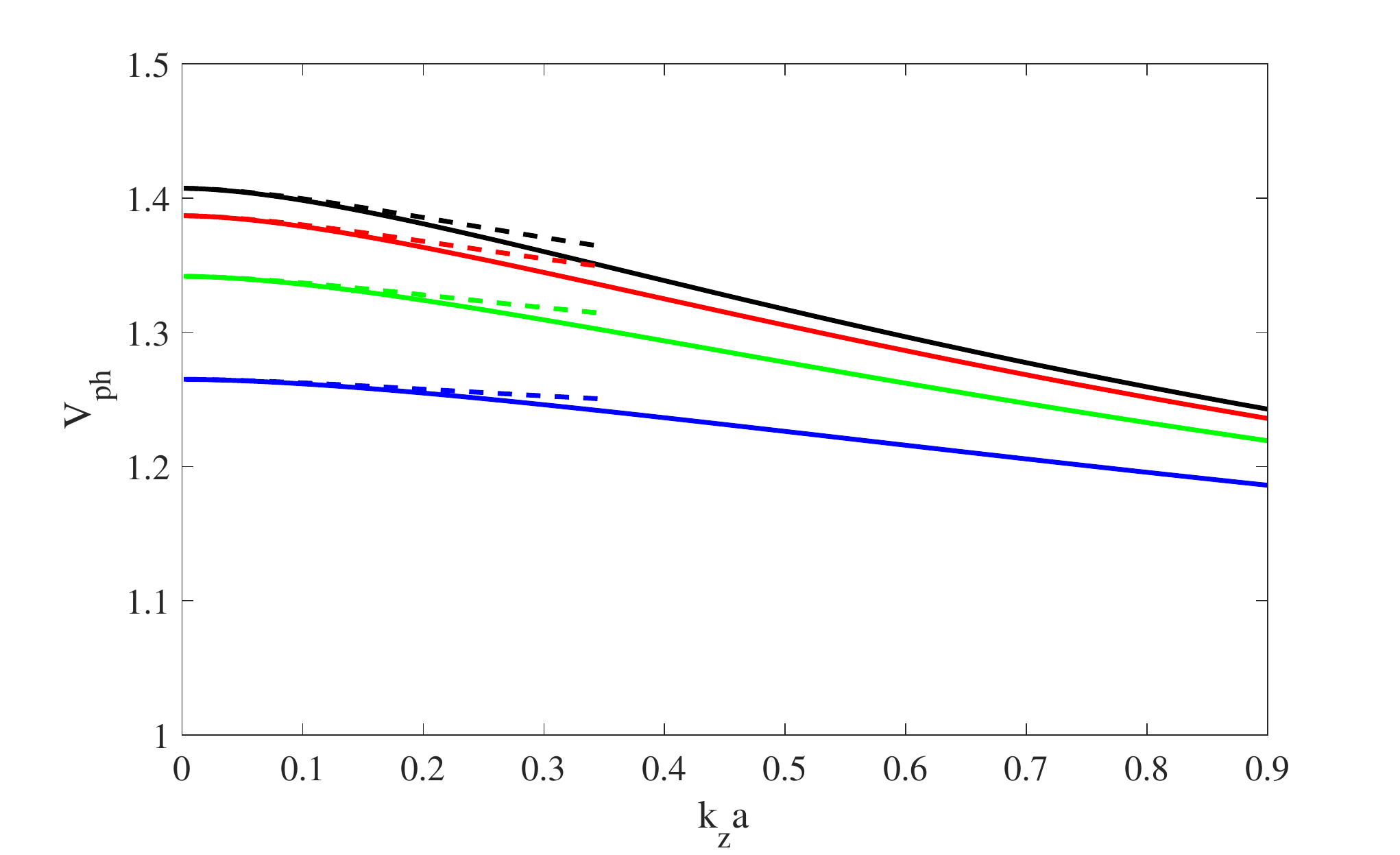}
\caption{The phase speed of the fundamental {radial harmonic of the} kink mode of a plasma cylinder as a function of the axial wave number, in the zero plasma-beta limit. The black, red, green and blue curves correspond to $C_\mathrm{Ae} = 10 C_\mathrm{Ai}$, $C_\mathrm{Ae} = 5 C_\mathrm{Ai}$, $C_\mathrm{Ae} = 3C_\mathrm{Ai}$ and $C_\mathrm{Ae} = 2C_\mathrm{Ai}$, respectively. The solid curves show solutions to the full dispersion relation, while the dashed curves show the asymptotic solutions. The phase speed is normalised to $C_\mathrm{Ai}$, while the axial wave number to the cylinder's radius.}
\label{kinkdisp}
\end{figure*}

Figure~\ref{kinkdisp} shows the phase speed as a function of the axial wave number for several different values of the density contrast in the cylinder. The full and asymptotic solutions are in a good agreement with each other. As in coronal conditions $m_\mathrm{i}^{2} < 0$, the perturbation inside the cylinder has an oscillatory structure, and hence the kink wave can be considered as a fast magnetoacoustic wave which propagates obliquely to the axis of the cylinder and experiences reflections at the boundary. At footpoints the kink speed experiences a very sharp and large jump, and the magnetic field is line-tied, which leads to the reflection of kink waves back to the coronal part of the loop. Thus, in loops, there could be standing kink waves with discrete values of the axial wave number $k_z$, referred to as parallel kink harmonics. 

If dispersive effects given by Eq.~(\ref{longwave}) are neglected, the oscillation period for the $n$th parallel harmonic standing kink mode is then
\begin{equation}
P_\mathrm{kink}^{(n)} \approx \frac{2L}{n C_\mathrm{k}}
\label{eq:pk}
\end{equation}
where $L$ is the length of the coronal loop. This definition of the kink period corresponds to the thin tube thin boundary (TTTB) approximation whereas parametric studies \citep{2004ApJ...606.1223V, 2014ApJ...781..111S, 2019FrASS...6...22P} find that the period of oscillation also depends on the width of the boundary layer when it is sufficiently large. The period of the fundamental parallel harmonic, with $n=1$, is $P_\mathrm{kink}$. One can introduce a corresponding cyclic frequency, $\omega_{\rm k} = 2 \pi/ P_\mathrm{kink}$. If the kink speed varies along the cylinder, ratios of the parallel harmonic periods could not be obtained from Eq.~(\ref{eq:pk}) (see Section~\ref{sec:p1_2p2}).

A model similar to ZSER has been constructed for a plane plasma slab \citep{1982SoPh...76..239E}, see also \citet{2015ApJ...814...60Y} and references therein for recent developments of the model. The key differences between the plane analogue and the ZSER model  are the trigonometric eigenfunctions instead of the Bessel ones, the phase speed of the kink wave in the long wavelength limit becoming the external Alfv\'en speed instead of the kink speed, and the exponential decrease in the perturbation amplitude in the external medium in contrast with the super-exponential decay. In a slab with a diffused boundary, the lower radial harmonic of the kink mode is trapped for all axial wavelengths, {similarly to the cylindrical case}. In the corona, the plane model is applicable to kink oscillations in a number of plasma non-uniformities, for example, in streamers and pseudo-streamers, plane jets, prominence slabs, etc. 

Exact analytical solutions describing kink oscillations in a cylinder with a smooth transverse profile, e.g., similar to the Epstein profile in the slab geometry, have not been found. The collective nature of kink oscillations suggests that the dependence of their periods on the specific shape of the transverse profile should be weak, at least in the linear regime. However, the effect of the transverse profile on the oscillation damping is crucial. The smoothness of the profile determines the effectiveness of the irreversible transfer of the kink oscillation energy into azimuthal movements of individual magnetic surfaces in the vicinity of a resonant layer where the local phase speed of the kink wave coincides with the local Alfv\'en speed (the \lq\lq Alfv\'enic resonance\rq\rq). This phenomenon is known as resonant absorption.

\section{Damping of kink oscillations by resonant absorption} 
\label{sec:ra}
\begin{figure*}
\centering
\includegraphics[width=0.45\textwidth]{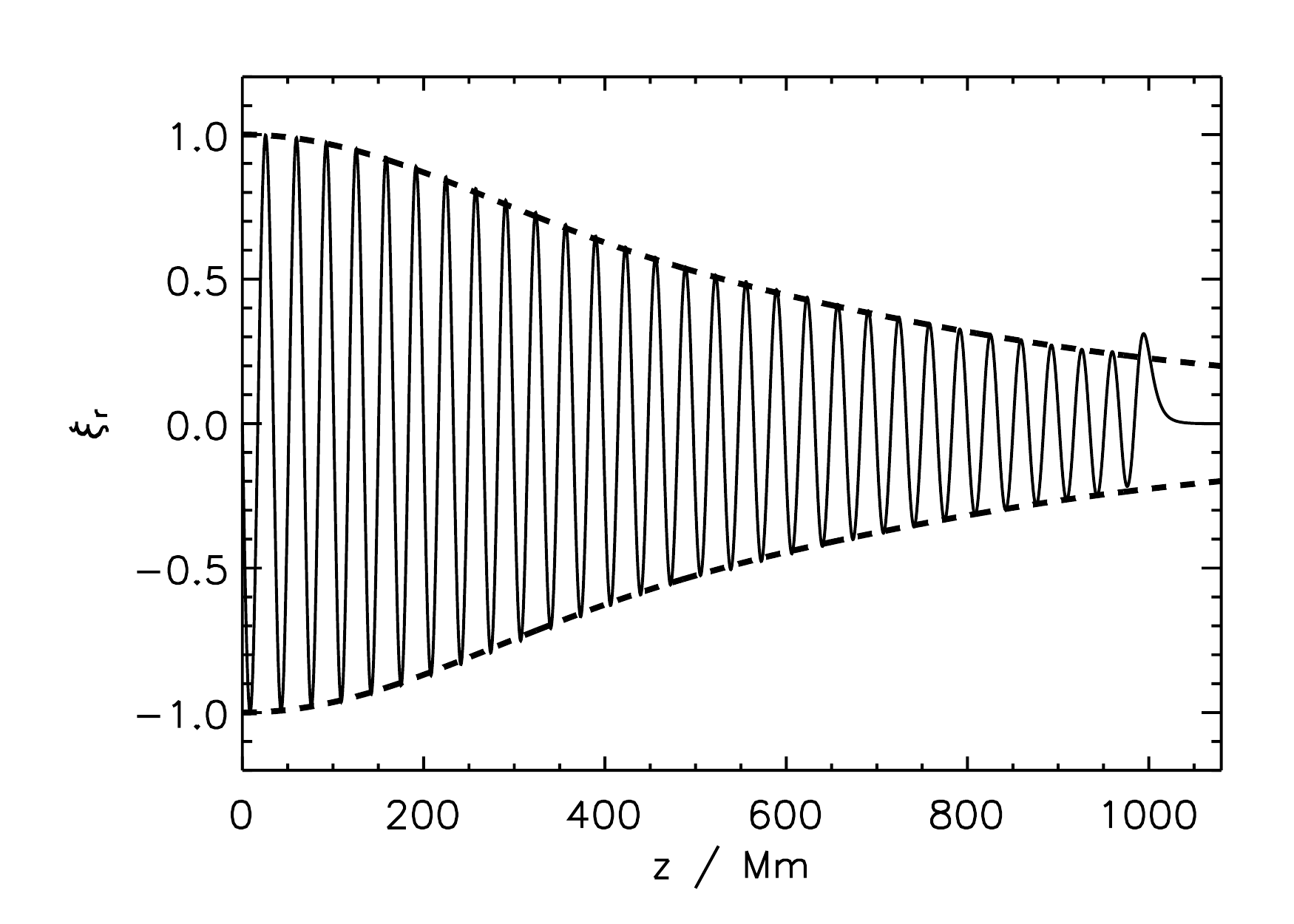}
\includegraphics[width=0.45\textwidth]{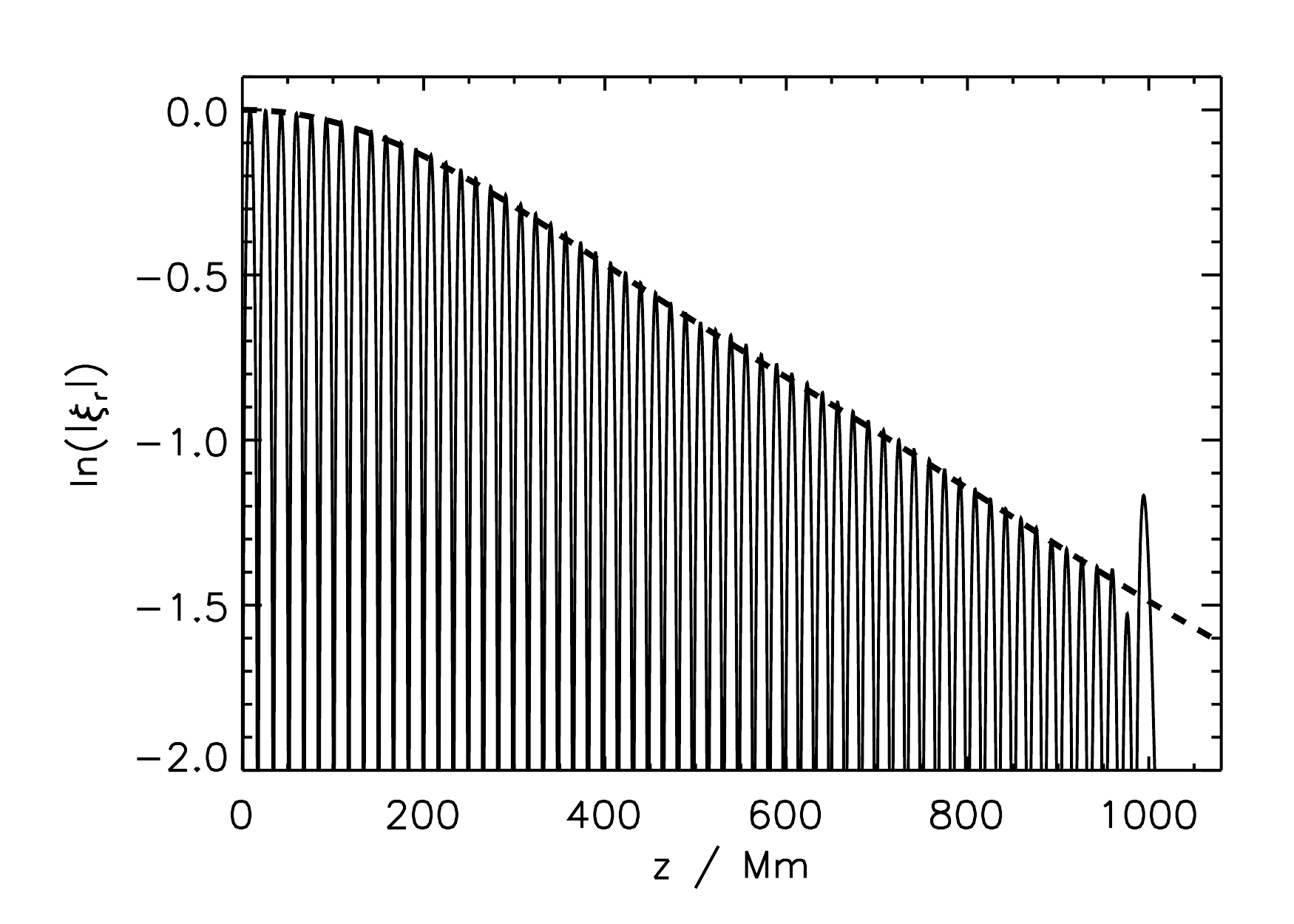}
\caption{Numerical simulation showing the variation of the amplitude $\xi_r$ (solid line) of a propagating kink wave in a coronal loop with density profile parameters the inhomogeneous layer width $\epsilon = 0.2$ and density contrast ratio $\zeta = 1.3$. The dashed line represents the analytical solution of \cite{2013A&A...551A..39H} which accurately reproduces the non-exponential (Gaussian) and exponential damping regimes of damping by resonant absorption.
Figure taken from \cite{2013A&A...551A..39H}.}
\label{fig:djp_regimes}
\end{figure*}

\begin{figure*}
\centering
\includegraphics[width=0.9\textwidth]{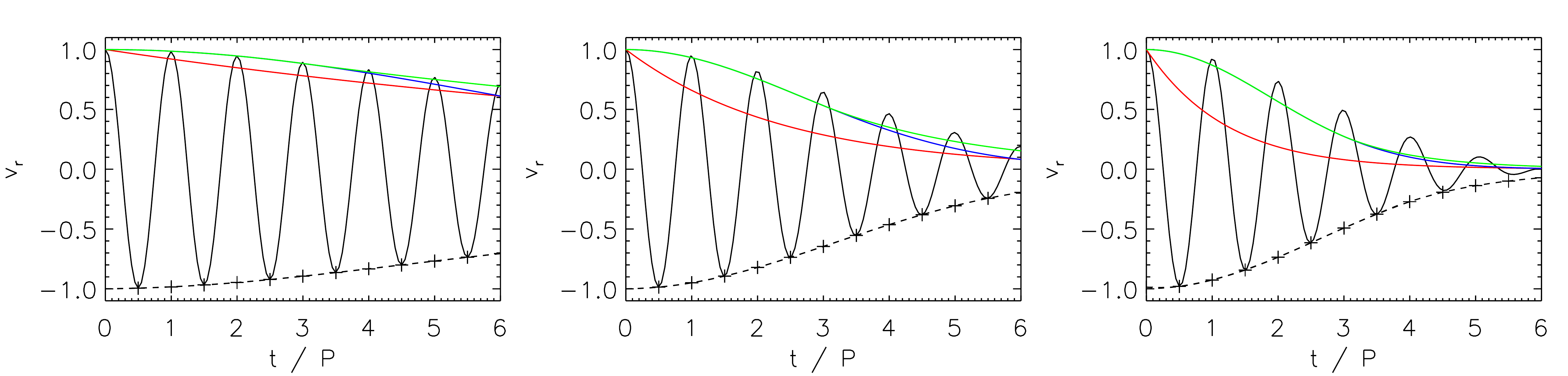}
\caption{Kink oscillation damping profiles calculated by numerical simulation (solid lines) compared with the analytical damping profiles corresponding to the exponential (red) and Gaussian (blue) damping profiles. Green lines represent the general damping profile of \citet{2013A&A...551A..40P} and dashed lines are fitted profiles used by \cite{2019FrASS...6...22P} to generate a lookup table.
Panels show the results for a coronal loop with a density contrast ratio $\zeta=2$ and inhomogeneous layer width $\epsilon=0.1$ (left), $0.5$ (middle), and $1.0$ (right).
Figure taken from \cite{2019FrASS...6...22P}.}
\label{fig:djp_profiles}
\end{figure*}

\begin{figure*}
\centering
\includegraphics[width=0.3\textwidth]{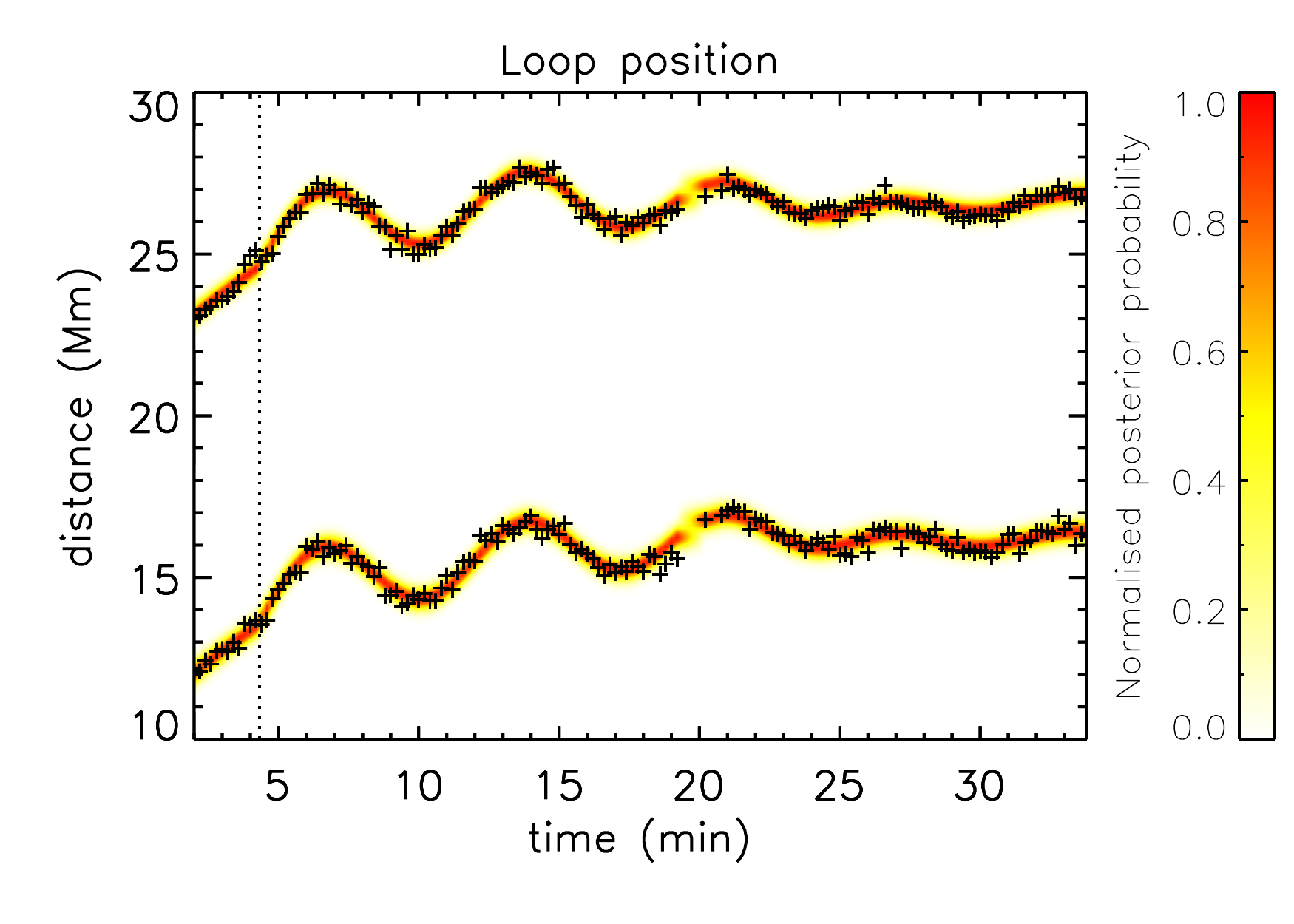}
\includegraphics[width=0.3\textwidth]{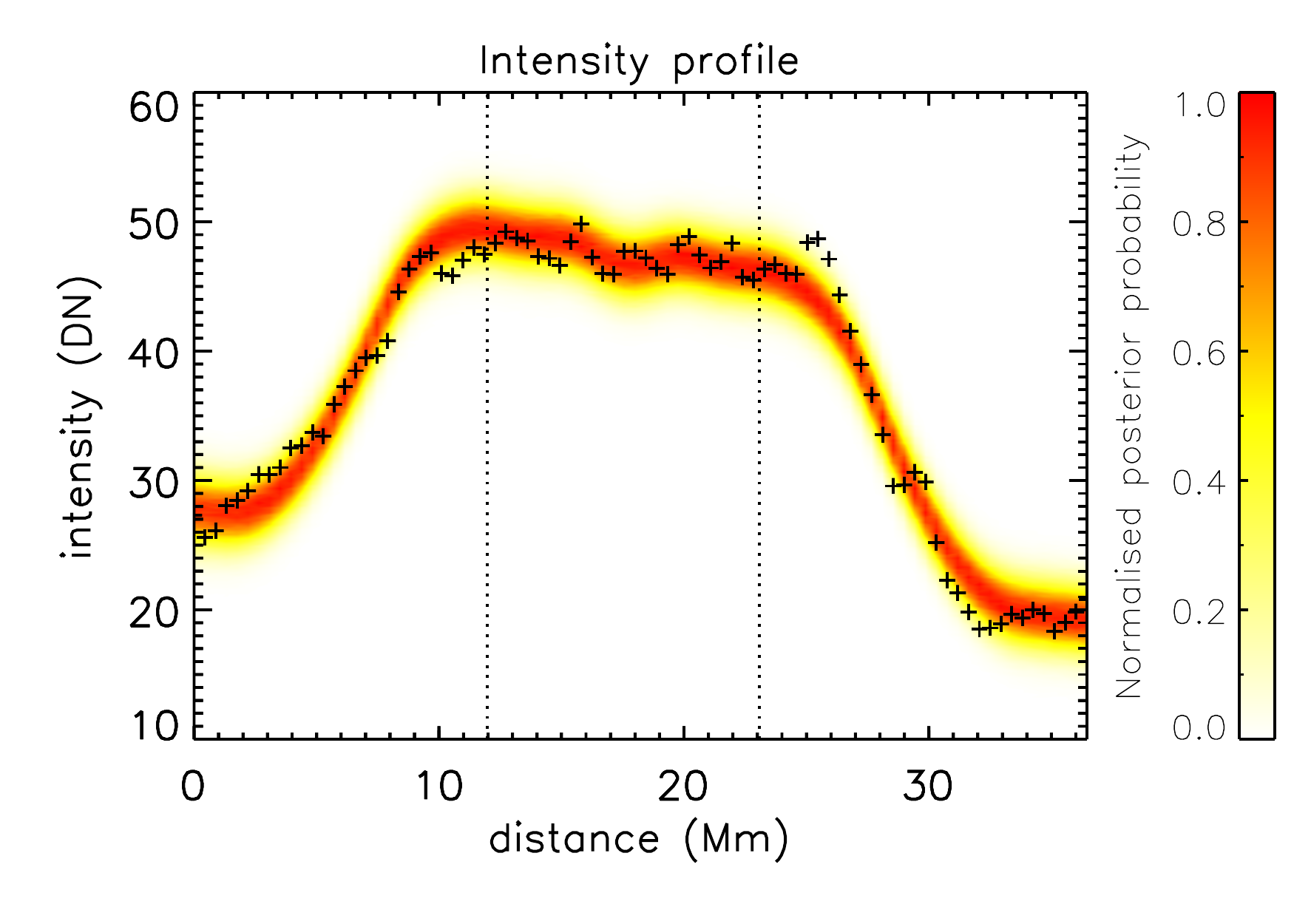}
\includegraphics[width=0.3\textwidth]{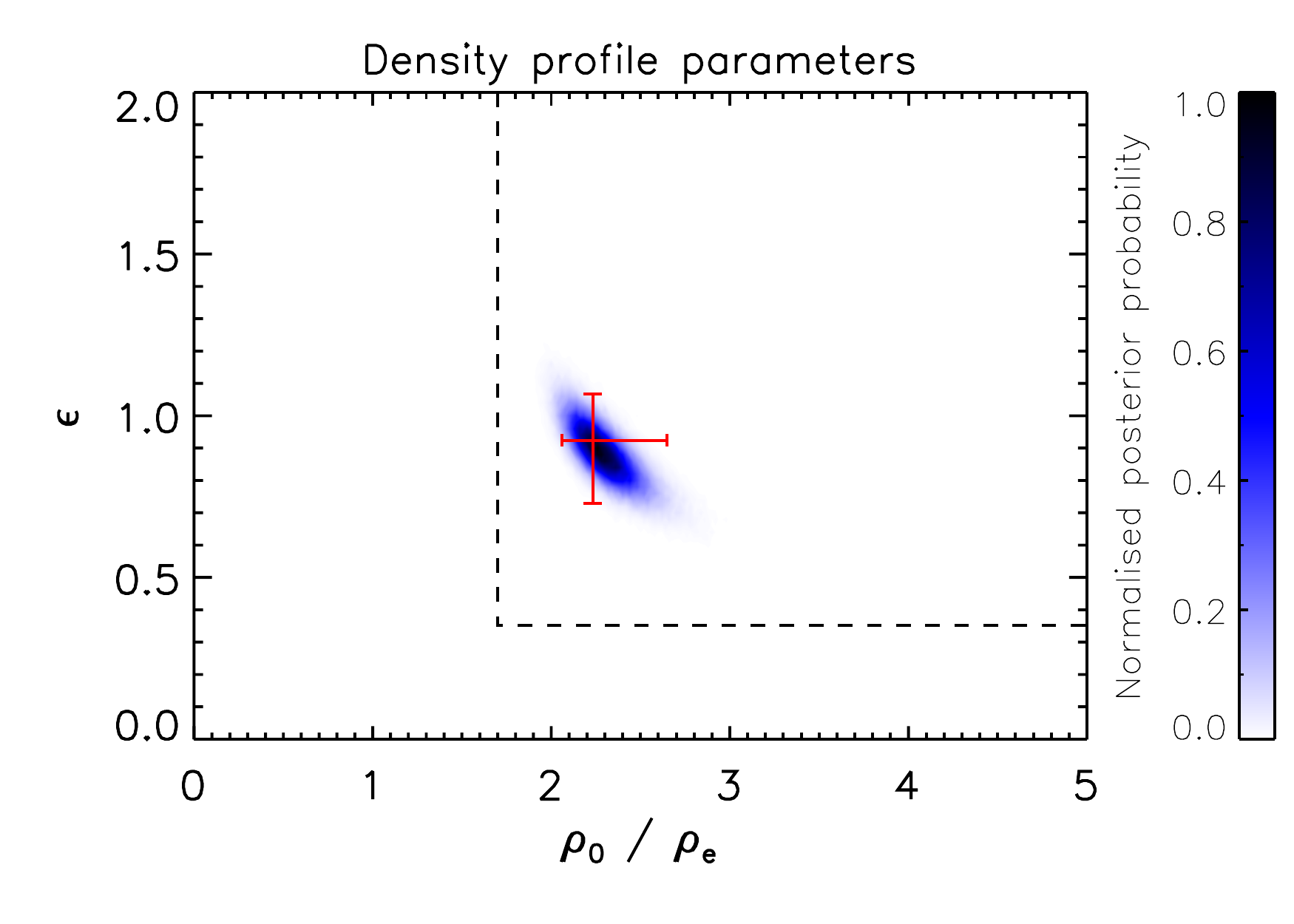}
\caption{Estimation of the transverse density profile parameters of a coronal loop based on the simultaneous analysis of the kink oscillation of the loop legs (left) and the transverse EUV intensity profile (middle). The right panel shows the normalised 2D histogram approximating the marginalised posterior probability density function, with red error bars corresponding to the maximum a posteriori probability value and 95\% credible intervals.
Figure taken from \cite{2018ApJ...860...31P}.}
\label{fig:djp_spatiotemporal}
\end{figure*}

The observed strong damping of impulsively excited kink oscillations is attributed to the process of resonant absorption \citep[e.g.,][and references therein]{2011SSRv..158..289G, 2019FrASS...6...20G, 2016PPCF...58a4001D}.
This has always been an appealing damping mechanism for kink waves due to its robustness in only requiring that the transition region between a higher density coronal loop and the lower density background plasma be diffused, i.e., occurs over a finite spatial scale.
Curiously, the only coronal loops which we would not expect to exhibit resonant absorption of kink oscillations would be those which have a discontinuous boundary, i.e., those which are described by the ZSER model (see Sec.~\ref{sec:zser}). From an observational point of view, this will appear as a damping of the kink oscillation, accompanied by the growth of the azimuthal motions manifesting as unresolved Doppler velocity perturbations due to line-of-sight integration of multiple waves and structures \citep[e.g.,][]{2012ApJ...746...31D, 2017ApJ...836..219A, 2019ApJ...881...95P}.
Subsequent phase-mixing of the Alfv\'en waves in the inhomogeneous layer can generate small spatial scales which enhance dissipative processes such as viscosity and resistivity \citep[e.g.,][]{2017A&A...601A.107P, 2018A&A...616A.125P}.

Initial applications of resonant absorption to account for the strong damping of large amplitude kink oscillations \citep{2002A&A...394L..39G,2002ApJ...577..475R} were based on analytical derivations for the asymptotic state of the system, subject to the TTTB approximation, and produced an exponential damping profile with the form
\begin{equation}
A \left( t \right) \propto \exp \left( - \frac{t}{\tau_\mathrm{D}} \right),
\label{eq:exp}
\end{equation}
\begin{equation}
\frac{\tau_\mathrm{D}}{P_\mathrm{kink}} = \frac{4}{\pi^2 \kappa \epsilon},
\label{eq:tau_d}
\end{equation}
where $\tau_\mathrm{D}$ is the exponential damping time and $\kappa = (\rho_\mathrm{0i} - \rho_\mathrm{0e}) / (\rho_\mathrm{0i} + \rho_\mathrm{0e})$.
In this geometry, the transition from $\rho_\mathrm{0i}$ to $\rho_\mathrm{0e}$ takes place over a cylindrical shell of thickness $l_\mathrm{tr}$ centred on the cylinder's radius $a$, with $\epsilon = l_\mathrm{tr} / a$ being the normalised inhomogeneous layer width.
The constant of proportionality depends on the shape of the density profile in the inhomogeneous layer, and here the factor $4/\pi^2$ corresponds to a linear smooth transition between $\rho_\mathrm{0i}$ and $\rho_\mathrm{0e}$.

Resonant absorption is a rather universal phenomenon, and depends weakly on the specific shape of the loop's cross-section. For example, this effect was observed in a numerical simulation of a kink oscillation of a bundle of ten closely packed plasma cylinders, with random positions and density contrasts \citep[e.g.,][]{2012ApJ...746...31D}. 
\citet{2015A&A...582A.120S} developed a mathematical formalism based on the T-matrix theory of scattering, allowing to compute the periods and damping times of kink oscillations of an arbitrary configuration of parallel plasma cylinders.

\citet{2012ApJ...753..111G} have focused on the role of vorticity in MHD waves. The authors have found that for a piecewise constant density profile the fundamental radial modes of non-axisymmetric modes, including kink modes, have the same properties as purely surface Alfv\'en waves at a true density discontinuity. But when the discontinuity is replaced with a continuous variation of density, vorticity is spread out over the whole in-homogeneous region. Along the same line, \citet{2020A&A...641A.106G} have used both compression and vorticity to characterise the spatial evolution of the kink MHD wave. The most surprising result is the huge spatial variation in the vorticity component parallel to the magnetic field. In the  non-uniform part of the tube, parallel vorticity increases to values that are several orders of magnitude higher than those attained by the transverse components {of the vorticity, i.e.,} in planes normal to the straight magnetic field. 

Note that the theoretical study of kink MHD waves in solar plasma waveguides is usually based on the simplification that the transverse variation of density is confined to a nonuniform layer much thinner than the radius of the tube, i.e., the TTTB approximation. \citet{2013ApJ...777..158S} have developed a general analytic method to compute the dispersion relation and the eigenfunctions of ideal MHD waves in flux tubes with transversely nonuniform layers of arbitrary thickness. Interestingly, the results for thick nonuniform layers deviate from the behaviour predicted in the thin boundary approximation and strongly depend on the density profile used in the nonuniform layer.

\citet{2014ApJ...788....9G} have shown that kink waves do not only involve purely transverse motions of the waveguiding plasma cylinders, but the velocity field is a spatially and temporally varying sum of both transverse and rotational motions. These rotational motions are not necessarily signatures of the classic axisymmetric torsional Alfv\'en wave alone, because of the contribution of the kink motion itself. This essentially means that in observations and depending on the line of sight, the interpretation of the Doppler velocity can be either very similar or very different to that from a purely torsional Alfv\'en wave. {Specifically, near the resonant surface, where the kink speed equals the local Alfv\'en speed, we expect the Doppler signal to be like that of an $m=1$ torsional Alfv\'en wave, while further from the resonant position the signal would look like a kink wave \citep[see][for the discussion of an observational manifestation of this effect]{2013ApJ...777...17S}.} 

Finally, \citet{2015ApJ...803...43S} have theoretically investigated the generation of small scales in nonuniform solar magnetic flux tubes due to phase mixing of MHD kink waves. Using a modal expansion these waves are written as a superposition of Alfv\'en continuum modes that are phase mixed as time evolves. This analysis describes both the damping of global kink motions and the building up of small scales due to phase mixing.


\subsection{Non-exponential damping regime of resonant absorption} 
\label{sec:2rRAthe}

\cite{2012A&A...539A..37P} performed numerical simulations of kink oscillations propagating in a coronal loop and found poor agreement with the classical damping behaviour, described by Eqs.~(\ref{eq:exp}) and (\ref{eq:tau_d}).
These simulations were carried out with a relatively wide inhomogeneous layer in order to reproduce the strong damping rates observed for both standing and propagating kink waves in the corona.
Since the analytical derivations for the resonant absorption damping rate are based on the thin boundary approximation, it would be reasonable to expect numerical simulations with a wide boundary to have a damping rate which differs from the analytical prediction, with differences of up to 25\% already reported by \cite{2004ApJ...606.1223V}.
However, \cite{2012A&A...539A..37P} also found that the shape of the damping profile was inconsistent with the analytical prediction, and proposed, empirically, that a Gaussian damping profile was more appropriate than an exponential one.

\cite{2013A&A...551A..39H} accounted for the existence of this Gaussian damping regime by producing an analytical description which considered the initial behaviour of the kink mode, not just its asymptotic state. Their Eq.~(47) describes the amplitude of the damping profile for all times, though again subject to the TTTB approximation. Figure~\ref{fig:djp_regimes} shows the variation of the amplitude $\xi_r$ (solid line) of a propagating kink wave in a coronal loop with density profile parameters $\epsilon = 0.2$ and $\zeta = 1.3$. The dashed line represents the analytical solution which accurately describes the variation of the damping profile. The use of a logarithmic scale (right panel) demonstrates that the damping profile is initially non-exponential, before switching to exponential after several cycles. The initial stage was characterised for propagating waves in terms of the normalised parameter $Z = \kappa k z /2$ being small, for which the non-exponential damping regime could be approximated with a Gaussian function, i.e., $\propto \exp ( - t^2)$, consistent with the empirical profile suggested by \cite{2012A&A...539A..37P}. This condition also implies an inverse dependence of the location of the switch on the density contrast ratio $\zeta$ and the wavelength of the oscillation. The parametric study by \citet{2013A&A...551A..40P} supported this dependence and proposed the general damping profile (GDP) as a combination of the two approximations for the Gaussian and exponential regimes, switching from one to another at a particular switch height $h$ for propagating waves, or equivalently a switch time $t_\mathrm{switch}$ for standing waves, given by
\begin{equation}
\frac{h}{\lambda} = \frac{t_\mathrm{switch}}{P_\mathrm{kink}} = \frac{1}{\kappa}.
\label{eq:switch}
\end{equation}
The observational detection of this switch in damping profiles was proposed as a means to estimate the density contrast ratio $\zeta$, whereas the particular damping rate for each regimes depends on both $\zeta$ and the inhomogeneous layer width $\epsilon$, and so neither rate alone could be used to constrain both density profile parameters \cite[e.g.,][and references therein]{2019A&A...622A..44A}.

While this Gaussian damping behaviour was initially simulated and derived in the context of propagating kink waves, its applicability to standing kink waves (with the appropriate change in variable) has been demonstrated in numerical simulations \citep[e.g.,][]{2013A&A...555A..27R,2016A&A...595A..81M,2018A&A...616A.125P, 2019FrASS...6...22P}.
For the exponential damping regime, the relationship between damping length scales and timescales for propagating and standing kink waves has been demonstrated explicitly, see, e.g., derivations by \cite{2002A&A...394L..39G} and \cite{2010A&A...524A..23T}.

The GDP proposed by \citet{2013A&A...551A..40P} combined both of the TTTB approximations for the Gaussian and exponential damping regimes into a single profile of the form

\begin{equation}
A \left( t \right) = \left\{ \begin{array}{rl}
\displaystyle \exp \left( - \frac{t^2}{2 \tau_\mathrm{g}^2} \right) \;\;\; &\mbox{$t \le t_\mathrm{switch}$} \\
\displaystyle A_\mathrm{s} \exp \left( - \frac{t - t_\mathrm{switch}}{\tau_\mathrm{D}} \right) \;\;\; &\mbox{$t > t_\mathrm{switch}$}
\end{array} \right. ,
\label{eq:gdp}
\end{equation}

\begin{equation}
\frac{\tau_\mathrm{g}}{P_\mathrm{kink}} = \frac{2}{\pi \kappa \epsilon^{1/2}}.\label{eq:tau_g}
\end{equation}
where the Gaussian damping time $\tau_\mathrm{g}$ here again corresponds to a linear transition, as for the exponential damping time in Eq.~(\ref{eq:exp}).

Figure~\ref{fig:djp_profiles} shows the results of numerical simulations of standing kink oscillations, performed as part of a parametric study by \cite{2019FrASS...6...22P} to investigate the damping behaviour for loops with wide inhomogeneous layers. Each panel corresponds to simulation with $\zeta = 2$ while $\epsilon$ increases from $0.1$ (left) to $1.0$ (right). For kink oscillations in low density contrast loops such as these, the Gaussian damping profile (blue curves) provides a much better description than the exponential damping profile (red curves), and the GDP (green curves) further improves the description for later times. Since the exponential, Gaussian, and general damping profiles are all based on the thin boundary approximation, they each become poorer as $\epsilon$ increases. The dashed curves correspond to damping profiles found by fitting the results of the numerical simulations using spline interpolation of the amplitude measured every half cycle of the oscillation (plus symbols). The results of $300$ simulations were compiled by \cite{2019FrASS...6...22P} into a lookup table to provide a convenient means of estimating the damping profile beyond the applicability of the thin boundary approximation.

\subsection{Observational evidence and seismological application of the two-regime damping} 
\label{sec:2rRAobs}

Investigations of the shape of the damping profile of kink oscillations by \cite{2002A&A...381..311D} and \cite{2002A&A...391..339I} suggested non-exponential behaviour, though the time resolution of TRACE was insufficient for conclusive evidence. In the statistical study of  $223$ standing kink oscillations by \cite{2016A&A...585A.137G, 2019ApJS..241...31N}, the authors attempted to classify visually whether the damping profile appeared to be non-exponential, exponential, or containing both profiles. Several clear examples of non-exponential damping behaviour were observed, motivating a follow up study by \cite{2016A&A...585L...6P} which aimed to quantitatively test whether a Gaussian or exponential damping profile best described the damping behaviour. For six of the highest quality oscillations observed, the instantaneous position of the oscillating loop as a function of time was fitted with a sinusoidal oscillation with an exponential and then a Gaussian damping profile, and the two models compared by their corresponding $\chi^2$ values. Three of the six cases favoured the Gaussian damping profile, two favoured the exponential, and the remaining one was inconclusive. Further analysis of one of the loop oscillations by \cite{2016A&A...593A..59M} also supported the damping profile being Gaussian rather than exponential, and demonstrated that the period of oscillation varies in time with the variation of the loop's parameters \citep[a feature of kink oscillations also observed by][]{2012ApJ...751L..27W, 2013ApJ...774..104W,2013A&A...552A..57N,2015A&A...581A...8R}.

The purely Gaussian or exponential profiles considered in the above studies represent the limiting cases of the two regimes of damping by resonant absorption, but demonstrate that kink oscillations can be measured with sufficient accuracy to discriminate between different damping behaviour. Thus, the regime switching time $t_\mathrm{switch}$ can be used as an additional observable for seismological diagnostics. The seismological method using both regimes (Eq.~\ref{eq:gdp}) was first applied by \cite{2016A&A...589A.136P} to estimate the density profile parameters $\zeta$ and $\epsilon$.
The seismologically-inferred density profile was also compared with the transverse EUV intensity profile of the loop to estimate the radius of the loop and hence the physical width of the transition region $l_\mathrm{tr} = \epsilon a$.
The method was subsequently improved by \cite{2017A&A...600A..78P} to include the additional effects of a time-dependent period of oscillation, the presence of additional longitudinal harmonics, and any decayless component of the oscillation.
The improved method also included the use of Bayesian inference \citep[e.g.,][]{2013ApJ...769L..34A, 2015ApJ...811..104A, 2018AdSpR..61..655A} to improve the calculation of parameter values and uncertainties.
The shape of the damping profile is sensitive to the level of the noise in the oscillation data \citep[see Figure 3 of][]{2018ApJ...860...31P}.
Furthermore, the dependence of the damping rate on both $\zeta$ and $\epsilon$ means the extent to which each of these parameters is constrained depends on the particular value, and in general a strong constraint on one parameter corresponds to a weak constraint on the other.
The use of Bayesian inference to calculate the joint posterior distribution is a convenient way of representing these uncertainties.

Since the damping behaviour of kink oscillations is not always sufficient on its own to strongly constrain the loop density profile parameters, it is desirable to complement the seismological method with additional information.
\cite{2017A&A...600L...7P} and \cite{2017A&A...605A..65G} used the EUV intensity profile of coronal loops to provide an independent estimation of $\epsilon$, and this method was combined in \cite{2018ApJ...860...31P} to produce a diagnostic method combining both seismological (kink oscillation) and spatial (EUV profile) information, as shown in Figure~\ref{fig:djp_spatiotemporal}.
One of the results of these studies is that coronal loops are observed to have a range of inhomogeneous layer widths such that the thin boundary approximation cannot generally be considered applicable.
In particular, the inferred value of $\epsilon \sim 0.9$ for the loop in Figure~\ref{fig:djp_spatiotemporal} is not consistent with the thin boundary approximation assumed by the GDP (c.f. Figure~\ref{fig:djp_profiles}).
This motivated the parametric study of \cite{2019FrASS...6...22P} to produce a seismological technique based on the results of numerical simulations which is not restricted by the thin boundary approximation (although it does not include nonlinear effects such as the modification of the coronal loop density profile by KHI, see Section~\ref{sec:nlgen}).

The above seismological studies are based on assumption of the linear density profile in the inhomogeneous layer. As previously mentioned, a different density profile would modify the constants of proportionality for the damping rates due to resonant absorption.
However, this effect represents a relatively small uncertainty in modern seismology \citep[see further discussion in Section 6.2 of][]{2018ApJ...860...31P}.


\section{Kink oscillations in twisted cylinders} 
\label{sec:twisted} 

{Eruptive phenomena occurring the corona, such as flares and mass ejections, release free, or non-potential magnetic energy stored in active regions. Coronal loops with a non-potential field have either a sigmoid shape or the field twisted around the loop's axis \citep[e.g.][]{2020ApJ...894L..23M}.}  Coronal loops with a free magnetic energy can be considered as plasma cylinders with an axially twisted magnetic field. In such loops the equilibrium magnetic field has axial and azimuthal components, with the latter referred to as  $B_\phi$. The rate of the twist should not be very high, as otherwise the equilibrium is unstable \citep[see, e.g.,][and references therein]{2018A&A...609A...2M}.  At the axis of the cylinder the component $B_\phi$ is zero and the magnetic field is purely axial. The main effect of magnetic twist on transverse kink oscillations  is to break  the symmetry with respect to the propagation direction. This property has relevant consequences for standing modes that require in general the superposition of propagating waves. For example, when there is no twist, the frequency of the mode with an axial wavenumber $k_z$ is the same as for the mode $ - k_z$ ($\omega_{\rm k}=\pm k_z\,C_\mathrm{k}$ in the thin tube limit). It is straightforward to construct the standing solution in this case. However, under the presence of twist the situation is more complicated since the modes with wave numbers $k_z$ and $ - k_z$ have different frequencies. To solve properly this problem we have to combine more than one wave to satisfy the boundary conditions at footpoints, as it was shown in detail by \cite{2015A&A...580A..57R} (see also references therein) for a particular choice of the dependence of the azimuthal component of the field on the transverse coordinate (see also \cite{2012A&A...548A.112T}).

Interestingly, the oscillation period of standing kink waves is unaffected by the presence of twist (for a weak twist and in the thin tube approximation). Therefore  the cylinder oscillates transversally at the characteristic kink frequency, $\omega_{\rm k}$. Second order modifications to this frequency have been calculated analytically in \cite{2015A&A...580A..57R}. Nevertheless, the most important effect of magnetic twist on transverse oscillations is related to the polarisation of the movements. It can be shown that the change in the direction of the polarisation is linearly proportional to the amount of twist. This was studied in detail by  
\cite{2012A&A...548A.112T, 2015A&A...580A..57R, 2018ApJ...853...35T}. For linear twist profile, $B_\phi = A_\mathrm{twist} r$ in $r < a$, the polarisation along the axis (pointing in $z-$direction) changes according to the following simple expression 
\begin{eqnarray*}  
\xi_x &\approx &
\cos\left(\frac{A_\mathrm{twist}}{B_0 a }\frac{1}{4} \left(2 z-L\right)\right)
\,\sin\left(\frac{\pi}{L} z\right),\label{twistpol1} \\  \xi_y &\approx&
\sin\left(\frac{A_\mathrm{twist}}{B_0 a }\frac{1}{4} \left(2 z-L\right)\right)
\,\sin\left(\frac{\pi}{L} z\right),\label{twistpol} 
\end{eqnarray*}    
where the variables $\xi_x$ and $\xi_y$ represent the displacements in the $x$ and $y-$directions, the axial coordinate $z$ varies from 0 to $L$, and $B_0$ is the axial magnetic field at the axis \citep{2015A&A...580A..57R} . 

Note that in those notations, in the zero-twist limit, $A_\mathrm{twist}=0$, the axis displacement is in the $x-$direction only and purely sinusoidal along the $z-$coordinate (satisfying line-tying conditions at the footpoints). Nevertheless, when $A_\mathrm{twist}\neq 0$ the polarisation is in general mixed and the previous expressions show that it depends on the position $z$ along the tube axis. The important result here is that a weak twist can produce displacements in any direction perpendicular to the unperturbed tube axis. In some cases the apparent displacement of a loop, produced by a fundamental mode, may resemble that of the second parallel harmonic, i.e., with a node near the apex. From the seismological point of view, the signatures of twist on observed standing kink oscillations could be used as a way to infer the value of the azimuthal component of the magnetic field. But since real coronal loops are in many cases non-planar and non-circular the comparison between theory and observations is not straightforward. These results are based on the assumption of very weak twist, allowing to avoid difficulties that appear when the $B_\phi$ component of the magnetic field is increased. Alfv\'enic resonances cannot be avoided for moderate twist even if the transverse profile of the density is infinitely sharp. The effect of magnetic twist on the nonlinear evolution of kink oscillations is discussed in Sec.~\ref{sec:nlgen}.

For propagating waves the effect of twist (and also flow) on the transverse kink modes has been investigated analytically and numerically in, e.g., \citet{2010SoPh..263...87K, 2017SoPh..292..110B, 2018ApJ...864....2B}.  \citet{2020ApJ...901...28B} have concluded that the asymmetry of the wave about the apex point is not affected much by the magnetic twist, but the magnetic twist causes an overestimation of both the flow speed and kink speed in the oscillating loop.

\section{Transverse oscillations of current-carrying loops due to the electromagnetic interaction with the external electric currents}
\label{sec:mirr}

\begin{figure}
	\centering
	\includegraphics[width=\linewidth]{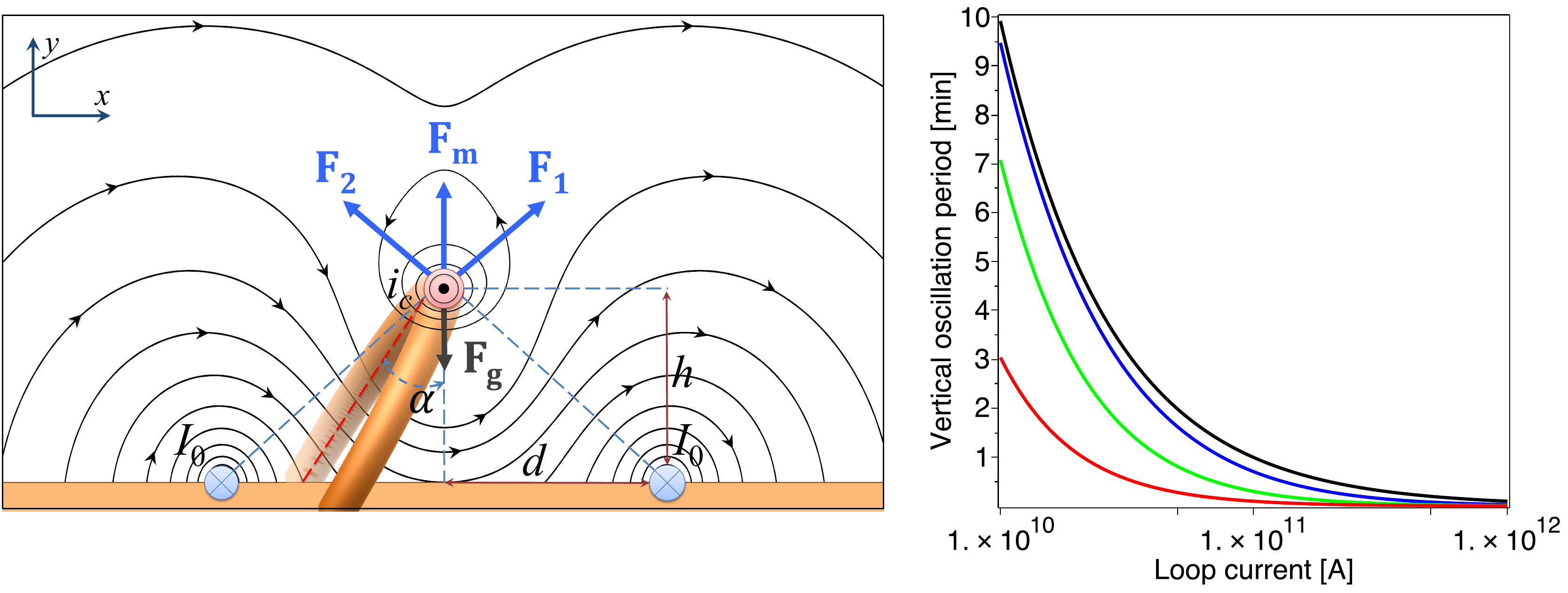}
	\caption{Left: A loop segment with a horizontal line-current $i_\mathrm{c}$, situated at the height $h$ above the surface of the Sun, in the magnetic environment formed by two external currents $I_0$ separated by the distance $2d$. A virtual mirror current describing the interaction of the loop current $i_\mathrm{c}$ with the electrically conductive surface of the Sun is situated at the height $2h$ strictly below $i_c$  (not shown in the sketch). The angle $\alpha$ shows the inclination of the apparent loop plane (the red dashed line) to the vertical axis (adapted from \cite{2016A&A...590A.120K, 2018JASTP.172...40K}).
	Right: Dependence of the vertically-polarised oscillation period $P_y$ (\ref{eq:knn16_period_y}) upon the loop current $i_c$ for the loop length $L=250$\,Mm, inclination angle $\alpha=\pi/6$, minor radius $a=1$\,Mm, particle concentration $n_e= 1.5\times10^9$\,cm$^{-3}$, and $I_0=10^9$\,A (red), $10^{10}$\,A (green), and $10^{11}$\,A (blue). The black solid line shows the characteristic period $P_\mathrm{mir}$ (\ref{eq:knn16_period_KR}), typical of vertically-polarised oscillations in a shallowed magnetic dip with $h\ll d$.
	}
	\label{fig:knn16}
\end{figure}

An alternative mechanism for transverse oscillations of current-carrying magnetic flux tubes was proposed in a series of works \cite{2016A&A...590A.120K, 2018JASTP.172...40K}, based on the interaction of the electric current inside the flux tube with the magnetic coronal environment and the electrically conductive photosphere. In this section we discuss on how this model could be adapted to transverse oscillations of coronal loops with electric current and assess, in particular, the applicability and importance of this effect in modelling the kink oscillations.

Consider a coronal loop segment with the horizontal electric current $i_\mathrm{c}$ embedded in the background magnetic field with a dip formed by two horizontal current sources $I_0$, situated at the lower layers of the solar atmosphere and directed oppositely to $i_\mathrm{c}$ (see the schematic sketch in Fig.~\ref{fig:knn16}). Also in the model, the loop segment is considered to interact with the electrically conductive photosphere through the inclusion of a so-called \emph{mirror current}, which is a virtual current situated strictly symmetrically with respect to $i_\mathrm{c}$, has the same magnitude and opposite direction (not shown in Fig.~\ref{fig:knn16}). Thus, the dynamics of the loop segment in such a low-dimensional model is governed by the mutual effect of the magnetic forces ${F}_1 = {F}_2 = k_1/\sqrt{d^2+h^2}$ (between the loop current $i_\mathrm{c}$ and two external currents $I_0$, with $k_1=\mu_0I_0 i_\mathrm{c}/2\pi$ and the characteristic spatial scales $h$ and $d$ shown in Fig.~\ref{fig:knn16} ) and ${F}_\mathrm{m}=k_2/2h$ (between $i_\mathrm{c}$ and the mirror current, with $k_2=\mu_0 i_\mathrm{c}^2/2\pi$), and the gravity force ${F}_\mathrm{g}={\cal R}g$ (with ${\cal R}$ being a linear mass density of the loop, measured in kg\,m$^{-1}$). All forces in this model are taken per unit length. We note that due to integrating over the loop cross-sectional area the internal structure of the loop does not affect the discussed transverse oscillations. Also, neglecting the line-tying boundary conditions for the guiding magnetic field of the loop does not allow for taking the effects of the magnetic hoop force and magnetic tension force into account, see e.g., \citet{1994ApJ...423..854C, 2008AnGeo..26.3089V}. On the other hand, a low-dimensional nature of the model allows for its straightforward analytical treatment and revealing the explicit relationships between the basic parameters of the oscillations and the loop.

The equilibrium state of such a loop segment is maintained by the vertical force balance
\begin{equation}\label{eq:knn16_equilib}
\frac{2k_1h}{d^2+h^2}+\frac{k_2}{2h}={\cal R}g,
\end{equation}
connecting the loop parameters ({${\cal R}$, $h$, and $i_\mathrm{c}$) with those of the external field ($I_0$ and $d$). For example, Eq.~(\ref{eq:knn16_equilib}) implies that increase in the external current $I_0$, keeping the loop parameters ${\cal R}$, $h$, and $i_\mathrm{c}$ constant, would lead to the corresponding increase of the distance $d$ thus decreasing the dimensionless parameter $h/d$ and mitigating the effect of the external field dip on the dynamics of the loop.

In the linear regime, i.e. for a small displacement of the loop from its initial equilibrium position, the oscillations in the vertical and horizontal directions are independent of each other and hence can be considered separately. In particular, the period of small-amplitude vertically polarised oscillations of the loop segment takes the following elegant form (cf. the expression for the oscillation period derived for curved current-carrying loops in Eq.~(B14) of \cite{1994ApJ...423..854C})
\begin{equation}\label{eq:knn16_period_y}
P_y=\frac{P_\mathrm{mir} \sqrt{k_1 k_2}}{[(2{\cal R}gh-k_2)^2-2k_2(2{\cal R}gh-k_2)+k_1k_2]^{1/2}},
\end{equation}
where
\begin{equation}\label{eq:knn16_period_KR}
P_\mathrm{mir}=2\pi\sqrt{\frac{4{\cal R}h^2}{k_2}}\approx 2.6\times 10^{-2}\times\frac{a[\mathrm{Mm}]L[\mathrm{Mm}]\cos\alpha\sqrt{n_e[10^9\, \mathrm{cm}^{-3}]}}{i_\mathrm{c}[10^{10}\,\mathrm{A}]}[\mathrm{min}]
\end{equation}
is the characteristic value of the vertical oscillation period for the regime of weak interaction with the external field dip, $h/d\ll 1$, independent of the external currents $I_0$ \citep{1974A&A....31..189K}.
In this regime, the magnetic field dip is shallowed and the vertically polarised oscillation is mainly sustained by the magnetic mirror force $F_\mathrm{m}$.
For practical purposes, the period $P_\mathrm{mir}$ (\ref{eq:knn16_period_KR}) was rewritten in terms of the loop length $L$, angle $\alpha$ between the loop plane and the vertical axis, loop volume number density $n_e$, minor radius $a$, and the loop current $i_\mathrm{c}$. As shown by Fig.~\ref{fig:knn16}, the values of the period $P_y$ are about a few to several minutes for typical loop parameters (see, e.g., \cite{1998A&A...337..887Z, 2009SSRv..149...83K} for seismological estimations of the loop current $i_\mathrm{c}$) and tend to $P_\mathrm{mir}$ for increasing $I_0$ (lessening $h/d$), as prescribed by equilibrium condition (\ref{eq:knn16_equilib}). For example, for the parameters used in Fig.~\ref{fig:knn16} and $i_\mathrm{c}=10^{11}$\,A, one obtains $P_\mathrm{mir}\approx 1$\,min. The possibility for the effective damping of oscillations in terms of a similar model with $h/d\ll 1$ was analytically demonstrated by \cite{2018JASTP.179..149Z} through accounting for the drag force between the oscillating flux tube and the ambient plasma. In the same regime with $h/d\ll 1$, the period of horizontal oscillations tends to infinity and hence is not discussed here. 

In the nonlinear regime with the loop displacements comparable to the characteristic spatial scales $h$ and $d$, the horizontal and vertical oscillations were found to be strongly coupled between each other, with the horizontal perturbation effectively exciting the vertical mode. The effect of coupling was shown to be more pronounced for smaller angles between the direction of the initial perturbation and the horizontal axis. As such, it demonstrates the lack of a simple elliptical trajectory of the loop segment in the nonlinear regime. Likewise, a metastable equilibrium of the loop was revealed, which is stable to small-amplitude perturbations and may become unstable if the oscillation amplitude exceeds a certain threshold. The nonlinear oscillation periods were shown to acquire a dependence on the oscillation amplitude, varying by up to 10--30\% with respect to the linear regime.

The presented model should be considered as a simple, essentially low-dimen\-sion\-al model, not taking the magnetic tension force typically associated with kink oscillations into account. Despite this, the obtained periods of vertically polarised oscillations of the loop segment, driven by the electromagnetic interaction of the loop current with the external field and electrically conductive photosphere, are seen to be well consistent with typical periods of kink oscillations. This indicates a clear need for accounting for this mechanism in modelling and interpreting their manifestations in observations. In particular, in the magnetic configurations without a dip (or with a sufficiently suppressed dip satisfying the condition $h\ll d$), the discussed vertically polarised oscillations may still occur due to the mirror current effect.

\section{Kink modes in the presence of parallel shear plasma flows} 
\label{sec:flow}

In the majority of observed plasma structures, stationary flows\footnote{Here we consider the flows which last longer than several oscillation periods as stationary.} are field-aligned, as a consequence of the low-$\beta$ plasma. The flow can be driven by pressure imbalances, as in the case of siphon flows between the two footpoints of a loop, heating or cooling, leading to coronal loop filling or draining, or be induced by magnetic reconnection \citep{2014LRSP...11....4R}. 

\subsection{Basic effect of flow on the kink eigen frequency and eigen function}
\label{sec:basflo}
 
 \citet{2010SoPh..267..377R} derived general expressions for the modification of the kink frequency in standing waves due to the presence of an axial stationary flow, $U_0$. Interestingly this author found that contrary to the static case, different positions along the tube oscillate with a different phase. This effect was further investigated by \citet{2011ApJ...729L..22T}.  Under the  assumption  that  the  flow  speed  is  much  smaller  than  the Alfv\'en speed in the cylinder, the kink frequency is modified as
\begin{eqnarray}\label{omegaeigen}
\omega=k_z\,C_{\mathrm{k}} {\left(
1-\frac{\rho_{\mathrm i}}{\rho_{\mathrm i}+\rho_{\mathrm
e}}
 \frac{U_0^2}{C_{\mathrm{k}}^2}\right)}
\left[{
1-\frac{\rho_{\mathrm i} \rho_{\mathrm e}}{\left(\rho_{\mathrm i}+\rho_{\mathrm
e}\right)^2}
 \frac{U_0^2}{C_{\mathrm{k}}^2}}\right]^{-1/2} ,
\end{eqnarray}
where  
\begin{eqnarray}
\label{k0} 
k_z=n\frac{\pi}{L}, \,\,\, n= 1, 2, \ldots.
\end{eqnarray} 
From the comparison with the eigen frequencies for the static case, $ \omega=k_z\,C_{\mathrm{k}}$, it turns out that flow always leads to a frequency reduction, i.e., an increase in the oscillation period.

In the presence of an axial stationary flow, the displacement along the axial axis has the following form
\begin{eqnarray}\label{eigenfuncts}
\xi(t,z)= A  \sin k_z z \cos \left(\omega t +k_\mathrm{U} z\right),
\end{eqnarray}
where $A$ is an arbitrary constant, and we have introduced the wavenumber modified by the flow,
\begin{eqnarray}\label{kU}
k_\mathrm{U}=k_z \frac{\rho_{\mathrm i}}{\rho_{\mathrm
i}+\rho_{\mathrm e}}\frac{U_0}{C_{\mathrm{k}}}\left[{{
1-\frac{\rho_{\mathrm i} \rho_{\mathrm e}}{\left(\rho_{\mathrm i}+\rho_{\mathrm
e}\right)^2}
 \frac{U_0^2}{C_{\mathrm{k}}^2}}}\right]^{-1/2}.
\end{eqnarray}
\noindent 
Therefore, there is a linear phase dependence of the standing kink mode along the loop, and in one full period, the eigenfunctions exhibit an asymmetry about the loop top. This information about the eigenfunction was used by \citet{2011ApJ...729L..22T} to infer, a stationary flow in an oscillating loop.

\citet{2017SoPh..292..110B} extended the theory of standing waves in the presence of a stationary flow to twisted magnetic tubes without resonant damping. \citet{2019A&A...631A..31R} have addressed the standing wave problem under the presence of a stationary flow and resonant damping. These authors obtained expressions for the ratio of decrements and oscillation frequencies for particular flow profiles, and concluded that for typical values of a stationary flow in coronal loops the effect of the flow on parameters of kink oscillations is weak.

For propagating waves the effects of stationary flows on resonantly damped kink oscillations have been investigated analytically and numerically by \citet{2010A&A...515A..46T, 2011ApJ...734...80S, 2018ApJ...864....2B}.

\subsection{Kink waves in jets}
\label{sec:jets}
Kink oscillations have been detected in plasma structures with collimated flows, for example in coronal jets and loops with siphon flows. 
Coronal jets are intensively studied in the context of impulsive energy releases and the mass supply to the solar wind \citep{2016SSRv..201....1R}. Coronal X-ray jets are the largest among jets of other types \citep{2007Sci...318.1591S}. Their widths are reported between 2 and 20~Mm. Hot jets constitute high-speed outflows which are observed to live from over a minute up to almost an hour. {Perpendicular} non-uniformity of the equilibrium flow leads to waveguiding effects on fast magnetoacoustic waves, in particular, on kink waves. As in static plasma structures, the oscillatory displacement of the jet axis is an indication of a kink oscillation. 
Kink oscillations of jets are an important tool for their diagnostics \citep{2009A&A...498L..29V, 2019ApJ...886..112K}. \cite{2007Sci...318.1580C} reported kink displacements of a soft-X-ray jet, with the oscillation period of 200~s and apparent propagation speeds about $800$~\kms. Similar oscillations with amplitudes around $800$ km and periods about $14$ min have been observed as Doppler oscillations of the Hi Ly$\alpha$ coronal emission line at 1.43~$R_\odot$ above the limb, and interpreted as kink oscillations of a narrow, jet-like ejection observed higher up in the white-light corona \citep{2015A&A...573A..33M}. {A related topic is kink oscillations of coronal streamers \citep{2020ApJ...893...78D}.} 

Under the assumption that the jet's life time is much longer than the kink oscillation period, \cite{2009A&A...498L..29V} described linear kink oscillations of hot coronal jets by adapting the ZSER model \citep[see Section~\ref{sec:zser}, and also][]{1992SoPh..138..233G}.
Dispersion relation (\ref{zser}) modified by a uniform axial stationary flow $U_\mathrm{i}$ becomes
\begin{equation}
\label{Ui}
\frac{\rho_\mathrm{0i}}{\rho_\mathrm{0e}}\,\frac{(C_\mathrm{Ai}^{2}-\Omega_{i}^{2}/k_z^2)}{(C_\mathrm{Ae}^{2}-\omega^{2}/k_z^2)} \frac{\kappa_\mathrm{e}}{n_\mathrm{i}}\frac{K^{\prime}_{1}\,(\kappa_\mathrm{e}\,a)}{K_{1}\,(\kappa_\mathrm{e}\,a)} = \frac{J^{\prime}_{1}\,(n_\mathrm{i}\,a)}{J_{1}\,(n_\mathrm{i}\,a)},
\end{equation}
where $\Omega_\mathrm{i} = \omega - U_\mathrm{i} k_z$, and the expression for $n_\mathrm{i}$ (see the text under Eq.~(\ref{zser}) has the Doppler-shifter phase speed $V_\mathrm{ph}-U_\mathrm{i}$ instead of $V_\mathrm{ph}$.
Using this model, \citet{2009A&A...498L..29V} concluded that if the only damping mechanism is due to resonant absorption, kink oscillations with the observed period of about 200~s in soft X-ray jets last several oscillation cycles before completely damping out. Regarding the origin of the oscillations, \citet{2009A&A...498L..29V} proved that neither KHI caused by the equilibrium flow shear at the jet's boundary, nor instabilities connected with negative energy waves (see Section~\ref{sec:new}) could be responsible for the generation of the oscillations.

\subsection{Negative energy wave effects on kink oscillations}
\label{sec:new}

Shear flows, for example, in plasma jets, could lead to an interesting phenomenon of negative energy waves \citep[e.g.,][]{1988JETP...67.1594R, 1995JPlPh..54..149R}. {When a shear flow speed, i.e., the difference between the flows internal and external to a plasma structure that hosts MHD waves exceeds the phase speed of any of guided waves, that wave becomes a backward wave \citep[e.g.,][]{1995SoPh..159..213N}.} In other words, a backward wave is the wave which propagates in the direction opposite to the direction in the absence of a flow. If the source of the stationary flow is excluded from the considered system, the energy of backward waves becomes negative \citep[e.g.,][]{1997SoPh..176..285J}. Any decrease in the energy, for example, due to dissipation or leakage, leads to the increase in the amplitude of a wave with the negative energy. It results in so-called negative energy wave instabilities. 

In the linear regime of a negative energy wave instability, the negative energy wave amplitude experiences the exponential growth, while in the nonlinear regime a much faster explosive instability is possible, when the amplitude reaches an infinite value in a finite time\footnote{Obviously, in reality, the infinite amplitude is not reached, as the amplitude growth above a certain level is counteracted by additional nonlinear effects.} An important feature of negative energy instabilities is their occurrence for flow shears much lower than the threshold of KHI \citep{1997SoPh..176..285J}.

To provide further insight, consider a plasma cylinder with a straight equilibrium magnetic field along the tube axis embedded in a viscous plasma medium with a straight magnetic field. The equilibrium is similar to the ZSER model (Section~\ref{sec:zser}), but, in addition, there is a field-aligned equilibrium flow inside the cylinder with the stationary speed $U_0$. In addition, the plasma viscosity $\nu_\mathrm{e}$ is taken into account in the external medium. The equilibrium flow creates a shear at the boundary of the cylinder. In the incompressible limit, the propagation of kink waves  guided by the cylinder is described by the dispersion relation 
\begin{eqnarray}
\left[\left(\frac{\rho_\mathrm{0i}I_1}{k_zI^{\prime}_1}-\frac{\rho_{0\mathrm{e}}K_1}{k_zK^{\prime}_1}\right)\omega^2-2k_zU_0\frac{\rho_\mathrm{0i}I_1}{k_zI^{\prime}_1}\omega+\frac{\rho_\mathrm{0i}I_1}{k_zI^{\prime}_1}(k_zU_0)^2\right.\nonumber\\
\left.-\frac{\rho_\mathrm{0i}I_1}{k_zI^{\prime}_1}\omega_\mathrm{Ai}^2+\frac{\rho_{0\mathrm{e}}K_1}{k_zK^{\prime}_1}\omega_\mathrm{Ae}^2\right]\nonumber\\
+i\omega\rho_{0\mathrm{e}}\nu_\mathrm{e} \left[\frac{K_1}{k_zK_1^{'2}}\left({\cal D}K^{\prime}_1-\frac{K_1^{\prime}}{a^2}\right)+\frac{2K_1^2}{k_z^2 a^3 K_1^{'2}}-2k_z\frac{K_1^{\prime\prime}}{K_1^{\prime}}\right]=0,
\label{DispersionFull}
\end{eqnarray}
where, $\omega_\mathrm{Ae} = k_z C_\mathrm{Ae}$ and $\omega_\mathrm{Ai} = k_z C_\mathrm{Ai}$.
As in the ZSER model, $I_1(k_za)$ and $K_1(k_za)$ represent the modified Bessel functions of the first order and of the first and second kinds, respectively. Note that the argument $k_za$ has been omitted, and ${\cal D}$ is a differential operator defined in \citep{2020ApJ...896...21Y}. We need to point out that in the considered incompressible limit, the internal structure of the perturbations is given by the modified Bessel function $I_1$ in contrast with the ZSER model which is essentially compressible. The dispersion relation is a quadratic equation with respect to $\omega$. Because of the viscosity,  coefficients of this equation include the imaginary unity. Thus, its solution could be written as a complex frequency constituted by real $(\omega_\mathrm{r})$ and imaginary $(\omega_\mathrm{i})$ parts.  The explicit expressions regarding the phase speed, and the damping/growth determined by $\omega_\mathrm{i}$ are
\begin{eqnarray}
\label{Phasespeeds}
V_{\mathrm{ph}\pm}= \left({U_0 \pm \sqrt{\frac{\rho_{0\mathrm{e}} I_1^{\prime}K_1}{\rho_\mathrm{0i} I_1K_1^{\prime}}U_0^2+{\cal A}{\cal B}}}\right) / {\cal B},
\end{eqnarray}
\begin{eqnarray}
\label{Growthrate}
\frac{\omega_\mathrm{i}}{\omega_\mathrm{r}}=\pm\frac{\nu_\mathrm{e} \frac{\rho_{0\mathrm{e}} I^{\prime}_1 K_1}{\rho_\mathrm{0i} I_1 K_1^{\prime}}\left[\frac{(a^2\nabla^2K_1^{\prime}-K_1^{\prime})}{K_1^{\prime}}-\frac{2k_z^2a^2K_1^{\prime\prime}}{K_1}+\frac{2K_1}{k_zaK_1^{\prime}}\right]}{C_\mathrm{Ai}k_z({\cal B}V_\mathrm{ph\pm}-U_0)},
\end{eqnarray} 
\begin{eqnarray}
{\cal A}=\left(C_\mathrm{Ai}^2-\frac{\rho_{0\mathrm{e}} I_1^{\prime}K_1}{\rho_\mathrm{0i} I_1K_1^{\prime}}C_\mathrm{Ae}^2\right),\,\,\,\,\, {\cal B}=\left(1-\frac{\rho_{0\mathrm{e}}I_1^{\prime}K_1}{\rho_\mathrm{0i} I_1K_1^{\prime}}\right).
\end{eqnarray}
Eqs.~(\ref{Phasespeeds}) and (\ref{Growthrate}) provide quantitative and qualitative information regarding the influence of the shear flow and viscosity on various aspects of the kink wave regarding its phase speed, damping, instability thresholds, and negative energy wave excitation. In particular, it could be noticed from Eq. (\ref{Growthrate}) that the sum of the three terms of the numerator, which dictates whether the growth rate is positive or negative, is positive for longer wavelengths as the denominator of the first and third terms (terms with positive sign) {possess} the square of the wave number that makes them large compared to the second term that has the wave number in its numerator. Negative energy wave instability appears for all axial wave numbers. In general, the onset of the negative energy wave instability of the kink perturbation occurs when the shear flow speed exceed the critical speed
\begin{equation}
\label{uc}
U_\mathrm{cr} = C_\mathrm{Ai} \sqrt{ 1 - \frac{\rho_\mathrm{0e}}{\rho_\mathrm{0i}}\frac{C_\mathrm{Ae}^2}{C_\mathrm{Ai}^2}\frac{I_1^{\prime} K_1}{I_1 K_1^{\prime}}}.
\end{equation}
More rigorously,  negative energy wave instabilities occur when the shear flow speed is between $U_\mathrm{cr}$ and the KHI
threshold. In particular, in the long-wavelength regime, the flow speed shear allowing for the onset of negative energy wave instabilities is several times lower than the KHI threshold. Hence the threshold of the negative energy wave instability tends to longer wavelengths. The lowest value of the shear flow, allowing for the negative energy wave instability, is reached for the axial wavelength comparable to the diameter of the cylinder. This result is readily applicable to standing waves, as \citet{2018JPlPh..84a9001R} showed that the growth rate of a standing wave is equal to the growth rate of the  propagating wave with negative energy minus the damping rate of the propagating wave with positive energy.

\citet{2016Ap&SS.361...51Z} carried out case studies on observed coronal jets and stated that the onset of KHI regarding kink waves depends on whether the plasma jet is incompressible or compressible, as the incompressibility elevates the KHI threshold speed. The magnetic twist of the jet decreases the KHI threshold speed. Nonetheless, a weak twist in case of an incompressible plasma jet still has a threshold higher than the compressible case. However, for the incompressible limit, the shear speed may not need to be be more than 40\% faster than the internal Alfv\'en speed to create negative energy waves. 
It is worth noting that the viscosity of the external medium itself provides information regarding the temperature of the external medium proving adequate for the background temperature to play a direct role on the negative energy kink wave excitation \citep{2020ApJ...896...21Y}.


\section{Kink oscillations in loops undergoing cooling} 
\label{sec:cool} 
Even though most coronal loops live for much longer than their characteristic cooling time \citep{2014LRSP...11....4R}, they appear to undergo continuous thermal evolution, for most of the time being in a cooling phase \citep{2012ApJ...753...35V}. There have been numerous observations of kink oscillations in loops undergoing evolution, manifesting in EUV variability, mostly consistent with cooling \citep[e.g.,][]{2008ApJ...686L.127A}. In some particular cases, undamped large-amplitude kink oscillations were observed in these apparently cooling loops \citep{2002SoPh..206...99A, 2011ApJ...736..102A}. Therefore, it seemed likely that cooling of the plasma during kink oscillations might have an impact on the oscillation properties. In what followed, numerous studies exploited the influence of time-dependent coronal loop models on the properties of kink oscillations. The first such studies incorporating loop cooling indicated that cooling contributes to the damping of kink oscillations \citep{2009ApJ...707..750M}. However, \citet{2011SoPh..271...41R} showed that neglecting the flow generated by the density decrease due to cooling led to wrong conclusions in \citet{2009ApJ...707..750M} about the effect of cooling, leading to an amplification of the oscillations instead. Additionally, the decrease in density leads to a higher kink speed of the loop, thus cooling reduces the kink oscillation period. Taking into account resonant absorption, \citet{2011A&A...534A..78R} showed that the amplification due to cooling is not very efficient in counterbalancing resonant damping of kink oscillations, except for very short cooling times. This was verified using 3D numerical simulations by \citet{2015A&A...582A.117M}, who found an even weaker amplification of the oscillations due to cooling, when accounting properly for the density evolution at the loop footpoints. This decrease in the efficiency of damping when considering time dependent density at the loop footpoints was confirmed theoretically by \citet{2017MNRAS.468.2781B}. It was concluded that cooling alone cannot explain the observed undamped amplitude kink oscillations (see Section~\ref{sec:decless}), especially when taking into account enhanced damping due to nonlinearity (see Section~\ref{sec:nonlin}) \citep{2016A&A...595A..81M}. An additional effect of loop expansion was considered by \citet{2017A&A...602A..50R} and \citet{2018A&A...619A.173S}. They showed that loop expansion reduces the damping rate due to resonant absorption, and accounting for the amplification due to cooling could result in this case in undamped oscillations, for specific parameters.  In Figure~\ref{one} the amplitude evolution of the fundamental kink oscillation of a coronal loop from \citet{2015A&A...582A.117M} (without expansion) is presented, compared with the analytical results described above.
\begin{figure}
  \centering
  \medskip
  \includegraphics[width=0.75\textwidth]{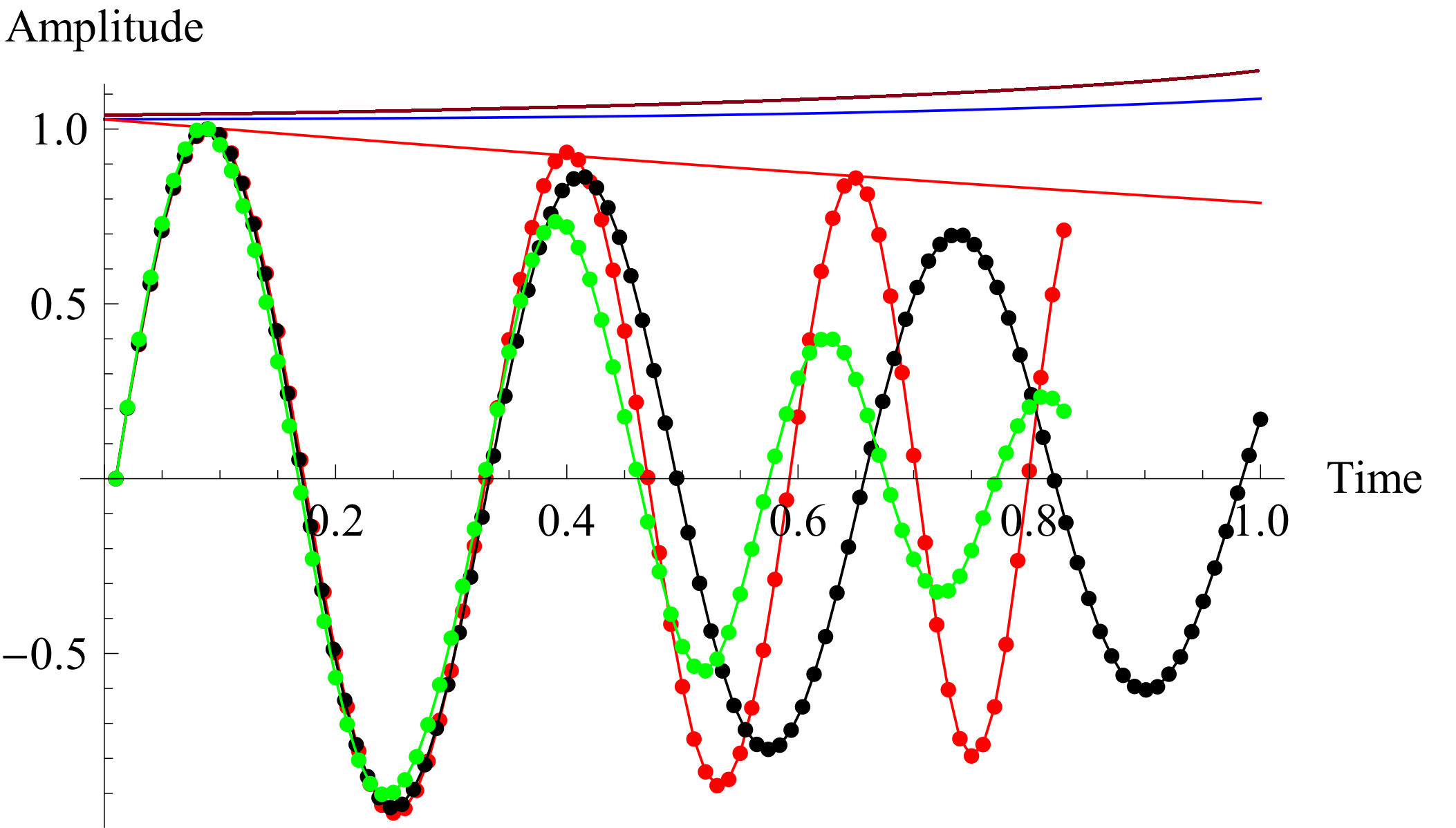}
  \caption{Normalized displacement amplitudes in a fundamental kink oscillation at the apex over time,
from three different numerical simulations: linear perturbation with cooling (red dots), without cooling (black dots) and nonlinear perturbation with cooling (green dots). The red curve represents the best-fit exponential decay, the blue curve is the analytically predicted displacement from \citet{2011A&A...534A..78R}, and the brown curve is accounting for expansion \citep{2017A&A...602A..50R}. Adapted from \citet{2015A&A...582A.117M}.} 
  \label{one}
\end{figure}
These theoretical models were tested and applied for coronal seismology on observed oscillations of cooling coronal loops by \citet{2019FrASS...6...45N}.


\section{Oscillation period ratios and their seismological inferences} 
\label{sec:p1_2p2}

\subsection{Theoretical modelling of the period ratio}
\label{sec:p1_2p2_the}

Theoretically, the ratios of the fundamental kink mode period $P^{(1)}_\mathrm{kink}$ to the $n$-th parallel harmonic period, $P^{(1)}_\mathrm{kink}/(nP^{(n)}_\mathrm{kink})$, have been shown to depend on the wave dispersion, i.e., the effects of finite $k_z$, gravitational stratification, axial and perpendicular density structuring, loop cross-sectional ellipticity and geometry, and the axial magnetic field non-uniformity due to field expansion \citep[e.g.,][and references therein]{2009SSRv..149....3A}. With regards to the density non-uniformity, this also depends upon the loop's temperature, curvature, inclination, etc., that affect the density scale-height and thus the harmonic period ratios \citep{2012A&A...537A..41O, 2013AN....334..948O}. For example, \citet{2016SoPh..291.1143R} showed that the period ratio $P^{(1)}_\mathrm{kink}/(2P^{(2)}_\mathrm{kink}$ is lower or greater than unity when the kink speed increases or decreases with height, respectively. 
{The former case represents loops where the plasma density is gravitationally stratified and the magnetic field is approximately constant (weakly expanding with height), while the latter case is applicable to loops with a significant enough expansion with height such that the rate of decrease in $B^2$ from the loop footpoints to apex dominates over the rate at which $\rho$ decreases.}  A similar approach has also been used to predict kink mode period ratios in prominence threads \citep[e.g.,][and references therein]{2015A&A...575A.123S}, where the plasma density is assumed to be greater at the apex than at the footpoints, i.e. the kink speed decreases towards the apex.

\citet{2012ApJ...757..186K} considered the effect of magnetic twist, considering the twisted field to be localised in an annulus region of the cylinder, which also resulted in the deviation of the period ratio of first two harmonics ($P^{(1)}_\mathrm{kink}/(2P^{(2)}_\mathrm{kink})$) from unity. As the twist parameter increases, it was found that the period ratio decreases from unity, and achieves a minimum value about 0.9 at approximately $B_\phi/B_z = 0.01$, and for stronger twist increases to unity again. 

Thus, the departure of the normalised kink period ratios  from unity can be employed as a diagnostics tool of coronal loops, and provide information crucial for understanding this enigmatic plasma structure \citep[e.g.,][]{2011A&A...534A..13B, 2013ApJ...777...17S, 2016A&A...593A..53P, 2017ApJ...842...99L, 2018ApJ...854L...5D, 2019A&A...632A..64D}.  

\subsubsection{Effect of perpendicular nonuniformity}
\label{sec:pernon}

The most basic reason for the departure of the $P^{(1)}_\mathrm{kink}/(nP^{(n)}_\mathrm{kink})$ ratios from unity is the effect of waveguide dispersion. Indeed, as the phase speed of the kink wave decreases with the increase in the axial wave number $k_z$, different parallel harmonics which have the wavelengths $2L/n$ have different phase speeds $\omega/k_z = 2L/(nP^{(n)}_\mathrm{kink})$ (see Figure~\ref{kinkdisp}), and thus the increase in the mode number does not decrease the oscillation period by an integer $n$.  
Exact expressions for dispersion relations could be obtained for a handful of transverse density profiles only.
\citet{2011A&A...526A..75M} studied the effect of waveguide dispersion on the period ratios in a zero-$\beta$ plasma slab with a symmetric Epstein density profile, and obtained
\begin{equation}
\displaystyle {\left({\frac{P^{(1)}_\mathrm{kink}}{2P^{(2)}_\mathrm{kink}}}\right)}^{2} =  \displaystyle {\frac{1}{4}}{\left( \frac{\displaystyle 2\frac{\pi^2 d^2}{4L^2}  - \frac{C^2_\mathrm{Ai}}{C^2_\mathrm{Ae}} + \sqrt{ \frac{C^4_\mathrm{Ai}}{C^4_\mathrm{Ae}} + 4\frac{\pi^2 d^2}{4L^2} - 4\frac{C^2_\mathrm{Ai}}{C^2_\mathrm{Ae}}\frac{\pi^2 d^2}{4L^2} } }          
{\displaystyle  \frac{\pi^2d^2}{8L^2}  - \frac{C^2_\mathrm{Ai}}{C^2_\mathrm{Ae}} + \sqrt{ \frac{C^4_\mathrm{Ai}}{C^4_\mathrm{Ae}} + \frac{\pi^2 d^2}{4L^2} - \frac{C^2_\mathrm{Ai}}{C^2_\mathrm{Ae}}\frac{\pi^2 d^2}{4L^2} } }     \right)},
\label{sri1}
\end{equation}
where $d$, $C_\mathrm{Ai}$, $C_\mathrm{Ae}$ are respectively the half-width, and the Alfv\'en speeds at the slab's axis and infinity. For thin and long or short and fat loops the period ratio ($P^{(1)}_\mathrm{kink}/2P^{(2)}_\mathrm{kink}$) is close to one. In particular, in the long wavelength limit, $k_zd \to 0$, the ratio tends to unity.  However, this ratio rapidly decreases with the increase in $k_zd$, and reaches 0.75 at $k_zd \approx 0.05$ in a loop with $C_\mathrm{Ae}/C_\mathrm{Ai} = 10$. This effect is even more pronounced for $C_\mathrm{Ae}/C_\mathrm{Ai} > 10$, the condition which can be seen in flaring loops. 

\citet{2015ApJ...814...60Y} performed a comprehensive study of dispersive properties of kink waves in a plasma slab with the linear, parabolic, inverse-parabolic and sine profiles of the density, sandwiched in between two regions with constant densities. Interestingly, for those profiles,  kink oscillation periods were found to differ by $<13$\% from each other. \citet{2018ApJ...855...47C} developed that study accounting for finite-$\beta$. It was established that for parameters typical for the coronal plasma, the finite-$\beta$ effect on kink oscillation is at most marginal. 

\subsubsection{Effect of axial non-uniformity} 
\label{theory1_abi}

For a coronal loop of length $L$, assuming a constant axial magnetic field and a gravitationally stratified plasma density with scale height $H$, if $L\ll H$, then the first three kink mode harmonic periods can be approximated as
\begin{eqnarray}
	P^{(1)}_\mathrm{kink}&=&P_\mathrm{kink}\left(1+\frac{L}{3\pi^2H}\right)^{-1}, \label{jesse1}\\ 
	2P^{(2)}_\mathrm{kink}&=&P_\mathrm{kink}\left(1+\frac{L}{15\pi^2H}\right)^{-1}, \label{jesse2}\\ 
	3P^{(3)}_\mathrm{kink}&=&P_\mathrm{kink}\left(1+\frac{L}{35\pi^2H}\right)^{-1}, \label{jesse3}
\end{eqnarray}
where $P_\mathrm{kink}$ is the fundamental kink mode period of a coronal loop of constant kink speed with a value equivalent to that of the kink speed at the footpoint of a gravitationally stratified loop \citep[e.g.][]{2009SSRv..149....3A}. For a loop inclined from vertical, the scale height $H$ should be multiplied by a cosine of the inclination angle. 

 Assuming the density to be constant along the loop, and taking magnetic field expansion with height into account, results in the period ratio,
\begin{equation}
\frac{P^{(1)}_\mathrm{kink}}{2P^{(2)}_\mathrm{kink}}\approx 1+\frac{(3\Gamma^{2}-1)}{2\pi^{2}},
\label{sri2}
\end{equation}
where $\Gamma=a_\mathrm{apex}/a_\mathrm{f}$ is the magnetic expansion factor such that $a_\mathrm{apex}$ and $a_\mathrm{f}$ are the minor radii at the loop apex and footpoints, respectively \citep{2008A&A...486.1015V}. Note that Eq.~(\ref{sri2}) was derived assuming a weakly expanding potential magnetic field with a straight axis, i.e., a \lq\lq magnetic bottle\rq\rq\ type configuration.  Since $\Gamma>1$ for a loop that expands in with height Eq.~(\ref{sri2}) gives a period ratio $P^{(1)}_\mathrm{kink}/(2P^{(2)}_\mathrm{kink})>1$.
Therefore, if the magnetic field strength is decreasing with height, i.e., there is an expanding magnetic flux tube, then it has the opposite effect on the period ratio to that of density stratification. This can be understood mathematically by considering the governing Sturm--Liouville problem in the long wavelength limit where both the density and magnetic field strength vary along the loop,
\begin{equation}
	\frac{d^2}{dz^2}\left(\frac{\xi_{\perp}}{a(z)}\right)+\frac{\omega^2}{C_\mathrm{k}^2(z)}\left(\frac{\xi_{\perp}}{a(z)}\right)=0, \; \; \xi_{\perp}=0 \;\; \mathrm{at} \; \; z=0 \;\; \mathrm{and} \;\; z=L,
	\label{ver1}
\end{equation}
where $\xi_{\perp}(z)$ is the displacement perturbation perpendicular to the magnetic field, $a(z)$ is the minor radius varying along the field, and $C_\mathrm{k}^2(z)=2B^2(z)/[\mu(\rho_{0\mathrm{i}}(z)+\rho_{0\mathrm{e}}(z))]$. Since the kink speed squared in Eq.~(\ref{ver1}) has the terms $B^2(z)$ and $\rho_{0\mathrm{i}}(z)+\rho_{0\mathrm{e}}(z)$ reciprocal to each other, if the magnetic field strength and plasma density both decrease with height then they will have an opposing effect on the solution. This is readily apparent when considering the effect of magnetic and density stratification on the eigenfunction of the second parallel harmonic. Loop expansion shifts the anti-nodes towards the apex of the loop but gravitational density stratification shifts them in the opposite direction towards the footpoints (see Figs. 6 and 7 in \cite{2009SSRv..149....3A}). Note also from Eq.~(\ref{ver1}), that in the particular case when $B^2(z) \propto \rho_{0\mathrm{i}}(z)+\rho_{0\mathrm{e}}(z)$ we will have $P^{(1)}_\mathrm{kink}/(2P^{(2)}_\mathrm{kink})=1$ since the kink speed is constant along the loop with both effects exactly cancelling each other. {Results of full-scale numerical simulations of kink oscillations in a stratified 3D coronal loop performed by \citep{2019ApJ...876..100A} agree well with the estimations made by \citep{2005ApJ...624L..57A, 2008A&A...486.1015V}.}

\subsubsection{Effect of stationary axial flow} 
\label{theory1_abi}

As it is discussed in Section~\ref{sec:jets}, a shear of an equilibrium steady flow leads to the modification of dispersive properties of kink waves, and hence the period ratios.
\citet{2013ApJ...767..169L} studied the effect of flows on the period ratio of standing kink modes in a plasma slab. It was found that the flow reduces $P^{(1)}_\mathrm{kink}/(2P^{(2)}_\mathrm{kink})$  up to 23\% compared with the static case modelled by \citet{2011A&A...526A..75M}.
\citet{2014SoPh..289.1663C} generalised this theory and found that the flow significantly reduces $P^{(1)}_\mathrm{kink}/(nP^{(n)}_\mathrm{kink})$ from unity for higher $n$ and weaker density contrast between the waveguide and its surrounding. In the high density contrast case, the period ratio has a substantial deviation from unity, up to 13.7 \%  when the Alfv\'enic Mach number reaches 0.8. \citet{2016RAA....16...92Y} confirmed that conclusion by showing that a steady flow can reduce $P^{(1)}_\mathrm{kink}/(2P^{(2)}_\mathrm{kink})$ by up to 17\% relative to the static case even when the density contrast approaches infinity.

In summary, the combined effects of field-aligned flow, longitudinal density and magnetic stratification and transverse structuring all affect kink mode period ratios \citep{2015ApJ...814...60Y}. Hence those effects could be estimated seismologically by this observable, provided there is a procedure allowing one to estimate them separetely. 

\subsection{Observational detection of higher parallel harmonics and their seismological applications}
\label{yuan_obs} 

Observations of higher parallel harmonics of the kink mode remain a challenging task. The detection of multiple parallel harmonics in the displacement signal requires the fitting of a background trend simultaneously with the oscillatory components, since detrending the time series would bias subsequent results. In particular, the assumption of an oscillation being equally distributed about the equilibrium position is satisfied by a weakly damped harmonic oscillation, but not necessarily for a strongly damped oscillation with or without additional harmonics present. \cite{2017A&A...607A...8P} have considered background trends describing with the change in equilibrium associated with the kink mode excitation mechanism for contracting and displaced coronal loops.

\cite{2012A&A...545A.129W} used the hot channels of SDO/AIA instrument and observed an oscillating coronal loop off the east solar limb in the 131 \AA{} and 94 \AA{} bandpasses suggesting a temperature in the range of 9--11~MK. They detected a kink mode period of $302\pm14$~s and a damping time of $306\pm43$ s, and confirmed that this transverse oscillation was more likely to be a higher harmonic mode since they detected a spatial phase shift of about $180^{\circ}$  in the opposite legs of the loop. It was suggested that the oscillation excitation mechanism was directly related with reconnection processes that resulted in the formation of a post-flare loop, and that the excitation mechanism was not due to a low coronal eruption, as in the majority of decaying kink oscillations (Section~\ref{sec:excit}).
 
Transequatorial loop systems are large-scale loop structures that are prone to be perturbed non-uniformly. \cite{2017A&A...603A.101L} used joint imaging and spectral observations and studied such a system as it was impacted by an EIT wave. They found that the transverse oscillation amplitude was very small at the loop apex, so they judged that this loop system oscillated in the second parallel harmonic mode. Using a seismological approach they estimated a magnetic field strength of $5.5\pm1.5$~G which was consistent with a Potential Field Source Surface (PFSS) magnetic field extrapolation. 

Since kink modes can be vertically or horizontally polarised, and observations have projection effects due to the line of sight and loop inclination, these cause ambiguities that bring additional difficulties in the accurate spatial identification of higher harmonics. \cite{2016ApJS..223...24Y} developed a forward model technique to address these issues. They modelled the manifestation of standing kink modes of coronal loops in specific EUV bandpasses, allowing the plasma parameters, loop widths and viewing angles to be adjusted to match observed events of kink loop oscillations (Figure~\ref{fig:yuan2016}). It was demonstrated that these forward models could be used to effectively identify kink oscillation harmonics, their polarisations, reproduce the general profile of oscillation amplitudes and phases, and also show signatures of multiple harmonic periodicities in the associated EUV emission intensity.

\cite{2017ApJ...842...99L, 2019ApJ...881..111L} analysed kink modes in a pair of coronal loops that were only about 10" apart from each other. It was found that these two loops oscillated differently after the excitation by a C-class solar flare. One loop oscillated with a fundamental mode of 4-min periodicity, whereas the oscillation of the second loop could be considered as a superposition of two oscillations, with periods of 4~min and 2-min.  According to the amplitude distribution along the loop (see Figure 4 in \cite{2017ApJ...842...99L}), the 2-min oscillation node was located at the loop top, so the authors interpreted this to be the second parallel harmonic. The 4-min oscillation was interpreted as a fundamental kink mode.  This observation shed light on the dynamics of the excitation and energy propagation of transverse loop oscillations, since the two oscillating loops were so very closely spaced. 

\cite{2015ApJ...799..151G} detected a fundamental kink mode and its second harmonic, with the periods $530.2\pm13.3$~s and $300.4\pm27.7$~s (or $334.7\pm22.1$ s), respectively. In this study, $P^{(1)}_\mathrm{kink}/(2P^{(2)}_\mathrm{kink}) < 1$, and it was concluded that the density stratification had a stronger effect on the period ratio than the axial non-uniformity of the magnetic field variation.

The inversion procedure which allows to deduce parameters of an oscillating loop from the period ratio measurement is rather subjective, and is based on certain hypotheses of the main mechanism affecting the ratio.  \cite{2013ApJ...765L..23A} suggested to use methods of Bayesian statistics to compare the plausibility of two competing hypotheses,  the density stratification and magnetic field expansion. If additional information is available, e.g., the damping profile, such a method could be extended to obtain more accurate seismology. \cite{2017A&A...600A..78P} analysed a kink oscillation with a reasonable damping function, and extracted a set of three parallel harmonics. This information was used for the estimation of the density ratio in the oscillating loop (Figure~\ref{fig:pascoe2017}). 

\begin{figure}
\centering
\includegraphics[width=\textwidth]{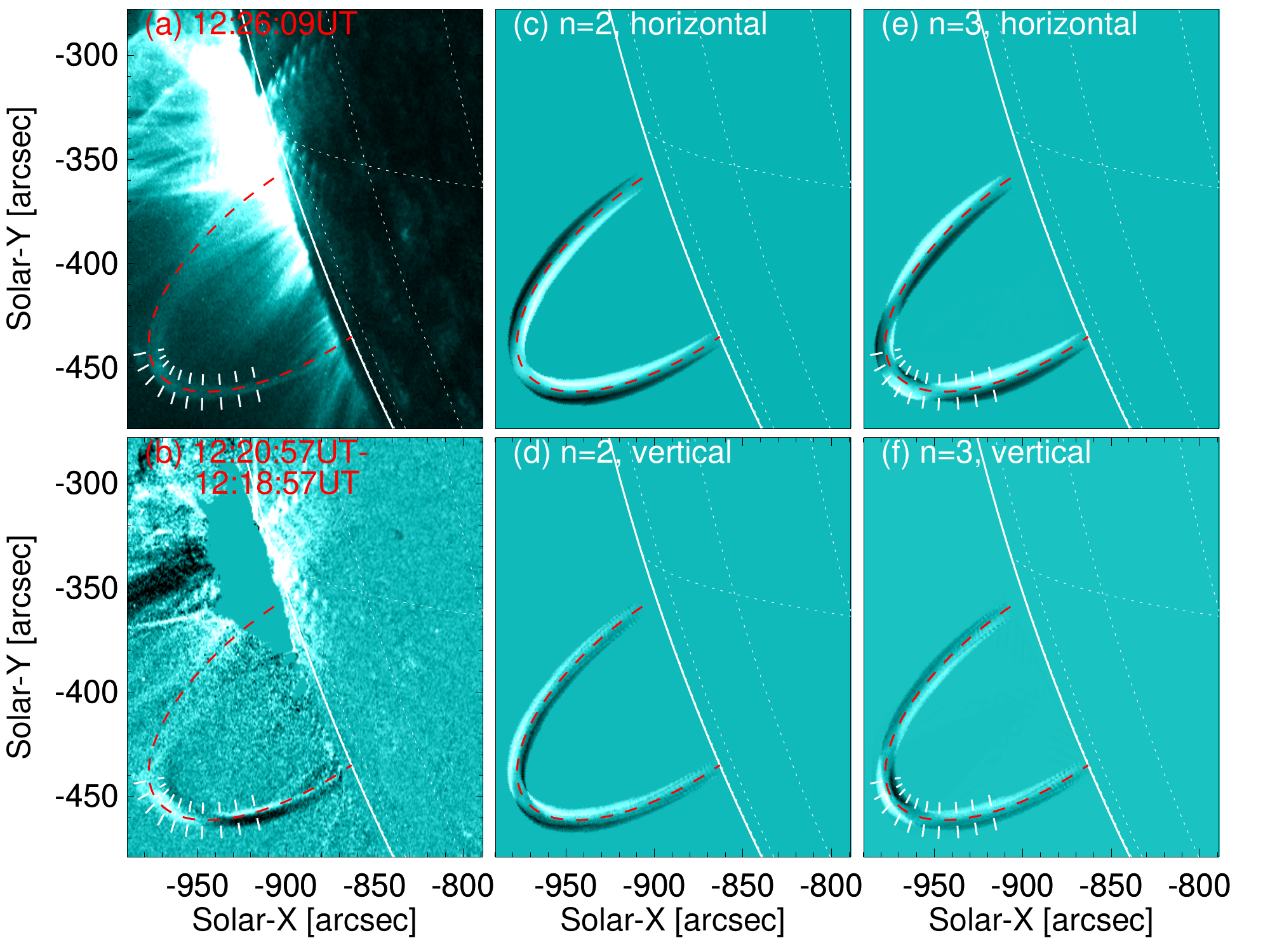} 
\caption{(a) Active region 11121 observed at the south-east solar limb by the SDO/AIA in the 131 \AA{} channel. (b) Difference images made by subtracting two images taken at about half an oscillation cycle apart. (c)-(f) Difference images of synthesised oscillating loop model for various modes. Adaptation of Figure 7 in \cite{2016ApJS..223...24Y} \label{fig:yuan2016}}	
\end{figure}

\begin{figure}
\centering
\includegraphics[width=0.48\textwidth]{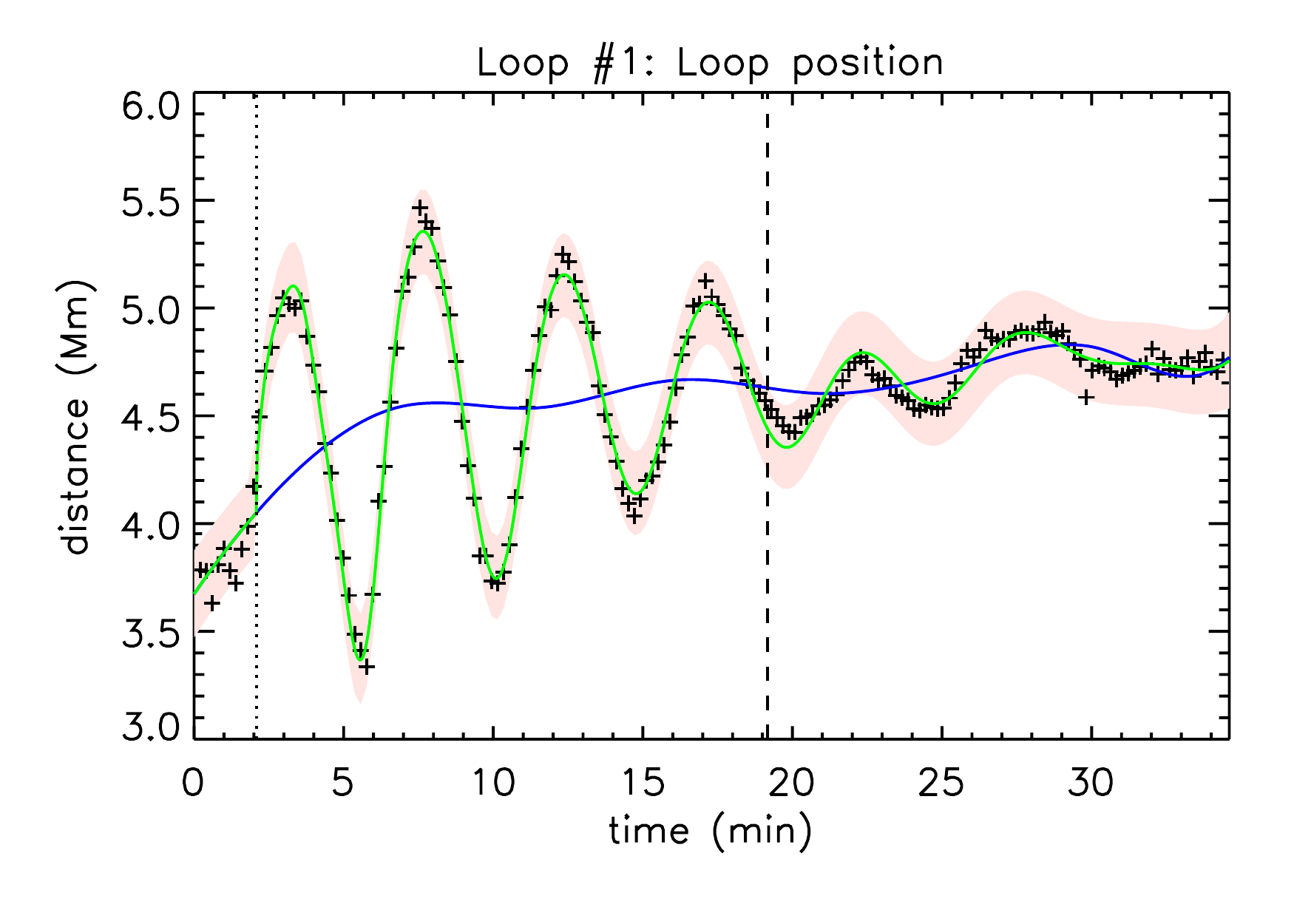}
\includegraphics[width=0.48\textwidth]{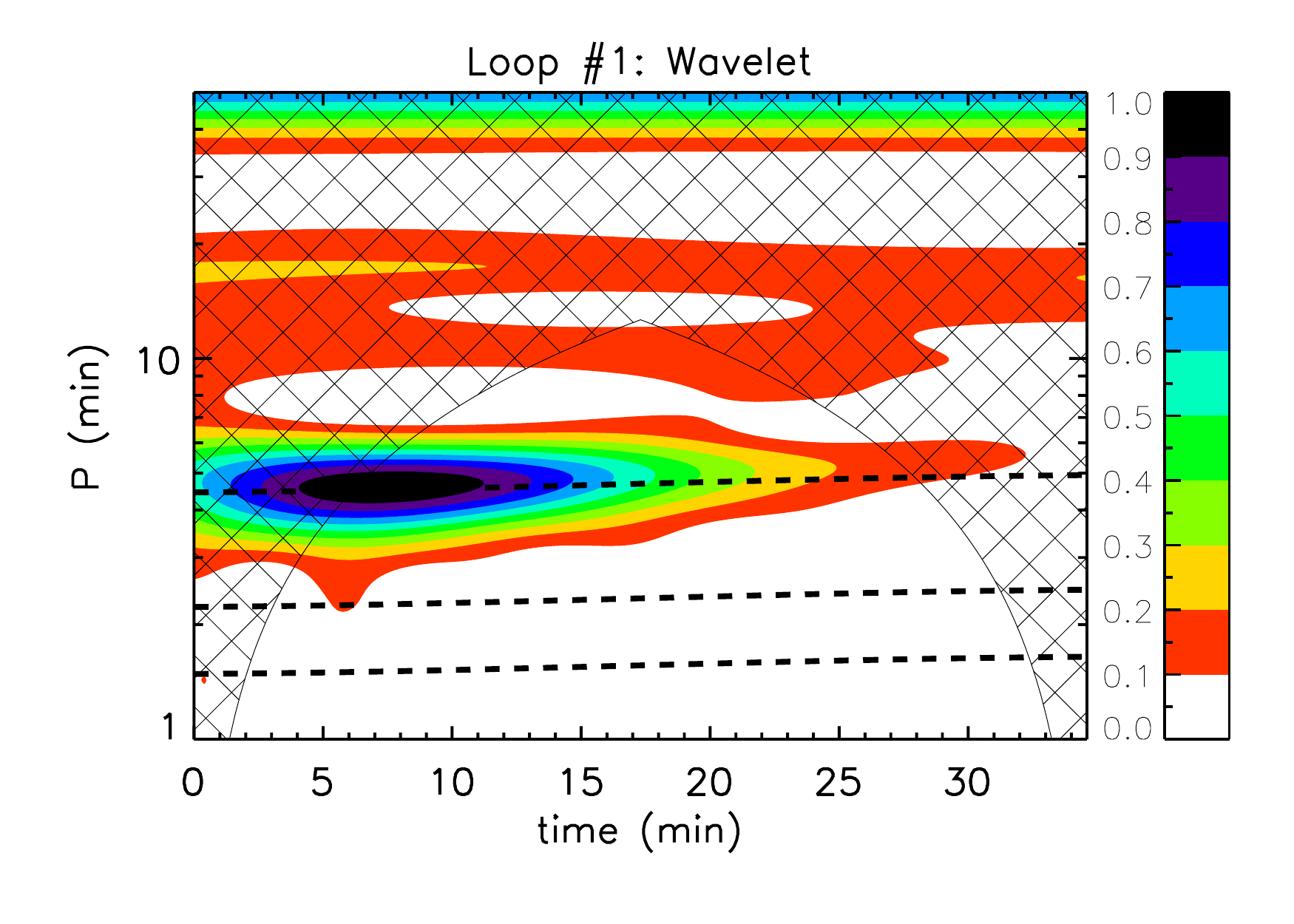}\\
\includegraphics[width=0.48\textwidth]{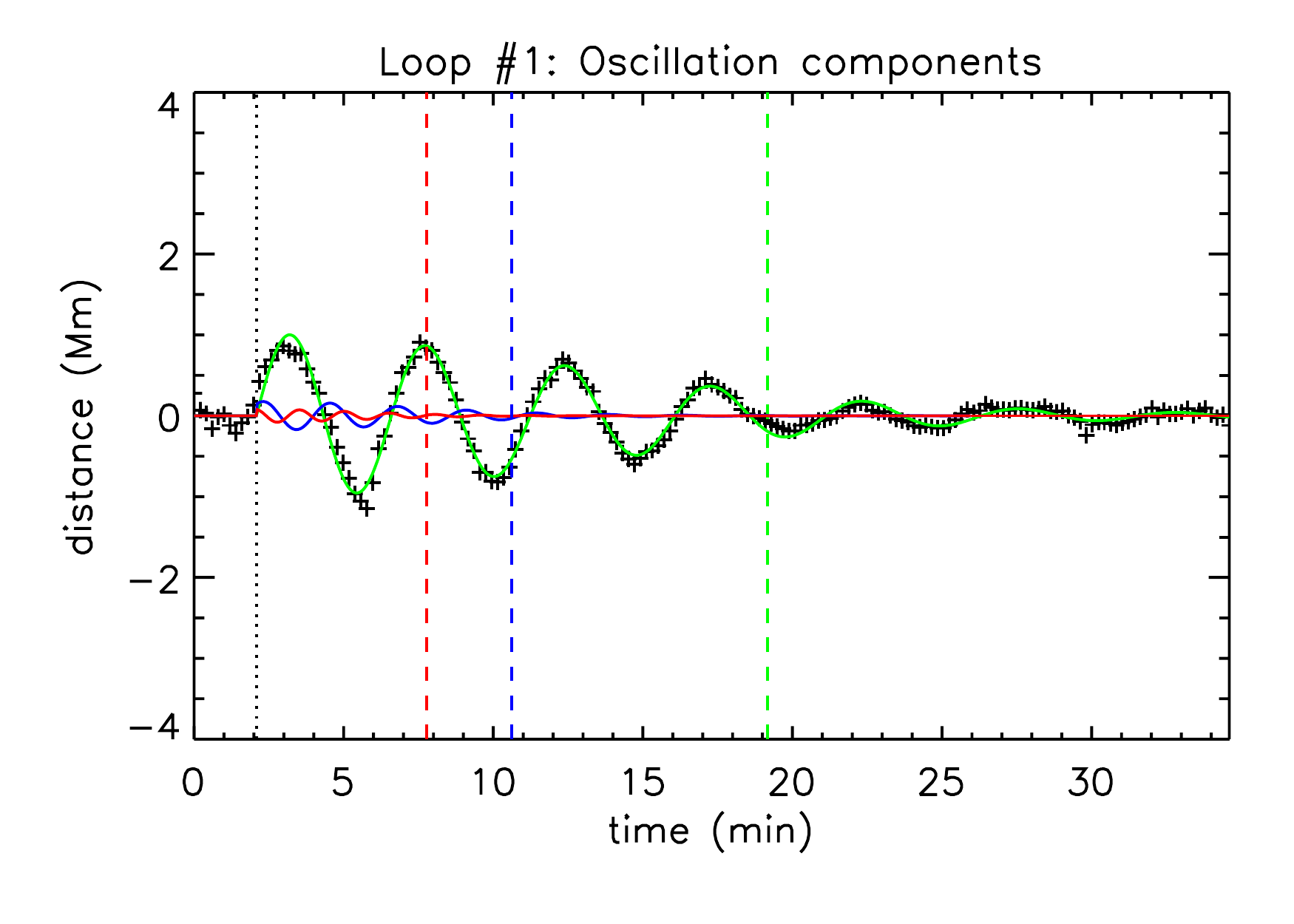}
\includegraphics[width=0.48\textwidth]{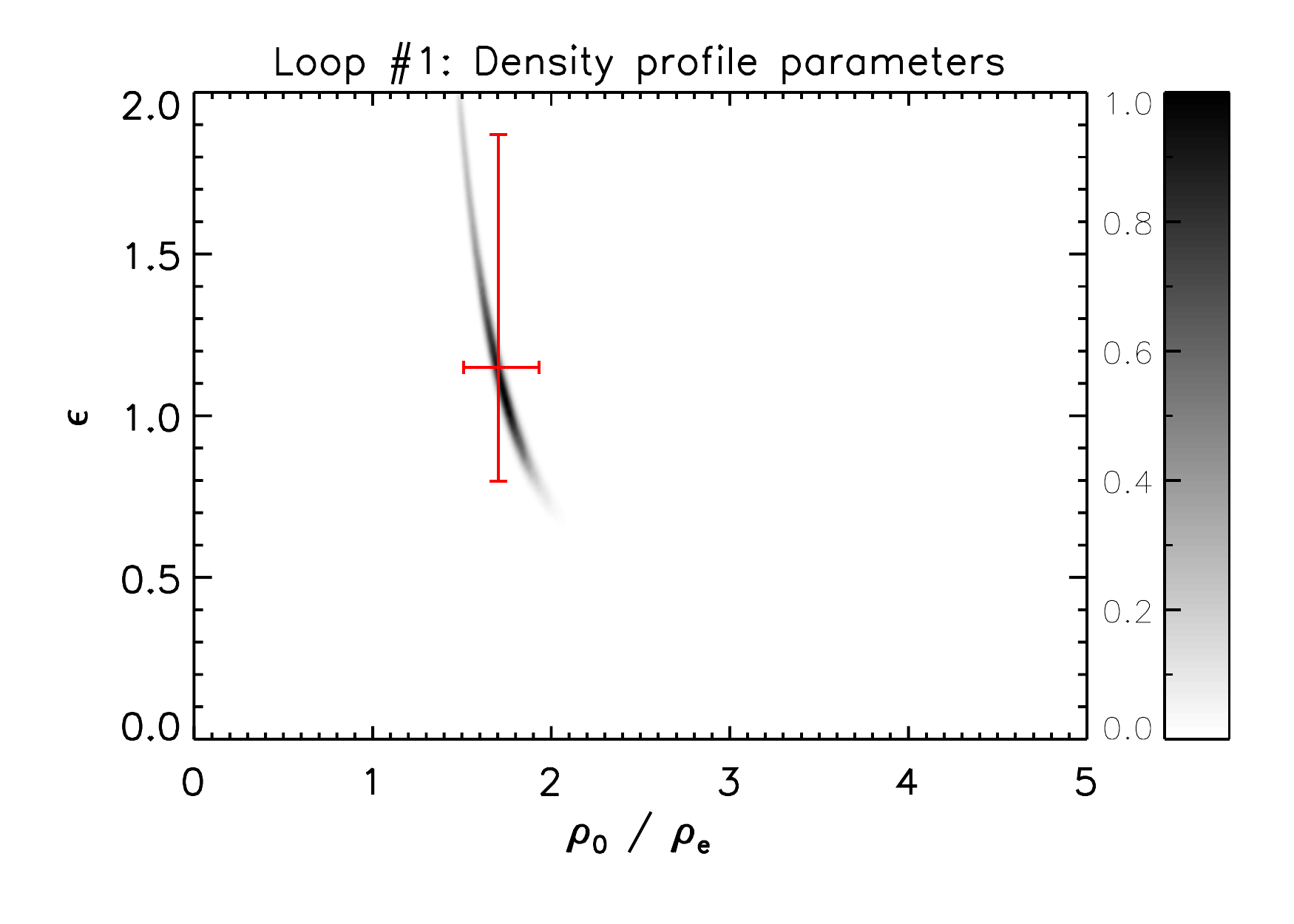}\\
\includegraphics[width=0.48\textwidth]{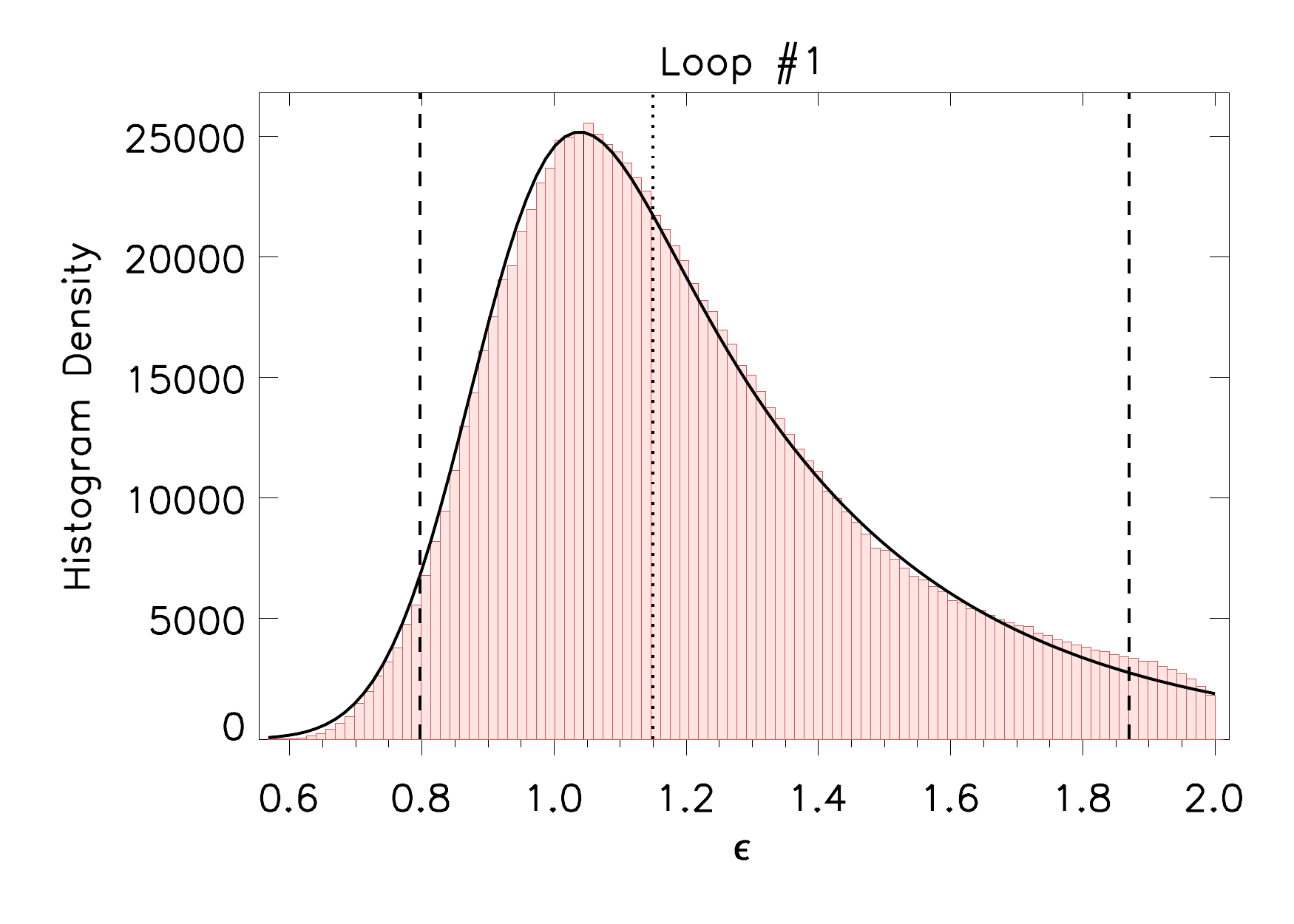}
\includegraphics[width=0.48\textwidth]{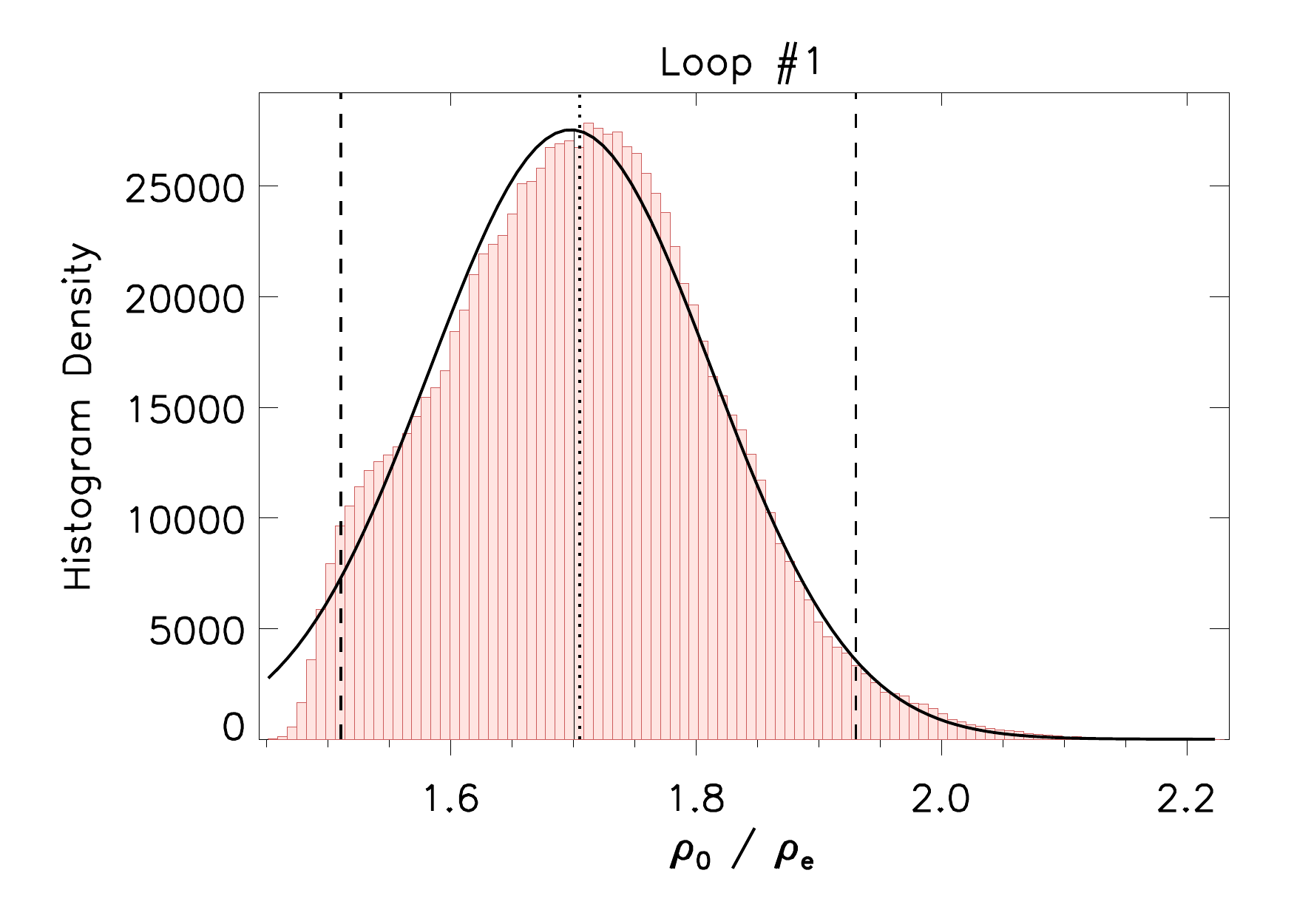}\\
\caption{Bayesian inference with higher harmonic modes.  
Top left: loop displacement as a function of time. The blue line represents a background trend, and the green is the model based on the assumption that the signal includes several decaying harmonic oscillations. The red shaded region represents the 99\% credible intervals for the loop position predicted by the model. 
Top right: wavelet analysis of the loop oscillation,  with colours representing the normalised spectral amplitude. The dashed lines show the time-dependent periods of oscillation described by our model. 
Middle left: detrended loop position (symbols) with the first (green), second (blue), and third (red) longitudinal harmonics. Middle right: density profile parameters determined by the oscillation damping envelope. The red bars are based on the median values and the 95\% credible intervals, indicated by the dotted and dashed lines, respectively, in the histograms (bottom panels). The solid curves are fits to the histogram data using the exponentially modified Gaussian function. Adaptation of Figure 6 in \citep{2017A&A...600A..78P}.}
\label{fig:pascoe2017}
\end{figure}

\section{Nonlinear effects in kink oscillations}
\label{sec:nonlin} 

\subsection{Nonlinear generation of fluting perturbations by kink-driven Kelvin--Helmholtz instability} 
\label{sec:nlgen} 

At high amplitudes, the kink mode evolution can be entirely different than in the small-amplitude linear regime. The nonlinear regime is well characterised by the nonlinearity parameter $\nu_\mathrm{NL}\approx {A L}/{a}$, where $A= {v_0}/C_\mathrm{Ai}$, is the normalised amplitude given by the ratio of the initial velocity amplitude $v_0$ to the internal Alfv\'en speed $C_\mathrm{Ai}$, and $L$ and $a$ correspond to the length and the radius of the loop, respectively \citep{2016A&A...595A..81M, 2014SoPh..289.1999R}. A nonlinear evolution is obtained for $\nu_\mathrm{NL}$ larger than $1$. We shall be focusing on this regime here\footnote{Another common measure of nonlinearity is the ratio $\xi/a$ of the displacement of the flux tube to its radius. This is related to our definition as $\xi/a = \nu_\mathrm{NL} / \pi \sqrt{(1+\rho_\mathrm{0e}/\rho_\mathrm{0i})/2} \approx 0.25\nu_\mathrm{NL}$, taking usual values for the external and internal densities $\rho_\mathrm{0e}$ and $\rho_\mathrm{0i}$.}. Nonlinear regime of the kink mode may lead to the development of KHI and the {Rayleigh--Taylor instability (RTI)}. The most common and important of the dynamic instabilities in terms of influence on the overall structure of the loop is the KHI. Recent analytical and numerical studies have expanded our understanding of the wave modes that accompany the kink mode in its nonlinear evolution. 

\begin{figure}[ht!]
\centering
\includegraphics[width=\linewidth]{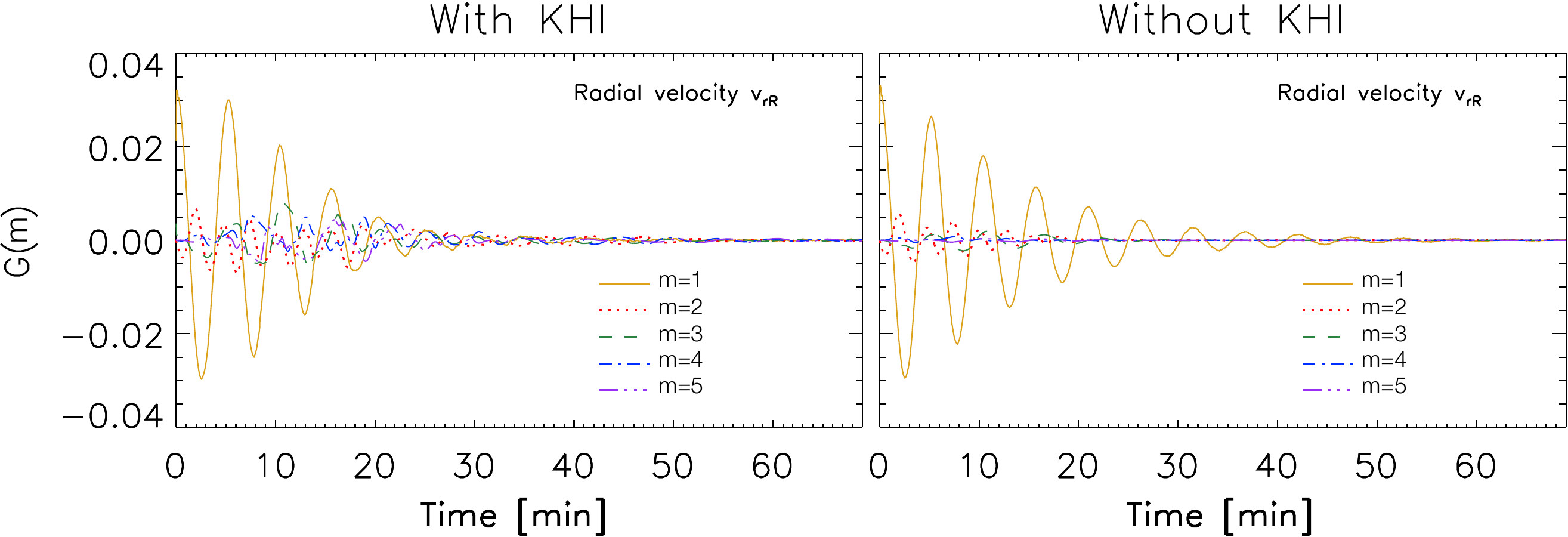}
\caption{Variation of different harmonics of the azimuthal wave number spectra $G(m)$ of the radial velocity in a kink oscillation, measured near the boundary of a simulated coronal loop, in time. Different colours and line styles (see legend) denote different azimuthal modes along a circle fitting the edge of the loop prior to the kink mode. The $m=1$ mode denotes the fundamental mode, and we show up to $m=5$. The numerical simulations corresponding to these results are described in \citep{2019FrP.....7...85A}.} \label{an_fig1}
\end{figure}

The flows which are strongly sheared in the radial direction, resulting from resonant absorption (see Sec.~\ref{sec:ra}) would lead to the rapid development of KHI. The instability is not stabilised by the magnetic field, as the shear flows appear across it. 
The KHI causes a mixing of internal and external plasmas and hence generates or broadens the transition region within which resonant absorption would take place.
The KHI associated with standing kink waves manifests in several different ways  \citep[see, e.g.,][]{2017ApJ...836..219A} for its observable signatures. The characteristic KH vortices occur first at the velocity shear region at the loop interface with the external corona and are therefore azimuthal and exist in the plane perpendicular to the magnetic field \citep{2008ApJ...687L.115T}. Since the amplitude of a standing kink oscillation changes gradually with height, the Kelvin--Helmholtz (KH) vortices become the so-called transverse wave induced Kelvin--Helmholtz (TWIKH) rolls in 3D, with a self-similar shape of slowly varying amplitude \citep{2014ApJ...787L..22A}. The distortion of the transition region in the azimuthal direction, caused by KHI could be considered as the nonlinear generation of fluting modes with increasingly growing azimuthal wave numbers $m$, i.e., the azimuthal nonlinear cascade. 

The number of TWIKH rolls excited at any one time is therefore determined by their azimuthal wave numbers and also by the thickness of the loop boundary layer at the time of the kink mode excitation. Small azimuthal wave numbers have high growth rate and therefore will be excited first if the boundary layer thickness (which determines the thickness of the velocity shear layer) is smaller than the wavelength of the unstable mode.  {One can understand why this is so with the following argument. A vortex is characterised by the top part of the crest moving within one of the layers generating the shear, while the bottom is anchored to the layer moving in the opposite direction. Hence, the size of the vortex cannot be smaller than the thickness of the shear layer. } After a few oscillation cycles the result is a mix-up of unstable modes, whose combined effect is a very efficient mixing of the plasma. Since TWIKH rolls are compressive, the mixing leads to a continuous perturbation of the boundary layer. We can therefore quantify the generation of the fluting modes by measuring the power of each azimuthal wave number in the radial velocity along the boundary of the loop \citep{2018ApJ...853...35T, 2019FrP.....7...85A}.
The distribution of the energy by the azimuthal wave numbers $m$ for the case of a coronal loop of length $L=200~$Mm,  radius $a=1~$Mm, with initial transition layer of width $l_\mathrm{tr}/a=0.4$ and subject to an initial velocity perturbation of $v_0=16.6~$km~s$^{-1}$, corresponding to a nonlinearity parameter $\nu=3.3$, is shown in Fig.~\ref{an_fig1}. The two simulations only differ on the viscosity in the numerical model, leading to a Lundquist number of $10^4$--$10^5$ for the low viscosity (and more realistic) case, and 10--100 for the highly viscous case \citep{2019FrP.....7...85A}. The high viscosity values in the second case effectively inhibit the KHI.
In the case without KHI, i.e., when the viscosity is high, we see the doubling of the frequency, and the fact that the amplitude of the $m=2$ mode is quadratically smaller than that of the $m=1$ mode, thereby matching the theoretical results \citep{2014SoPh..289.1999R}. All other, higher azimuthal wave modes have negligible amplitude. The case with KHI presents a very different picture. In this case, the KHI occurs at $t\approx6~$min, which matches with the time of significant excitation of all azimuthal wave numbers. The amplitudes of the $m=2$ and $m=3$ become comparable to the amplitude of the fundamental mode. 

The increase of amplitude of the $m=2$ mode in the presence of the KHI can be understood as follows. At times of maximum displacement the squashing of the waveguiding cylinder leads to the acceleration of material in the boundary layer in the opposite direction (i.e. backwards), a flow that is further enhanced by resonant absorption {(see the top row in Fig.~\ref{figsquash})}. When the KHI is triggered, since this happens at both sides of the cross-section with respect to the axis of the (linearly polarised) oscillation, we end up with 2 azimuthal flows that collide with each other and generate two compressive vortices. Because of the compression, a magnetic pinch is produced on the plasma at the centre of the loop, which is then accelerated in the opposite direction when the loop starts moving backwards {(bottom row in Fig.~\ref{figsquash})}. This produces a faster leading edge that further enhances the squashing effect, leading to a stronger amplitude for this mode.

\begin{figure}
\centering
\includegraphics[width=7cm]{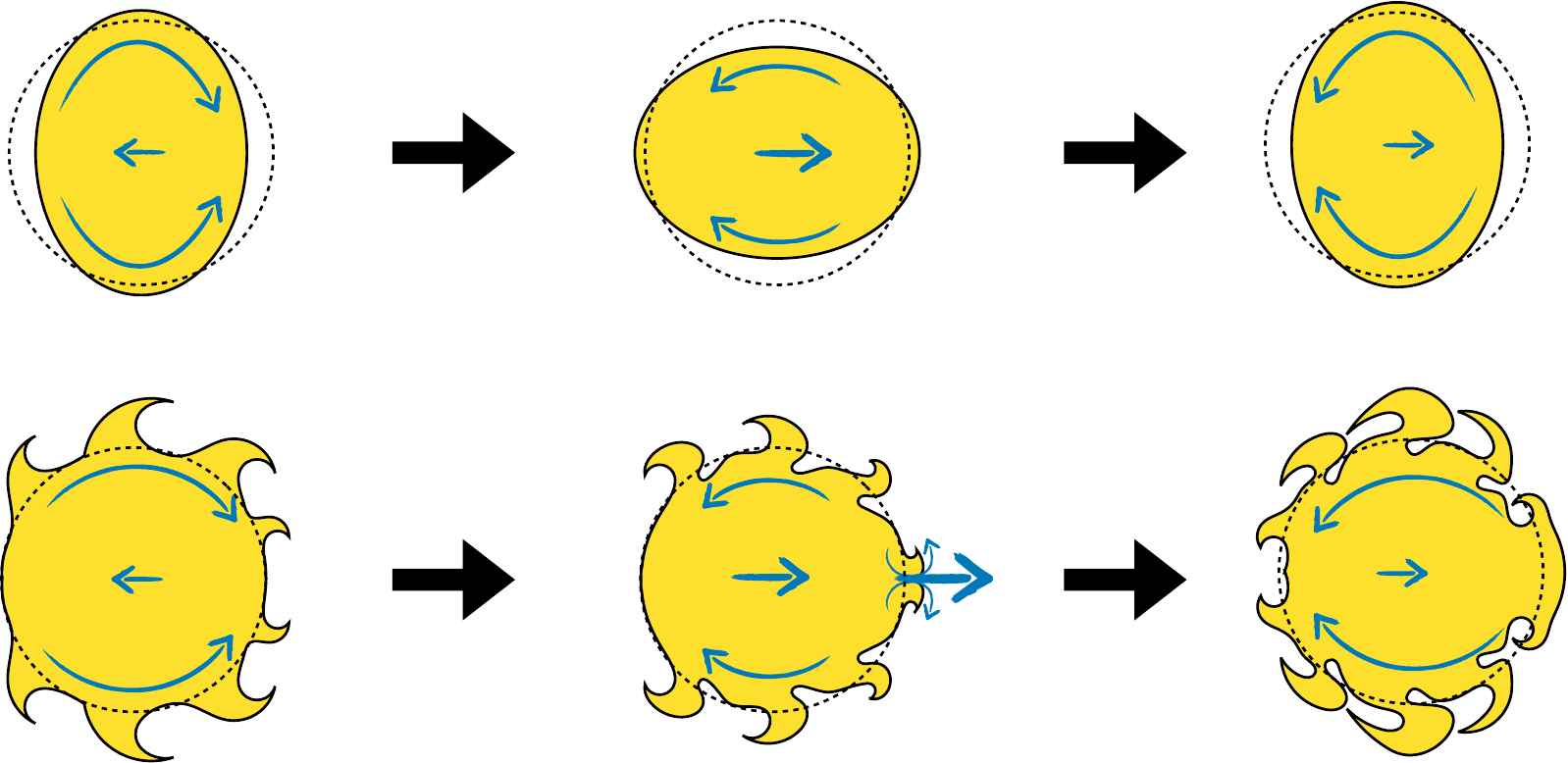}
\caption{Sketch of the cross-section of a cylinder with the m=2 mode with (bottom row) and without (top row) KHI. During the transverse oscillation ($m=1$ mode, in the horizontal direction in the sketch), field lines in the leading edge of the tube have slightly stronger magnetic tension than those at the trailing edge, leading to a stronger (nonlinear) deceleration at the leading edge compared to to the trailing edge, thereby squashing the flux tube (dashed circles in the sketch denote the original shape of the cylinder's cross-section). Since the motions are incompressible, this leads to a slightly elongated cross-section in the vertical direction. Due to the squashing, the magnetic pressure increases, leading to a fast acceleration in the opposite direction. This motion (which is partly azimuthal and symmetric with respect to the oscillation axis) is further enhanced by resonant absorption. When the KHI is triggered, the combined effect of the azimuthal resonant flow and the KHI flow (compressive vortices) leads to a compression at the trailing edge of the flux tube (magnetic pinch) which accelerates the plasma in the opposite direction, thereby increasing the amplitude of the $m=2$ mode.}
\label{figsquash}       
\end{figure}

It is important to note that, because of resonant absorption at the boundary, the perturbations and TWIKH rolls are not confined to the boundary layer. Indeed, as shown in \citep{2019FrP.....7...85A}, the resonant flow produces a velocity shear in a neighbouring shell that is closer to the axis of the cylinder, which will, in turn, become KH unstable. This process repeats inwardly until most of the loop is covered by TWIKH rolls, even in the case of a single initial perturbation.

Besides the KHI, the RTI can also manifest in nonlinear kink mode oscillations. 
When during a linearly polarised kink oscillation, the loop moves into the background plasma, the front edge of the cylinder, which is locally perpendicular to the direction of the motion, is subject to RTI.
The characteristic finger-like RTI structures have been detected observationally by \cite{2018ApJ...856...44A, 2019FrP.....7...85A}, and attributed to the higher magnetic tension force and lower density region of the external medium pressing unto the lower magnetic field strength but higher density region of the loop. The RTI perturbations would also produce compression at the wake, thereby contributing to fluting mode generation. 

\begin{figure}
\centering
\includegraphics[width=7cm]{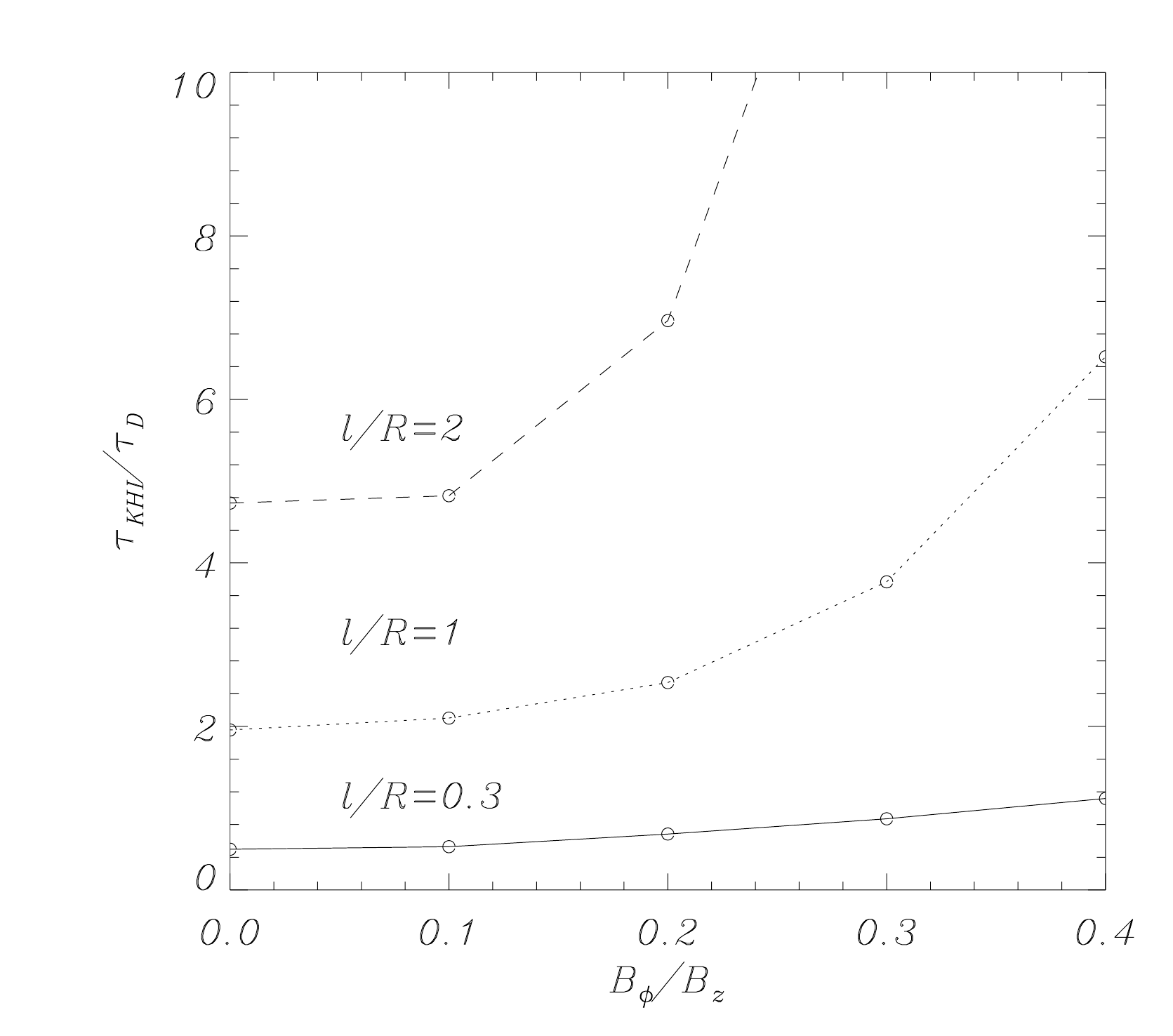}
\caption{Ratio of KHI onset time to the damping time  as a function of
twist for three different widths of the inhomogeneous layer ($l_\mathrm{tr}/a$).}
\label{figtwist}       
\end{figure}

Magnetic twist has significant effects inhibiting KHI induced by a nonlinear kink oscillation. This has been investigated numerically in \cite{2017A&A...607A..77H,2018ApJ...853...35T}. The results indicate that magnetic twist delays the onset of the KHI instability but the thickness of the inhomogeneous layer also plays an important role (as in the untwisted situation). It has been established that in the regime of weak damping, when the inhomogeneous layer is thin, a weak twist does not delay significantly the onset of the instability. On the contrary,  when the inhomogeneous layer is wide, and therefore the damping time small, then twist has a strong stabilising effect, see
\cite{2018ApJ...853...35T}. It is therefore interesting to compare the timescales of the attenuation, $\tau_\mathrm{D}$ with the onset times of the KHI in the presence of the twist. The results are shown in Fig.~\ref{figtwist}. The dependence $\tau_\mathrm{KHI}/\tau_\mathrm{D}$ with
twist is weak for thin layers but significantly strong for thick layers ($l_\mathrm{tr}/a$ large).

\begin{figure*}
\centering
\includegraphics[width=0.45\textwidth]{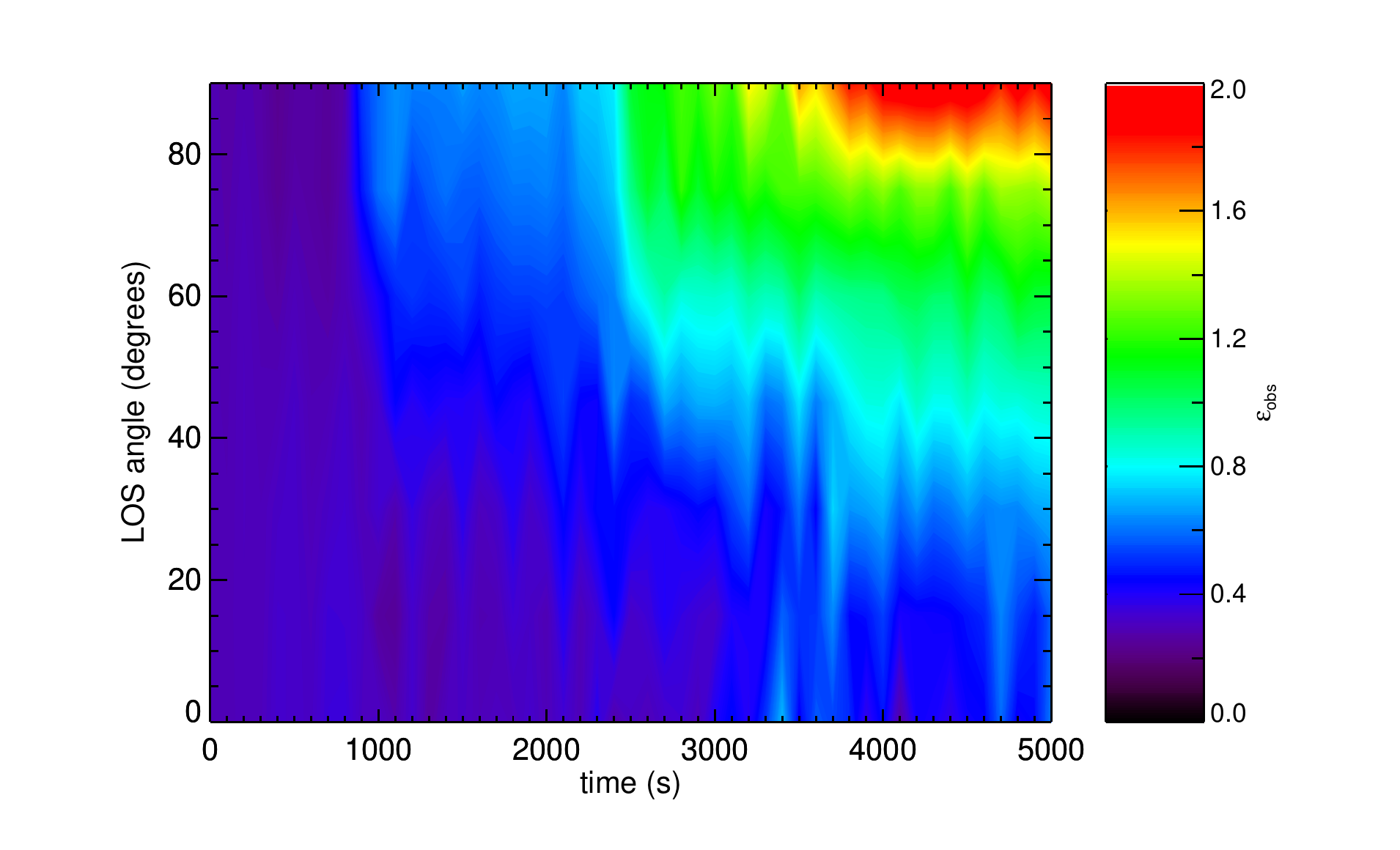}
\includegraphics[width=0.45\textwidth]{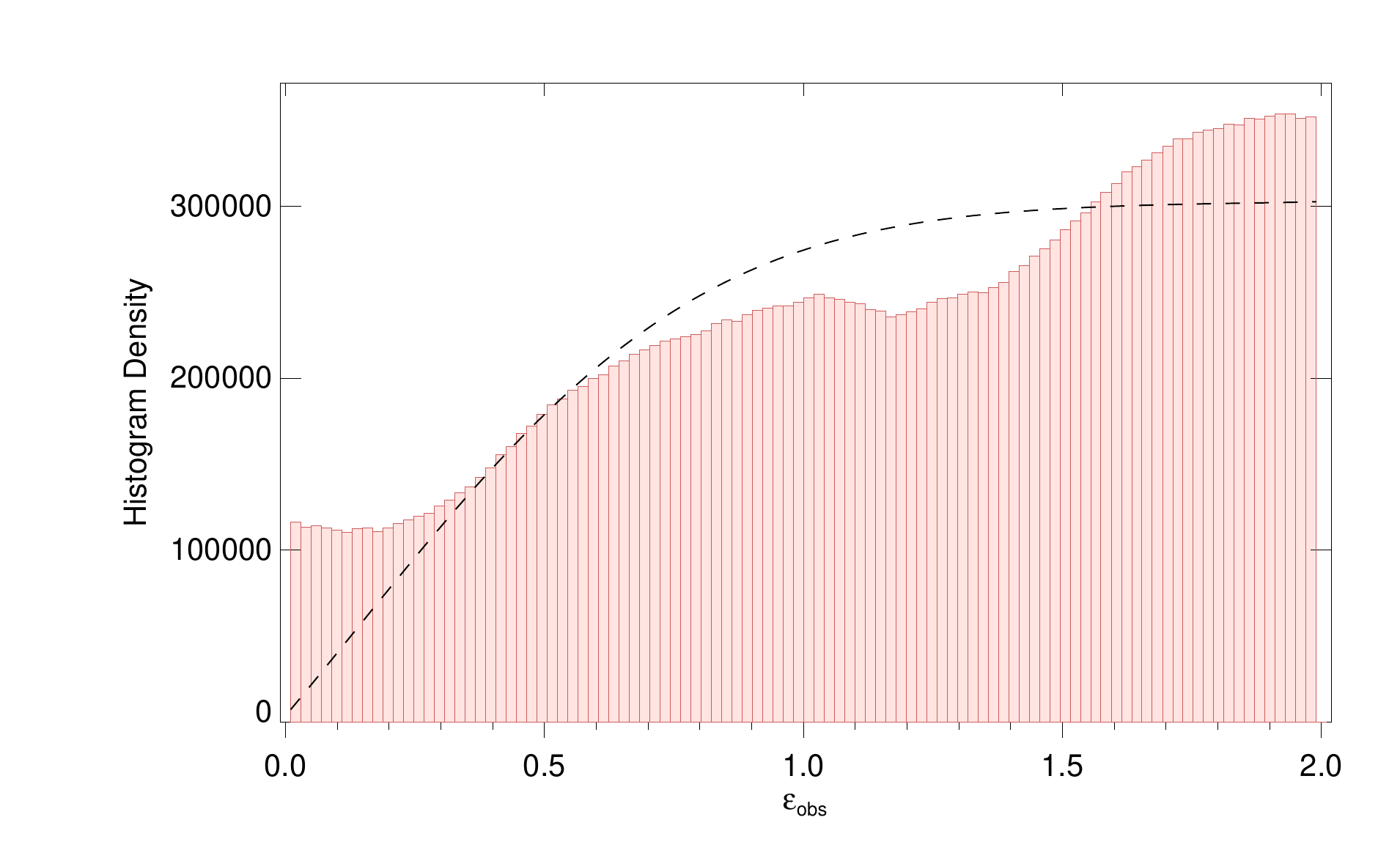}
\caption{The left panel shows the inhomogeneous layer width calculated from the transverse intensity profile of a simulated coronal loop undergoing KHI.
The inferred value increases with time and line-of-sight angle.
The right panel is a histogram of observed values of inhomogeneous layer width based on the statistical study of \cite{2017A&A...605A..65G}. The dashed line represents a model based on the assumption that loops initially have a thin inhomogeneous layer which is broadened due to KHI from ubiquitous decayless kink oscillations.
Figure adapted from \cite{2020FrASS...7...61P}.}
\label{fig:djp_khi1}
\end{figure*}

\begin{figure*}
\centering
\includegraphics[width=0.45\textwidth]{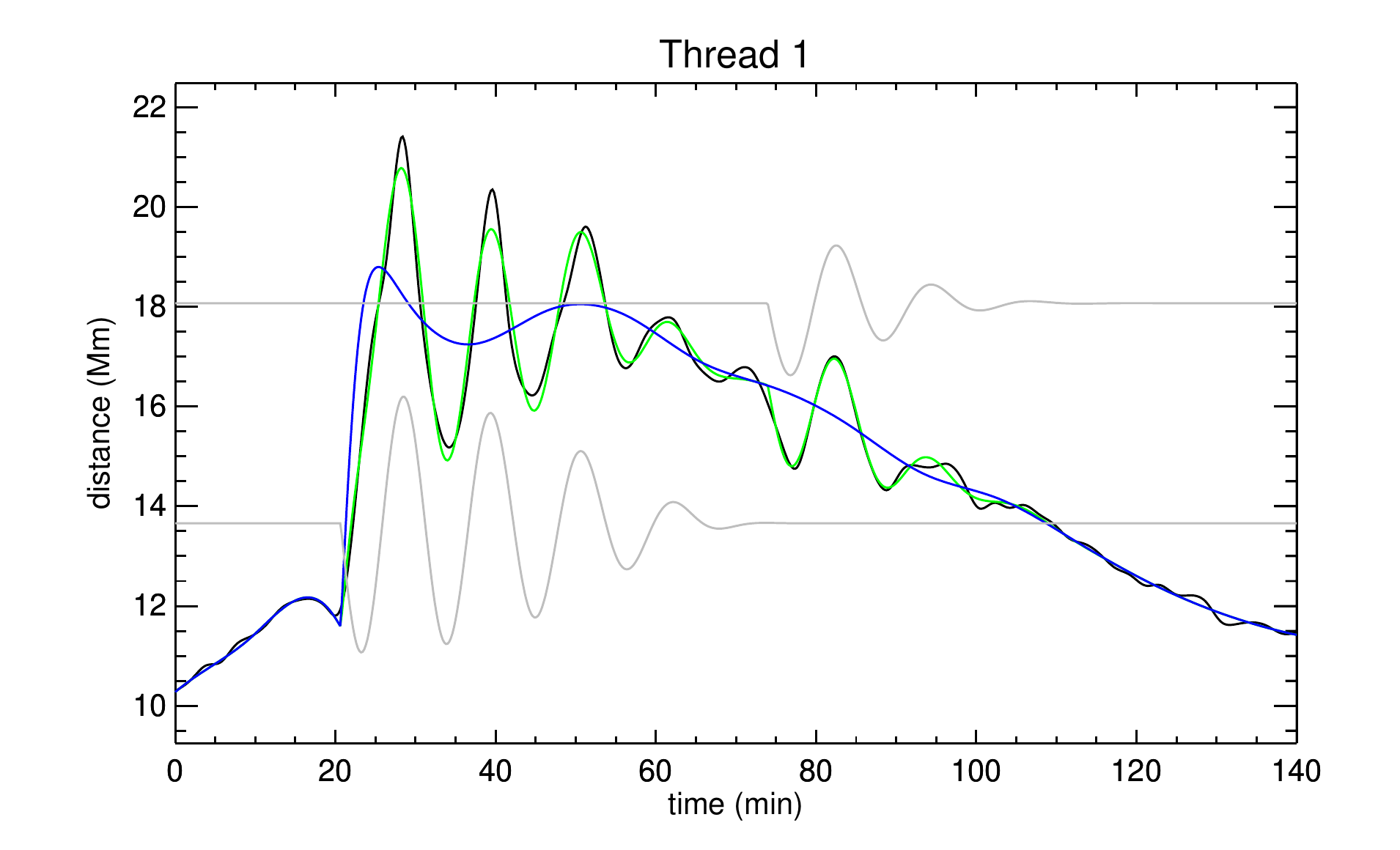}
\includegraphics[width=0.45\textwidth]{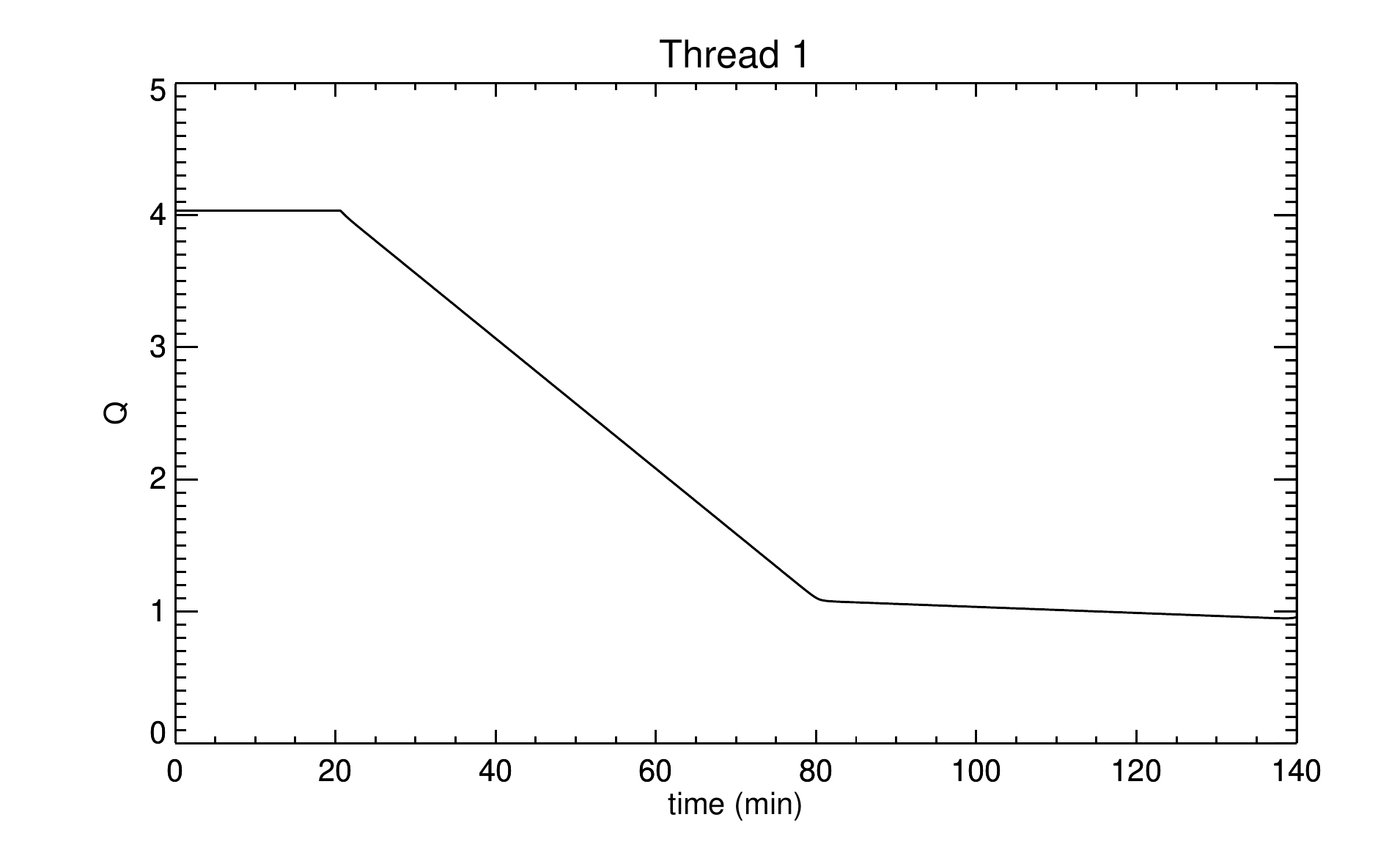}
\caption{
The left panel shows the position of a loop (black line) perturbed by two flares approximately 1 hour apart. The modelled position (green line) is composed of a background trend (blue line) and two damped oscillations (grey lines).
The time dependence of the signal quality (right panel) is consistent with evolution of the loop due to KHI.
Figure adapted from \cite{2020ApJ...898..126P}.}
\label{fig:djp_khi2}
\end{figure*}

\citet{ 2017A&A...602A..74H} demonstrated that the increase in the viscosity and resistivity acts to suppress the KHI in kink oscillation, delaying its onset or completely preventing it. Viscosity was found to have a greater effect on the development of the KHI than resistivity.

\cite{2018ApJ...863..167G} modelled the evolution of the loop transverse intensity profile during the decay of a kink oscillation, as manifested in EUV imaging observations. The results of numerical simulations of kink oscillations with KHI were used as input parameters of the forward modelling. The strongest observational evidence for the KHI is found to be a widening of the loop boundary.
With SDO/AIA this effect may be inferred for sufficiently broad loops, e.g., with more than fifteen data points across the loop. However, such a theoretically predicted widening of the loop boundary has not been detected in direct observations of decaying kink oscillations \citep[see e.g. Fig. 9 in][]{2018ApJ...860...31P}. Additional signatures of KHI are a decreasing intensity, an unchanged minor radius, and visible fine transverse structuring. The appearance of these signatures is delayed for loops with broader inhomogeneous layers, and quicker for kink oscillations of higher amplitudes.
\cite{2020FrASS...7...61P} demonstrated that the widening of the loop boundary also depends on the LOS angle of the observation (Figure ~\ref{fig:djp_khi1}), and how the distribution of widths measured in the statistical study of non-oscillating loops by \cite{2017A&A...605A..65G} is consistent with KHI due to ubiquitous decayless oscillations acting to increase the detected values.
\cite{2020FrASS...7...61P} also calculated the increase in damping due to resonant absorption owing to the broadening of the boundary layer by KHI and found the actual damping to be stronger, indicating that the fine structures generated by KHI also have a role in attenuating kink oscillations.
\cite{2020ApJ...898..126P} analysed oscillations in a coronal loop perturbed by two solar flares approximately 1 hour apart, noting that such multiflare observations are more suited to revealing the evolution of the loop profile since the time required for KHI to develop is comparable to, or greater than, the damping time of kink oscillations. The second oscillation was found to have a lower signal quality (Figure~\ref{fig:djp_khi2}), consistent with KHI generated in the loop by the earlier large amplitude oscillation.

Finally, it is worth mentioning that \cite{2017A&A...607A..77H} have concluded that the presence of twist may increase the ohmic dissipation due to the larger currents that are produced in this case. However, the presence of magnetic twist will likely make the instability more difficult to detect in the corona, but at the same time it will increase its contribution to heating of the solar atmosphere. 

\subsection{Nonlinear coupling of kink and fluting oscillations}

As the generation of TWIKH rolls is a nonlinear cascade in the azimuthal direction, it is of interest to consider nonlinear interaction of a mother kink mode with the daughter fluting modes. The first analytical studies of the kink mode assumed no boundary perturbation, which is equivalent in assuming that the cross-section remains circular. In turn, this means that there is no interaction between the kink modes and the fluting modes. Instead, the first studies on nonlinearity focused on the interaction between the transverse and field-aligned perturbations \citep{1993JETPL..58..507R}.  

For the case of a uniform cylinder with uniform magnetic field, the deformation of the flux tube's cross-section due to the fluting modes can be helical in the case of propagating waves \citep{2010PhPl...17h2108R}, but is symmetric in the case of standing waves. Another main characteristic is that their frequency is double that of the kink mode, regardless of whether the wave is propagating \citep{2010PhPl...17h2108R, 2014SoPh..289.1999R} or standing \citep{2017SoPh..292..111R}. As explained by \citet{2018ApJ...853...35T}, the deformation of the cross-section due to the first harmonic and the doubling of the frequency can be easily understood from a physical point of view by taking the standing wave case. As the cylinder moves sideways it is decelerated due to the magnetic tension from the line-tying condition. Due to nonlinearity, this deceleration is slightly stronger at the head of the cylinder than at the rear. Also, the higher inertia for the denser, middle parts of the tube implies a less strong deceleration. These effects lead to a squashing of the cylinder along the direction of the oscillation at times of the maximum displacement. Because of the negligible compressibility of the kink wave in the long-wavelength limit, this squashing necessarily means a proportional enlargement of the flux tube's width in the direction perpendicular to the oscillation in a way to keep the cross-sectional area unchanged. Similarly, at times of zero displacements coinciding with the maximum velocity, the front of the cylinder travels at a slightly faster speed than the rear, leading to an overall oval shape of the cross-section of the cylinder at those times. 

The generation of fluting modes  enhances the resonant damping of the wave. This is because in the thin-tube approximation the kink mode and the fluting modes propagate at the same speed (the kink speed), and thereby resonate strongly (even in the absence of a resonant layer in the case of an inhomogeneous boundary layer). This resonance transfers energy from the fundamental mode into the higher azimuthal harmonics (fluting modes), which damp faster due to their shorter wavelengths \citep{2010PhPl...17h2108R}. The transfer of energy is increasingly more efficient for larger nonlinearity. Hence, the stronger the nonlinearity is, the stronger is the damping. Still, the amplitude of the overtones is quadratically dependent on the amplitude of the fundamental mode, which means that the amplitude of the fundamental mode is always expected to be much larger than that of the overtones regardless of the nonlinearity \citep{2014SoPh..289.1999R, 2018ApJ...853...35T}. Further characteristics include a sausage mode perturbation in addition to the fluting perturbation and a loop apex location for the maximum of the fluting perturbation \citep{2017SoPh..292..111R}.

On the other hand, in the case of a longitudinally stratified loop (and as long as the amplitudes are not too high\footnote{Formally, $\nu_\mathrm{NL} \lesssim |2-\omega_2 / \omega_1 |$, where $\omega_1$ and $\omega_2$ correspond to the frequencies of the fundamental mode and first harmonic.}), the nonlinearity does not cause damping \citep{2014SoPh..289.1999R}. This is because a change in the oscillation frequency is obtained, which destroys the resonance between the kink mode and the fluting modes, thus preventing the transfer of energy.

\subsection{Field-aligned flows driven by the ponderomotive force in kink waves} 

Apart from siphon or reconnection-driven flows, there also could be field-aligned flows induced by MHD waves. 
In the case of kink waves, axial flows of the plasma could be driven because of nonlinear effects,  due to the ponderomotive force. A distinction must be made between \lq\lq genuine\rq\rq\ flows that transfer the plasma in a certain direction and do not return it back, and velocity components of waves along the field. Slow magnetoacoustic waves are well-known to manifest in field-aligned velocity perturbations. However, when propagating through a plasma which is inhomogeneous across the field, any transverse wave has field-aligned velocity perturbations, as the components of velocity perturbations are linearly coupled \citep[see, e.g.,][]{2019ApJ...882...50M}. The distinction between flows and waves is clear in theory, but it is less so in observations \citep[see][]{2010ApJ...724L.194V, 2015SoPh..290..399D}.

The ponderomotive force arises nonlinearly whenever there is a gradient of magnetic pressure, which generates a plasma flow. The force density vector is defined by
\begin{equation}
\mathbf{F} = - \nabla \left(B^2_\perp\right)/\mu_0,
\end{equation}
where $B_\perp$ is the magnetic field perturbation
perpendicular to the background magnetic field. The nonlinear nature of this force is evident, given the quadratic dependence of magnetic pressure on field perturbations. Kink waves as other transverse waves readily induce magnetic pressure variations, thus inducing field-aligned flows. An exception is the case of circularly-polarised and propagating kink waves, for which the magnetic pressure perturbation is constant along the field. Nevertheless, circularly-polarised standing kink waves still induce a ponderomotive force, as magnetic field perturbation nodes are still present (at the apex in the case of the fundamental mode). In coronal loops, the ponderomotive force of standing kink waves is known to lead to flows towards the apex, causing density enhancement there \citep{2009SSRv..149..255T}. However, for typically observed standing kink wave amplitudes, this effect is small (less than 3\%), explaining the lack of its detection \citep{2004ApJ...610..523T}. Gas pressure counteracts the continuous mass accumulation, even in the limit $\beta \ll 1$, leading to the saturation of this effect \cite[see][for some quantitative estimates]{1994JGR....9921291R}. The flows induced by the ponderomotive force are therefore periodic (with double the period of the inducing wave) and point in the same direction in both phases of the wave. This direction is the kink wave propagation direction \citep{2017ApJ...840...64S, 2018ApJ...869...93M}, and from magnetic field antinodes to nodes in the case of standing kink waves. Figure~\ref{two} shows the flow induced by the ponderomotive force in a numerical simulation of a fundamental standing kink mode of a plasma cylinder.
\begin{figure}
  \centering
  \medskip
  \includegraphics[width=0.5\textwidth]{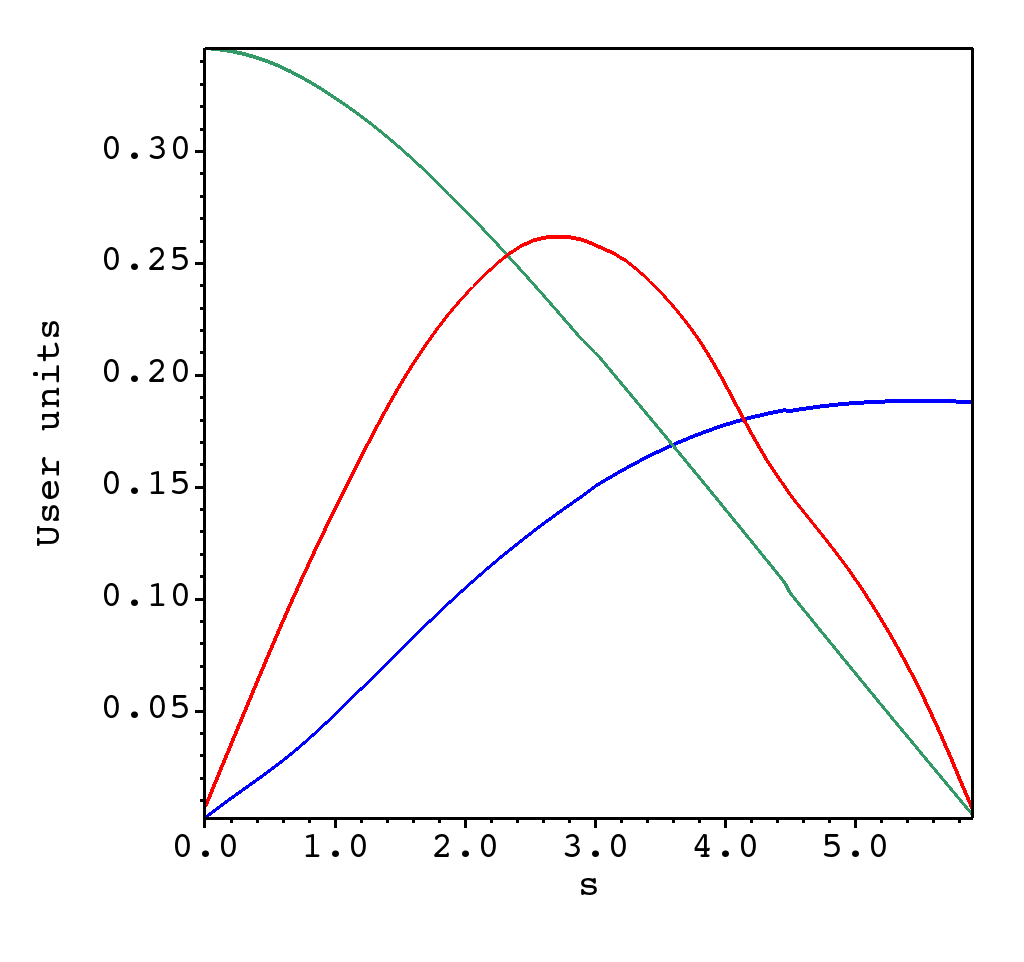}
  \caption{The variation of transverse velocity perturbation (blue), transverse magnetic field perturbation (green), and axial flow (red) along the middle of a plasma cylinder, from the footpoint ($s=0$) to the apex ($s=6$), at some intermediary displacement. The initial velocity perturbation amplitude is 10\% of the internal Alfv\'en speed. The axial flow speed is multiplied by 10. Data from a numerical simulation presented in \citet{2016A&A...595A..81M}.}
  \label{two}
\end{figure}
 For propagating waves, the flow induced is also referred to as \lq\lq Alfv\'enic wind\lq\lq. 
 
In large loop oscillations, the ponderomotive force could affect the phenomenon of the coronal rain \citep{2017A&A...598A..57V}, but it is probably not the primary reason for the observed sub-ballistic motions of the plasma blobs \citep{2012ApJ...745..152A}. The ponderomotive force could be a possible cause of the first ionisation potential (FIP) effect, related to the enhancement or depletion, respectively, in coronal abundance with respect to photospheric values of elements with FIP below about 10~eV \citep{2015LRSP...12....2L}. Besides its effects in the case of kink oscillations, the ponderomotive force has been shown to be omnipresent in numerical simulations of turbulent coronal loops, and it acts as a source of flows, alongside reconnection \citep{2016ApJ...831..160D}. Additionally, it was shown that spicules might be driven by propagating Alfv\'en and kink waves via the steepening and ultimately shock heating of the flows driven by the induced ponderomotive force in the upper chromosphere \citep[e.g.,][]{2016ApJ...829...80B}.

\section{Excitation of kink oscillations}
\label{sec:excit}

Decaying kink oscillations are usually detected as an impulsively excited oscillation with the decay phase being longer than the growing phase.
The initial amplitude of kink oscillations is typically several Mm or several minor radii of the loop.  
The excitation has been associated with impulsive releases of energy and shown to be caused by several mechanisms. Analysis of 58 kink-oscillation events, performed by \citet{2015A&A...577A...4Z}, showed that 57 events (98\%) were accompanied by lower coronal eruptions/ejections. Thus, it is natural to link the onset of the oscillation with an impulsive\footnote{Here, by \lq\lq impulsive\rq\rq, we mean a process with a characteristic time much shorter than the period of the excited oscillation.} deposition of the energy associated with the eruption. On the other hand, it remains unclear whether the initial perturbation provides the loop with a push, i.e., kinetic energy, or a displacement from the equilibrium, i.e., the potential energy, or their combination. The estimation of the speed of the agent which carries the energy from the site of its release to the loop is usually lower than  500\,\kms. This value is at least two times lower than the expected fast magnetoacoustic speed in a coronal active region. Thus, most likely this agent is not a fast magnetoacoustic wave or a fast blast wave. The latter option is also inconsistent with the observed appearance of Type-II radio bursts in only 40\% of kink oscillation events, which shows that coronal shock waves appear in less than a half of kink oscillation events. A remaining option is a displacement of the loop from an equilibrium by a slowly moving plasma blob or a magnetic {rope}. In this case, as the main displacement is experienced by the loop top, the preferentially excited kink mode is the fundamental one. For example, \citet{2018MNRAS.480L..63S} studied an observation of a simultaneous excitation of a  kink oscillation with the period of 428~s by an impingement of a coronal jet on the loop. If the eruption is sufficiently localised in space and touches a segment of the loop in one of its legs, the second harmonic would be excited too. The preferential polarisation of the oscillations excited by this mechanism seems to be horizontal, as the erupting plasma pushes the plasma in the direction perpendicular to its motion. However, no theoretical modelling of this process has been performed yet, and the efficiency of this mechanism remains unknown. 

Another possibility is association of the initial displacement of a loop with a sudden destruction of the magnetostatic equilibrium in a form of the reduction of the magnetic pressure nearby the loop, e.g., an expansion or implosion of a loop system \citep[see, e.g.,][for observational examples]{2012ApJ...749...85G, 2013ApJ...777..152S}. For example, the equilibrium achieved by the magnetic pressure gradient force acting at the loop from the magnetic field under it, and the magnetic tension force directed from the loop downwards, which depends on the major radius of the loop, could be violated by a sudden decrease of coronal magnetic energy, caused by magnetic reconnection under the loop \citet{2015A&A...581A...8R}. In this scenario, the kink oscillation takes place around a new equilibrium, which may explain the mismatch between the initial displacement of the loop, and the following up oscillation. 

Kink oscillations of individual loops could be induced by a collision of counter-streaming upflows along the loop, suddenly generated at the footpoints \citep{2018ApJ...861L..15A}. In this scenario, the excitation of kink oscillations is effective if the colliding fronts are offset each other, for example, are oblique. Oscillations with the amplitudes up to several minor radii of the loop have been successfully excited in a numerical experiment for the plasma $\beta$ ranging from 0.09 to 0.36. This mechanism produces kink oscillations polarised in an arbitrary  plane, which does not allow us to explain the preferential excitation of horizontally polarised oscillations. Vertically polarised kink oscillations could also be excited by a centrifugal force caused by an unsteady flow of a plasma along a bent magnetic field in the loop  \citep{1989SvAL...15...66Z}. This effect is most pronounced in short loops, and proportional to the square of the flow speed. In a loop with the major radius of 90~Mm and the density contrast of 10, modelled by a bent plasma slab, an axial flow pulse with the amplitude of 80~\kms has been numerically shown to excite a kink oscillation with the displacement amplitude of 1~Mm \citep{2017A&A...606A.120K}. 

In some cases, the increase in the amplitude of decaying kink oscillations occurs not suddenly, but in one or two oscillation cycles \citep{2009A&A...502..661N}. A possible excitation mechanism could be associated with a build-up of the oscillation amplitude in response to a periodic driver acting in resonance with the natural frequency of the loop. Interestingly, effective excitation occurs even when the driving force has only one or a few oscillation cycles, see Fig.~2 of \citet{2009A&A...502..661N} who suggest shedding of Alfv\'enic vortices as the driving force. In addition, the periodic driver could be 5-min or 3-min oscillations {producing transverse displacements of the footpoints}, i.e., associated with p-modes or chromospheric oscillations, respectively. If the monochromatic driver operates continuously, the change of the driven oscillation behaviour from a rapidly growing oscillation to a gradually decaying one could occur when either the loop is sufficiently deformed by KHI (see Section~\ref{sec:nlgen}) or some dynamic process changes the natural frequency, destroying the resonance. However, acting on a bundle of loops with different resonant frequencies, this mechanism should preferentially excite oscillations in the loops with the natural frequencies close to the driver frequency, which has not been observed. The driver which periodically pushes the loop in the transverse direction could be a periodic centrifugal force associated with periodic slow magnetoacoustic waves propagating along the loop. Such, quasi-monochromatic slow waves are often detected in coronal loops \citep[see][]{2012RSPTA.370.3193D}. However, this mechanism has not been studied yet. 

Nevertheless, neither of those mechanisms explains the relative rareness of decaying kink oscillations, as in the whole cycle 24, only a few hundred kink oscillations of coronal loops was detected. This number is a small fraction of all impulsive energy release events in this period of time. Perhaps, the rareness of kink oscillations could be attributed to the saturation of CCDs  in the vicinity of low coronal eruptions during an energy release event, as the oscillations could be excited in a narrow expansion cone around the eruption. The investigation of kink oscillation excitation mechanisms remains at the cutting edge of modern research. 



\section{Decayless kink oscillations} 
\label{sec:decless}

A careful inspection of AIA data has led to the detection of a puzzling decayless regime of kink oscillations with an amplitude $<$ 1 Mm, i.e., near the very threshold of the instrumental resolution, and a period of several minutes similar to that of the large-amplitude decaying kink oscillations. The key features of this mode are that the oscillations do not exhibit a systematic decaying trend and the oscillation amplitude remains almost constant, or sometimes increases, and the relatively stable oscillation phase. Therefore, to distinguish them from the standard large-amplitude rapidly-decaying kink mode, these oscillations have been termed as \lq\lq {\it decayless}\rq\rq \citep{2013A&A...552A..57N}.

\subsection{Observations of decayless oscillations }  
\label{s:first_decayless}

First observations of decayless kink oscillations are given in \citet{2012ApJ...751L..27W}, who analysed oscillations growing in time in a multi-stranded loop system, visible in the AIA field-of-view on 8 March 2011. The loop strands were apparently multi-thermal, as they were simultaneously observed in the 171, 193, and 211~\AA\ passbands of AIA. The growing oscillations of the loop strands were mainly observed in the 171~\AA\ channel having similar periods, while the oscillations in 193~\AA\  exhibited almost constant amplitude,  and a quarter-period phase delay between two close strands. The growing oscillations were assumed to be caused by a driver providing a continuous energy supply at a rate faster than the damping. The phase difference in the oscillations of nearby strands confirms those obtained in simulations of collective kink modes in a loop bundle \citep[see, e.g.,][]{2008ApJ...676..717L}. 

\begin{figure}[htpb]
\centering
		\includegraphics[width=1.0\textwidth]{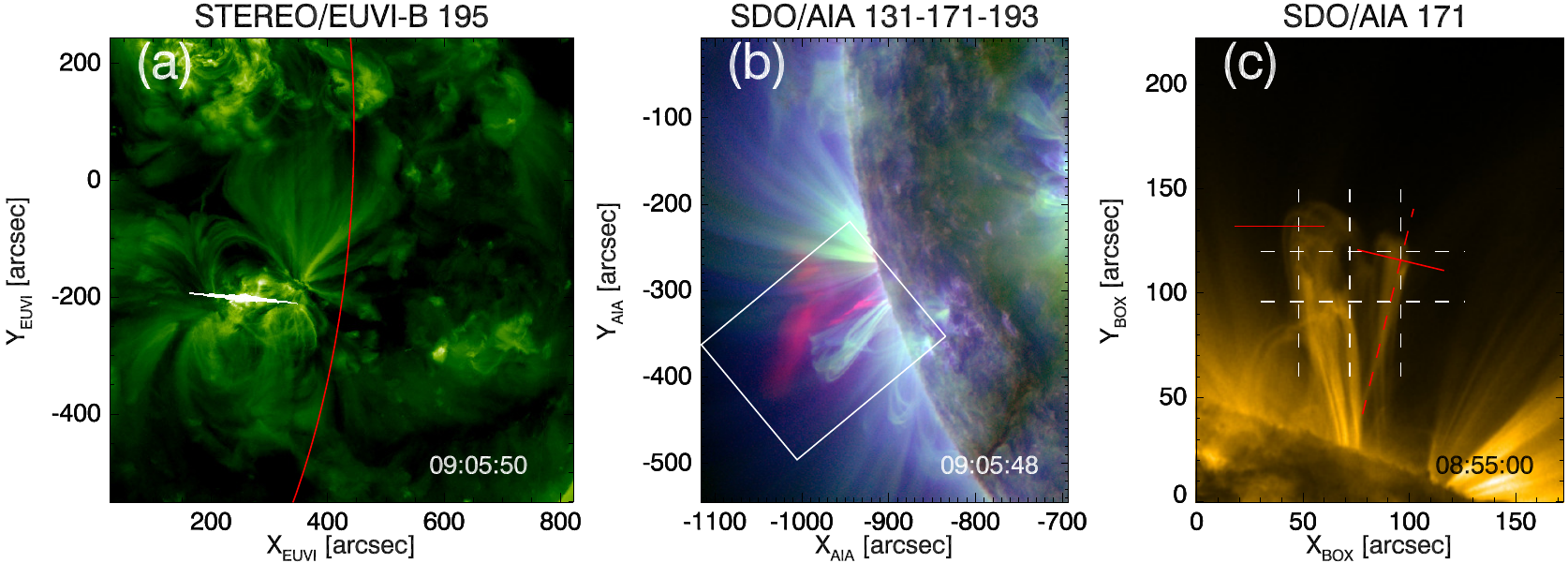} \\
	\caption{Coronal loops exhibiting decayless oscillations. {a)} View of the region of interest from STEREO-B at 195~\AA~with the flare signature, in red the solar limb as seen from SDO. {b)} Multi-wavelength image from SDO/AIA with the coronal loop system within the white square. The plasma in the red channel highlights hot plasma erupted after the flare and triggering transverse oscillations. {c)} Coronal loops in the 171~\AA\ channel with lines in red and white marking artificial slits to construct time-distance maps. Adapted from \citet{2013A&A...552A..57N}.}
	\label{nis_fig1}
\end{figure}

It is in \citet{2013A&A...552A..57N}, however, that the term \lq\lq decayless\rq\rq~was used for the first time to characterise the appearance of persistent kink oscillations. They studied a system of coronal loops on the East limb of the Sun, observed on 30 May 2012, which hosted a decaying large-amplitude transverse oscillation event, triggered by a local flare-related eruption, as also shown in the panels a) and b) of Fig. \ref{nis_fig1}. These observations are in agreement with the conclusions reached in \citet{2015A&A...577A...4Z} (see also Section~\ref{sec:excit}). 
However, analysis of the time-distance maps 
before and well after the decaying large-amplitude oscillation, permitted to detect some periodic transverse movements of the loops. The amplitude of these oscillations was found to be $<$1 Mm, while the periods were  the same as of the large-amplitude oscillation, i.e., 3--5 min, but with no obvious damping. Often, the same loop is observed to oscillate in both those regimes in different periods of time (see Fig.~\ref{nis_fig2}). The oscillation period remains the same in both these regimes.

\begin{figure}[htpb]
\centering
		\includegraphics[trim= 0 145 1 1,clip, width=0.7\textwidth]{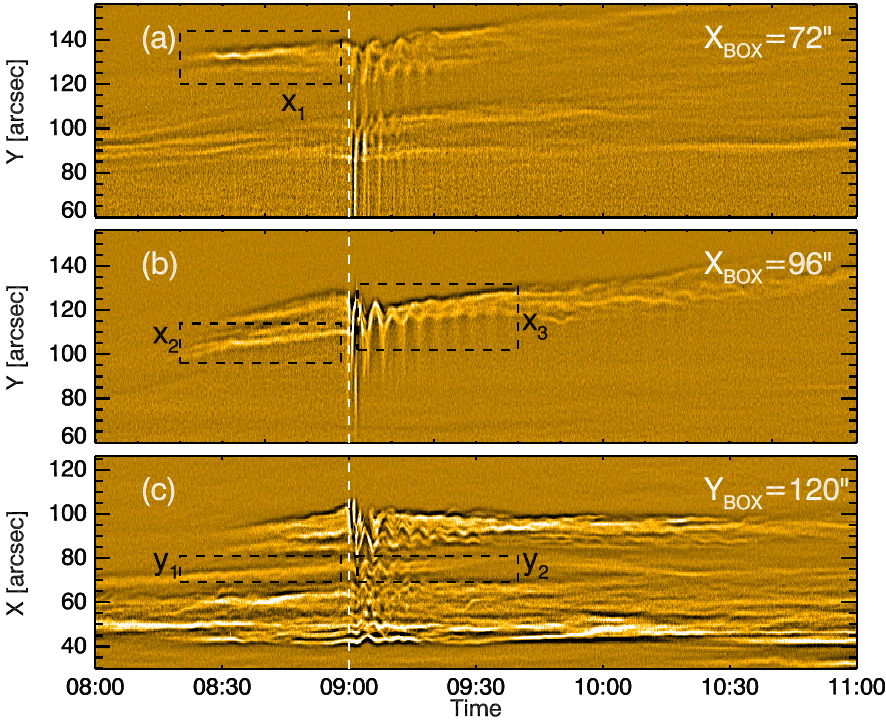} \\
	\caption{Time-distance maps from \citet{2013A&A...552A..57N} showing low-amplitude decayless kink oscillations within the dashed black squares $x_1, x_2, y_1$ and large-amplitude decaying oscillations in $x_3,y_2$.}
	\label{nis_fig2}
\end{figure}

\citet{2013A&A...560A.107A} analysed dynamics of off-limb loops in the active region NOAA 11654 on 22 January 2013, with no flare-energy releases at the time of the observations. Time-distance maps constructed from slits taken at different locations along the loops have shown the presence of decayless kink oscillations. The beginning and end times of these oscillations were different for different loops, as well as the oscillation periods that ranged between 2--11 minutes. The observed duration of the oscillations was between 3--4 to more than 10 oscillation cycles. The duration of the oscillations was associated with the varying observational conditions rather than being a sign of real physical damping.  Cross-correlation analysis of periodic intensity signals, extracted close to the boundary of a loop as an imprint of the transverse movement and at different locations along the loop path, was used to measure any possible phase delay between the oscillations at different segments of the oscillating loop. However, no phase delay was found: the maximum of the cross-correlation function locates at a lag-time of 0 minutes and does not depend on the loop position, hence confirming that the spatial structure of these transverse oscillations corresponds to the fundamental standing mode of a kink wave. 

The study of \citet{2014A&A...570A..84N} also proved the presence of decayless oscillations in a coronal loop bundle made of several multi-thermal strands, observed in the \lq\lq warm\rq\rq\ channels of AIA (i.e.,  171, 193, and 211 \AA) on 21 January 2013. The oscillations tracked at the top of the loop bundle have periods between 3--15 minutes, and showed an intermittent behaviour, i.e., with temporal changes both in the period and amplitude. The irregularities in the oscillations were associated with a possible stochastic driver operating at the footpoints of the loop, while it could also be attributed to the effect of insufficient resolution. 

The same event was analysed by \citet{2018ApJ...854L...5D}. In addition to the standard technique of time-distance maps, the authors used a motion magnification algorithm \citep{2016SoPh..291.3251A} to artificially increase the amplitude of the transverse displacements. Fourier analysis of the time series revealed the presence of two oscillation periods: one of about 10 min associated with a fundamental standing mode, the other of about 7 min. The longer period oscillation was more prominent at the loop top, whereas the shorter period one dominated at the loop two legs which oscillated in anti-phase with each other.   \cite{2018ApJ...854L...5D}  calculated the $P^{(1)}_\mathrm{kink}/(2P^{(2)}_\mathrm{kink})  = 0.69\pm0.16$. Assuming a constant magnetic field this yielded a plasma density scale height range of 7--45~Mm,

The discovery of co-existing multiple harmonics in the decayless regime opens up interesting perspective for coronal MHD seismology, in particular, for the estimate of the density scale height in the oscillating loop.

\begin{figure}[htpb]
\centering
	\includegraphics[width=1.0\textwidth]{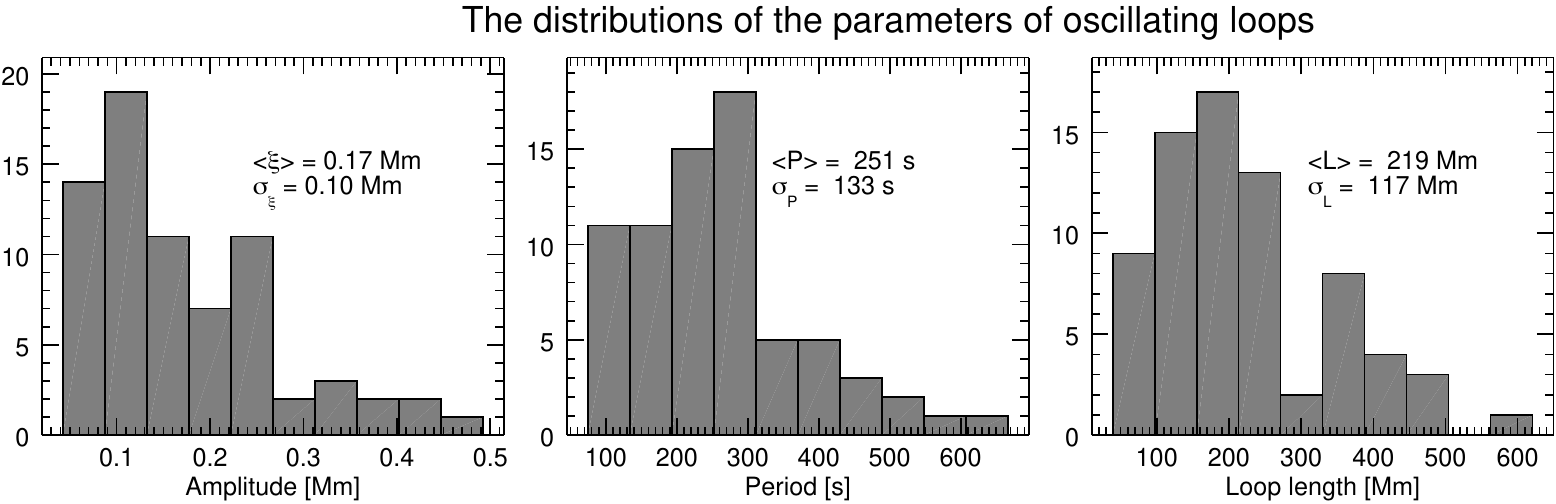}
	\caption{Histograms of parameters of decayless kink oscillations and associated loops. (From \cite{2015A&A...583A.136A}).}
	\label{nis_fig3}
\end{figure}
\begin{figure}[htpb]
\centering
	\includegraphics[width=0.5\textwidth]{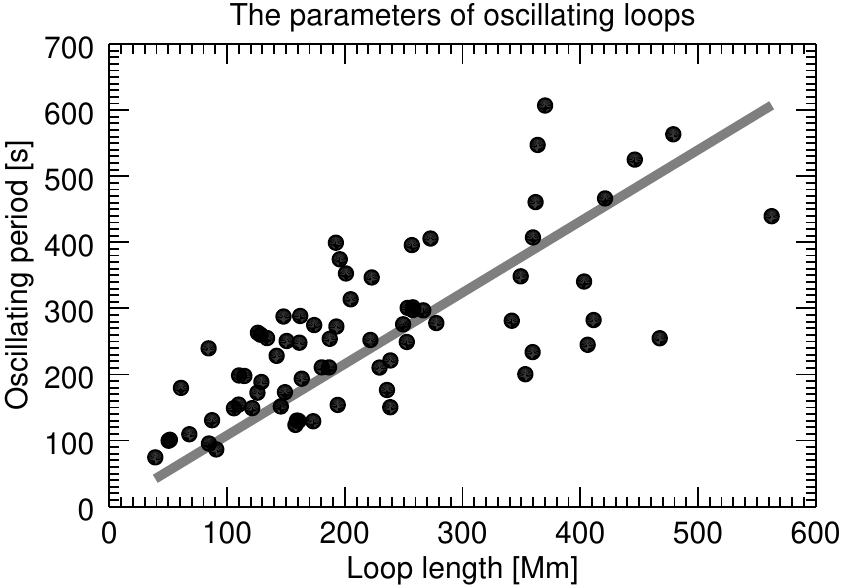}
	\caption{Empirical scaling of decayless kink oscillation periods with the lengths of the oscillating loops. (From  \cite{2015A&A...583A.136A}). }
	\label{nis_fig4}
\end{figure}

Similarly to the statistical study for decaying kink oscillations undertaken by \citet{2015A&A...577A...4Z} (see also Section~\ref{sec:scal}), a similar investigation addressing decayless oscillations has been accomplished in \citet{2015A&A...583A.136A}. Analysis of 21 active regions resulted in the detection of low-amplitude decayless oscillations in the majority of the analysed loops. Distributions of the detected parameters are shown in Fig. \ref{nis_fig3}. The average oscillation period was  $251\pm133$~s. The average displacement amplitude was $0.17\pm0.10$ Mm, i.e., smaller than the pixel size of AIA. Lengths of the oscillating loops were estimated to be between 50 and 600 Mm. Main results of this study are as follows:
\begin{itemize}
	\item[$\bullet$] decayless kink oscillations appear to be a common phenomenon in the solar corona;
	\item[$\bullet$] detected oscillation periods are found to be in the same range as those of decaying large-amplitude oscillations;
	\item[$\bullet$] as for decaying kink oscillations, the period of decayless oscillations scales linearly with the loop length (Fig.~\ref{nis_fig4}), which is consistent with the interpretation in terms of standing kink waves which are natural modes of the kink wave resonators constituted by coronal loops;
	\item[$\bullet$] amplitudes of decayless kink oscillations are about one order of magnitude lower than those in the decaying regime.  
 \end{itemize} 

Some studies have also pointed out a possible connection between decayless kink oscillations and other coronal phenomena, such as coronal rain. Coronal rain consists of dense and cold plasma blobs guided along magnetic field lines and resulting from a process of thermal instability. Observations of a coronal loop collected on 27 August 2014 with IRIS, Hinode/SOT, and AIA showed that coronal rain, forming at the loop top by a process of catastrophic cooling, moved downwards along the loop in conjunction with small-amplitude, vertically polarised decayless kink oscillations \citep{2017A&A...601L...2V}.

Decayless kink oscillations of coronal loops have also been detected in association with flares.
\cite{2020A&A...642A.159Z} reported a small-amplitude (such as 0.3$\pm$0.1~Mm) transverse oscillation of a coronal loop in active region 12,434, which was induced by a circular-ribbon flare on 2015 October 16. The oscillatory pattern consisted of four clear oscillation cycles without significant damping. In another event, both decayless and decaying kink oscillations were detected in the same active region, 11,991, apparently triggered by two homologous flares on 2014 March 5 \citep{2020A&A...638A..32Z}.

\begin{figure}
\centering
\includegraphics[width=\textwidth]{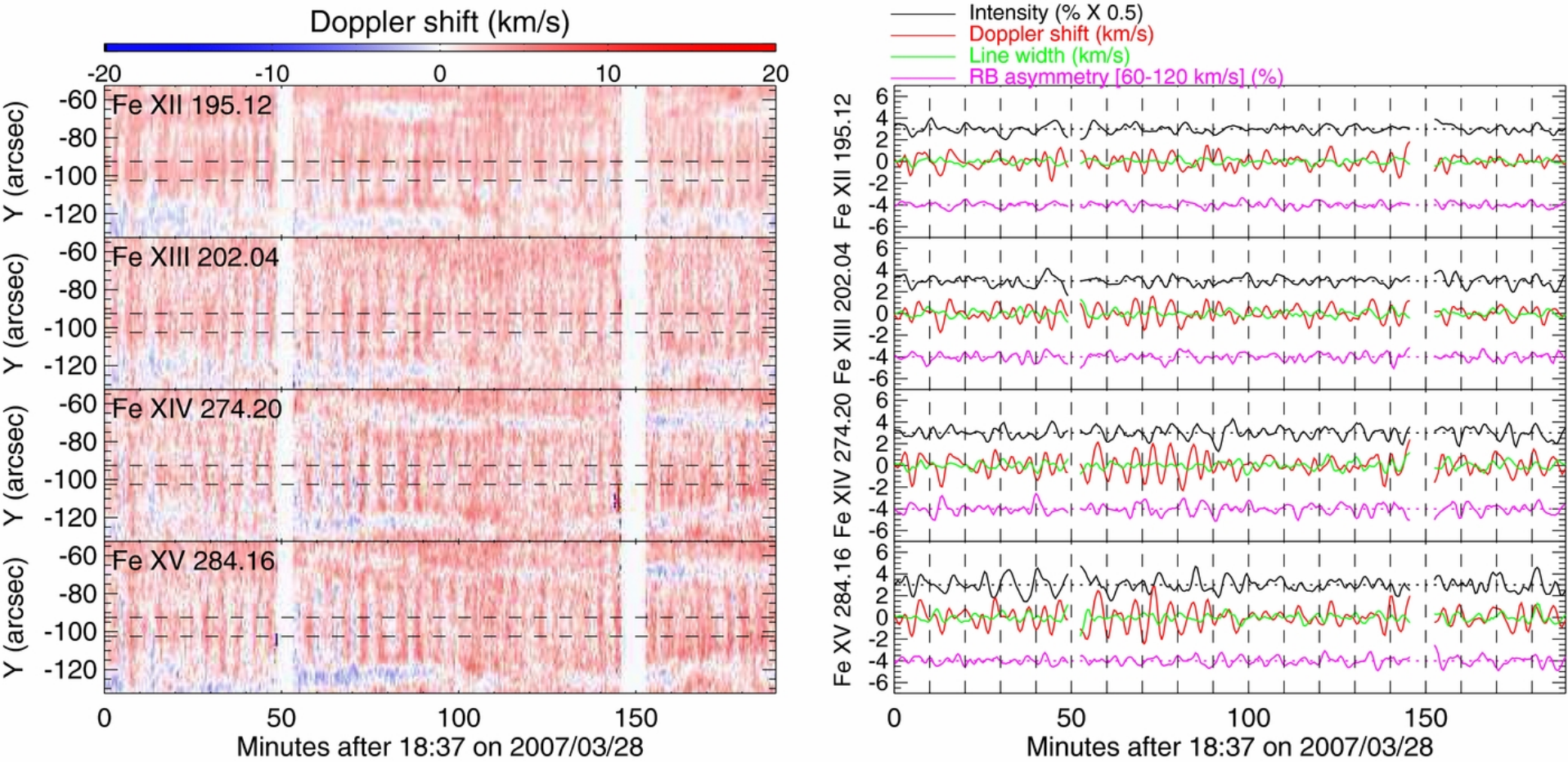}
\caption{Left: Temporal evolution of Doppler velocities of four coronal lines. Right: the detrended line intensity (black), Doppler velocities (red), line width (green), and red--blue asymmetry (violet) averaged over the regions between the two dashed lines shown in the left panels. Adaptation of Figure 4 in \cite{2012ApJ...759..144T}
\label{fig:tian2012}}
\end{figure}

Perhaps, a similar phenomenon was revealed by the analysis of spectroscopic measurements with some coronal lines, which demonstrated persistent 3--6~min Doppler-shift oscillations at the loop apexes. Using the Hinode/EIS data, \cite{2012ApJ...759..144T} detected a group of decayless oscillations in coronal loops from February to April in 2007. These long-lasting oscillations were mostly pronounced in Doppler velocities of warm coronal lines (e.g., \ion{Fe}{XII}--\ion{Fe}{XV}), which have formation temperatures between 1.3~MK and 2.0~MK (Figure~\ref{fig:tian2012} left). They found the oscillation amplitudes to be very small, i.e., $<$2~km s$^{-1}$ for the Doppler velocities, and $<$2\% for the line intensities (Figure~\ref{fig:tian2012} (right)). These oscillations were interpreted as signatures of kink or Alfv\'{e}n waves.  A similar decayless oscillation was detected in hot flare loops on 2014 September 6 by \cite{2018A&A...617A..86L}. Using the IRIS observation, the long-lasting oscillation with a period of $\sim$40~s was clearly seen in Doppler velocities of a hot line with a formation temperature of $\sim$11~MK, such as \ion{Fe}{XXI}, while the oscillation was very weak in the line width and peak intensity.  Based on combined techniques from the coronal seismology and the magnetic field extrapolation, the authors estimated the magnetic field strengths between 120--170~G which is consistent with the value expected in flare loops.

These observational works have posed the basis for the development of theoretical models on the nature of the decayless regime of kink oscillations.

\subsection{Theoretical and numerical models for decayless oscillations}  
\label{sec:decthe}

Decayless oscillations and, especially, their coexistence with decaying oscillations, pose new challenges about the identification of the energy source which counteracts the damping.
This issue is not trivial, as resonant absorption and the distortion of the loop's boundary by KHI seem to be intrinsic features of kink oscillations. 

In specific events, one or more external factors can contribute to shaping the oscillations. For example, the growing oscillations described by \citet{2012ApJ...751L..27W} could be influenced by the eruption or the observed local surge. In this context, the plasma flow in the direction perpendicular to the segment of the loop near its apex could cause the formation of Alfv\'enic vortices.
This effect is well-known in hydrodynamics, and is known as vortex shedding. When a laminar flow interacts with a bluff body, {i.e., an obstacle,} vortices are generated downstream the cylinder. The vortices are shed from the alternate sides of the body. This process is periodic, with the period of vortex shedding determined by the flow speed and the geometrical size of the body. In MHD, if the bluff body is a cylinder stretched along the magnetic field, Alfv\'enic vortices with the vorticity vector parallel to the axis of the cylinder are excited \citep{2009A&A...502..661N}. The back reaction on the cylinder leads to its rocking in the plane perpendicular to the flow. This external force could resupply the energy lost by resonant absorption, and even amplify the oscillation. First imaging evidence of vortex shedding {in the solar corona} is reported in \citet{2019PhRvL.123c5102S} where an oscillating plasma flow event observed in the wake of a hot loop with SDO/AIA, is discussed. As another option, \citet{2012ApJ...751L..27W}  proposed that the external driver could be quasi-periodic fast magnetoacoustic wave-trains \citep{2011ApJ...736L..13L, 2014A&A...569A..12N}.

However, to keep the decayless oscillations lasting for several periods or longer, the external driver should operate \lq\lq continuously\rq\rq~in time and not being of impulsive nature.
Some insight into this phenomenon could be obtained by considering an oscillating loop as a pendulum, possibly excited by some kind of an external driver. In this approach, the displacement $\xi$ of the top of the loop is described by the harmonic oscillator equation  
\begin{equation}
\frac{\mathrm{d}^2 \xi}{\mathrm{d}t^2} + 2\delta_\mathrm{diss} \frac{\mathrm{d}\xi}{\mathrm{d}t} +{\Omega}_\mathrm{k}^2\xi = f(t),
\label{eq_motion}
\end{equation}
where ${\Omega}_\mathrm{k}$ is the natural frequency of the pendulum, which corresponds to the kink oscillation period, $P_\mathrm{kink}=2\pi/{\Omega}_\mathrm{k}$, and $\delta_\mathrm{diss}$ a parameter accounting for damping (e.g., due to resonant absorption\footnote{Strictly speaking, this simple form of the damping term does not allow for frequency-dependent damping typical of resonant absorption, see, e.g., Eq.~(\ref{eq:tau_d}), and needs to be generalised in the future \citep[see, also][]{2021MNRAS.501.3017R}}). The term $f(t)$ on the right-hand side represents the external force. Defining the functional form of $f(t)$ is crucial for the best reproduction of the observations.

In \citet{2013A&A...552A..57N}, the observations of a loop oscillating subsequently in damping and decayless regimes were modelled with 
\begin{equation}
	f(t) = A_0 \exp(-i \omega_0 t) + A_1 \delta(t-t_0),
	\label{ft1}
\end{equation}
with a harmonic non-resonant driver of amplitude $A_0$ and frequency $\omega_0$ and the Dirac's delta function $\delta(t-t_0)$ representing an impulsive excitation of the oscillation with the amplitude $A_1$ at the instance of time $t_0$. The first term is responsible for the sustainability of low-amplitude decayless oscillations, and the impulsive driver excites a large-amplitude and damped oscillation. The solution of Eq.~(\ref{eq_motion}) with the driving term  given by Eq.~(\ref{ft1}), as shown in Eq.~(6) of \citet{2013A&A...552A..57N}, consists of a superposition of the harmonic low-amplitude signal and the impulsively induced decaying oscillation. Such a simple model provides some important insights. Firstly, the observed damping time, $\tau_\mathrm{D}$ does not necessarily correspond to the physical damping time $\delta_\mathrm{diss}^{-1}$: the higher the amplitude $A_0$ is, the longer the oscillations looks to be. Secondly, when the driving ($\omega_0$) and the natural ($\mathrm{\Omega}_\mathrm{k}$) frequencies are close, a beat appears in the oscillatory pattern causing a considerable underestimate of the damping time of the oscillations (see Fig.~\ref{nis_fig5}). In addition, a feature of this scenario is the sudden change of the oscillation phase at the time of the impulsive excitation. 

\begin{figure}[htpb]
	\centering
	\includegraphics[width=0.7\textwidth]{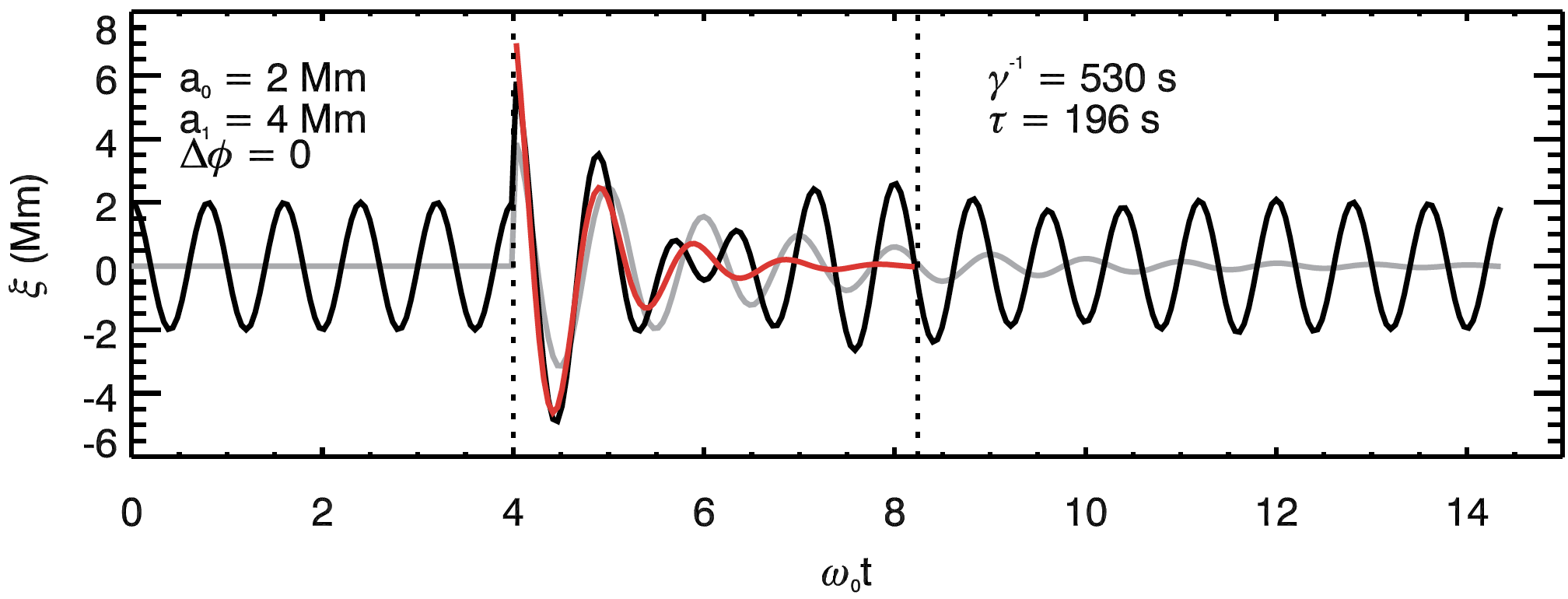}
	\caption{Simulated displacement $\xi(t)$ (black line) for a loop excited by a harmonic driver of amplitude $A_0=$2 Mm and an impulsive driver of amplitude $A_1=4$~Mm. The oscillations with the only impulsive driver are shown with the grey line. The red line is a fit of the oscillations enclosed between the vertical dashed lines. Because of the beating in the oscillatory pattern, the estimated damping time $\tau_\mathrm{D} = 196$~s is underestimated with respect to the physical damping time $\delta_\mathrm{diss}^{-1}=530$~s. From \citet{2013A&A...552A..57N}.}
	\label{nis_fig5}
\end{figure}

Quantifying the energy flux taken by decayless oscillations from the lower layers of the solar atmosphere is interesting for unveiling their role in coronal heating. \citet{2019FrASS...6...38K} have studied the energy carried out by decayless oscillations by simulating loop oscillations via a periodic driver that shakes the footpoints. They found no clear correlation between the oscillation amplitude and the energy driver, consequently, it is difficult to link the measured amplitudes from observations to any possible energy source. The energy flux is found enough for feeding the quiet Sun region. Therefore, decayless oscillations could be considered as a viable mechanism to transport energy in the quite corona.

The model with a harmonic non-resonant driver is not able, however, to reproduce statistical properties of decayless kink observations presented in \citet{2013A&A...560A.107A, 2015A&A...583A.136A, 2014A&A...569A..12N}. In the case of the harmonic driver, the amplitude of the oscillations should depend on the difference between the natural frequency of the loop $\Omega_\mathrm{k}$ and the driving frequency $\omega_0$, reaching a maximum value at the resonance, i.e., when $\omega_0\approx \Omega_\mathrm{k}$. This behaviour is not consistent with the observed distribution of detected amplitudes with oscillation periods (Fig. \ref{nis_fig6}), which do not show a peak. On the contrary, results presented in \citet{2015A&A...583A.136A}, in addition to the fact that a loop was observed simultaneously in both regimes (decaying and decayless) with no big difference in the periodicity, favour the interpretation of decayless oscillations in terms of the eigenmodes or natural modes of individual loops or bundles of loops. 
 
\begin{figure}[htpb]
\centering
	\includegraphics[trim= 0 600 300 0, width=1.0\textwidth]{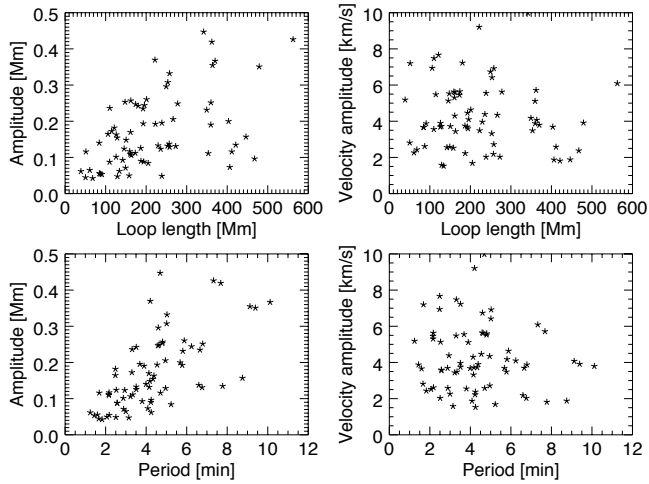}
	\caption{Scatter plots of the oscillation amplitudes (left) and velocity amplitude (right) against the loop length and the period. Adapted from \citet{2016A&A...591L...5N}.}
	\label{nis_fig6}
 \end{figure}
 
Decayless oscillations with a natural frequency of the oscillator could occur if $f(t)$ is a random function, possibly associated with the granulation motion that persistently shuffles the loop footpoints \citep{2014A&A...570A..84N, 2016A&A...591L...5N}. In this case, the loop oscillates at the natural frequency, and the damping by, for example, resonant absorption, is compensated by the random driver which does not have a preferential frequency. This scenario has been investigated numerically by \citet{2020A&A...633L...8A}. By solving a 1D wave equation in a gravitationally stratified loop with boundary conditions consistent with a random excitation, it was demonstrated that the loop would respond to such a driver with multi-harmonic oscillations with the natural kink periods of the loop (Fig. \ref{nis_fig7}). The ratio between the oscillation periods, $P^{(1)}_\mathrm{kink}/2P^{(2)}_\mathrm{kink}$ is found to be less than unity, reflecting the effect of the density stratification in the loop \citep[see ][and Sec.~\ref{sec:p1_2p2}]{2009A&A...497..265A}. This effect was confirmed by the observational detection of the fundamental and second parallel harmonics of decayless kink oscillations  \citep{2018ApJ...854L...5D}. The main difficulty of this mechanism is the need to reproduce an almost monochromatic oscillatory pattern observed in the majority of oscillations. For example, numerical simulations performed by \citet{2016A&A...591L...5N} showed that the response of the randomly driven oscillator  has chaotic variations of the oscillation phase. On the other hand, results of this modelling can be relevant to the interpretation of the rather chaotic kink oscillations detected by \citep[][see also Section~\ref{s:first_decayless}]{2014A&A...570A..84N}. 
\begin{figure}[htpb]
\centering
	\begin{tabular}{c c}
		\includegraphics[width=0.7\textwidth]{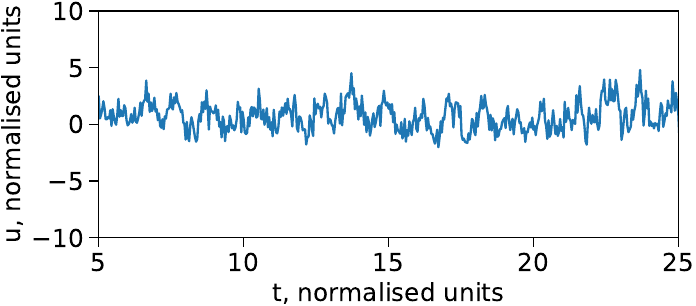}\\
		\includegraphics[width=0.7\textwidth]{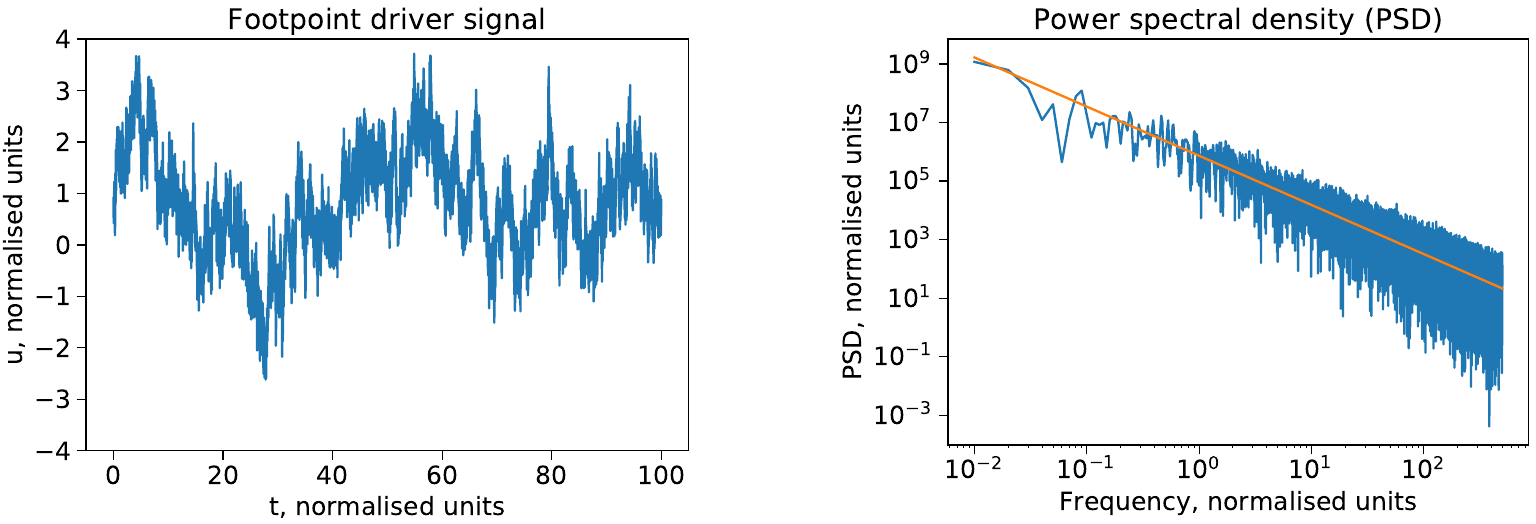}\
	\end{tabular}
	\caption{Oscillatory response of a loop (top) to random footpoint motions (left bottom). The power spectrum of the driver, with the orange line indicating the spectral power-law fall-off with the index 1.66 (right bottom). Adapted from \citet{2020A&A...633L...8A}.}
	\label{nis_fig7}
\end{figure}

\begin{figure}[htpb]
\centering
	\includegraphics[trim= 0 680 300 0, width=1.0\textwidth]{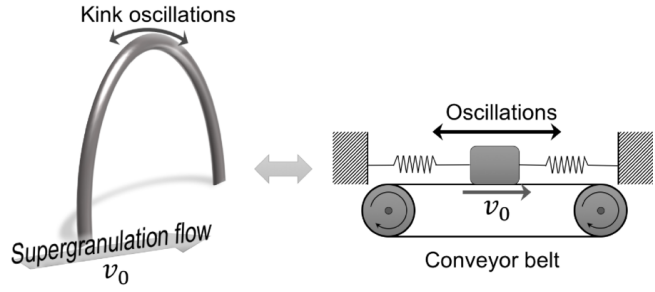}
	\caption{Sketch of the self-oscillatory mechanism for decayless kink oscillations, based on the interaction between a supergranulation flow and the loop (left); and a mechanical analogue of this scenario, a spring pendulum on a conveyor belt moving steadily at a speed $v_0$. From \citet{2016A&A...591L...5N}.}
	\label{nis_fig8}
 \end{figure}

Another mechanism was studied in \citet{2016A&A...591L...5N}, interpreting decayless oscillations with self-oscillations caused by the slippery interaction of the loop with some steady external medium flow, such as supergranulation (Fig. \ref{nis_fig8}). The forcing term in Eq.~(\ref{eq_motion}) is connected with the friction between the loop and the external flow,
\begin{equation}
	f(t) = F\left( v_0 - \frac{\mathrm{d}\xi(t)}{\mathrm{d}t}\right), 
\end{equation}
where $F$ is a function which depends on the difference between the speed of the loop's displacement and the speed of the flow.  When the difference between the speeds is zero, the function $F$ is maximum. By Taylor-expanding this term up to the third order, Eq.~(\ref{eq_motion}) reduces to a second-order nonlinear differential equation, known as the Rayleigh oscillator equation,
\begin{equation}
\frac{\mathrm{d}^2 \xi}{\mathrm{d}t^2} - \left[ \mathrm{\Delta} - \alpha\left(\frac{\mathrm{d}\xi}{\mathrm{d}t}\right)^2 \right]\frac{\mathrm{d}\xi}{\mathrm{d}t} + \mathrm{\Omega}_\mathrm{K}^2\xi = 0, 
\label{eq_ray}
\end{equation} 
where $\mathrm{\Delta}$ is the difference between the linear friction with the medium and the regular damping, $\delta_\mathrm{diss}$, and $\alpha$ is a coefficient depending on the medium speed $v_0$. It is important that, from the point of view of the loop, the friction between the flow and the loop is negative, as the loop gains energy from the flow by means of this friction. Mechanical analogues of this model are a spring pendulum on a conveyor belt moving at a constant speed (see Figure~\ref{nis_fig8}), or,  a violin string with a bow moving across it. In the latter example, the friction acting between the string and the bow, when it is greater than other dissipation losses in the system, acts in phase with the direction of the flow velocity (the movement of the bow), and it is positive. This will introduce negative damping, which has the effect to make the amplitude of the oscillations exponentially growing up (or decreasing, depending on the initial conditions) to a fixed value. In the phase portrait of Eq.~\ref{eq_ray}, this solution representing decayless oscillations manifests as a limit cycle. The ability of the self-oscillatory scenario to reproduce observed properties of decayless kink oscillations has been confirmed by numerical simulation in \citet{2020ApJ...897L..35K}, demonstrating that it is a viable mechanism for the generation of multi-harmonic persistent oscillations in coronal loops.

Several other models explaining the excitation of decayless kink oscillations have been proposed. \citet{2017A&A...598A..57V} investigated the connection between coronal rain and kink waves. They developed an analytical model where a loop is  described by a unidimensional magnetic field line guiding a moving rain blob. The model includes gravity, the effect of the ponderomotive force generated by transverse oscillations and acting on the rain mass, and the feedback of the inertia of the rain blob on the oscillations. They found that the rain inertia in bent loops can excite persistent small-amplitude oscillations. In this context, the decayless oscillation would not be caused by the action of a driver acting at the loop footpoints, but would be a result of the catastrophic cooling process that creates coronal rain.

\begin{figure}[htpb]
\centering
	\includegraphics[ width=1.0\textwidth]{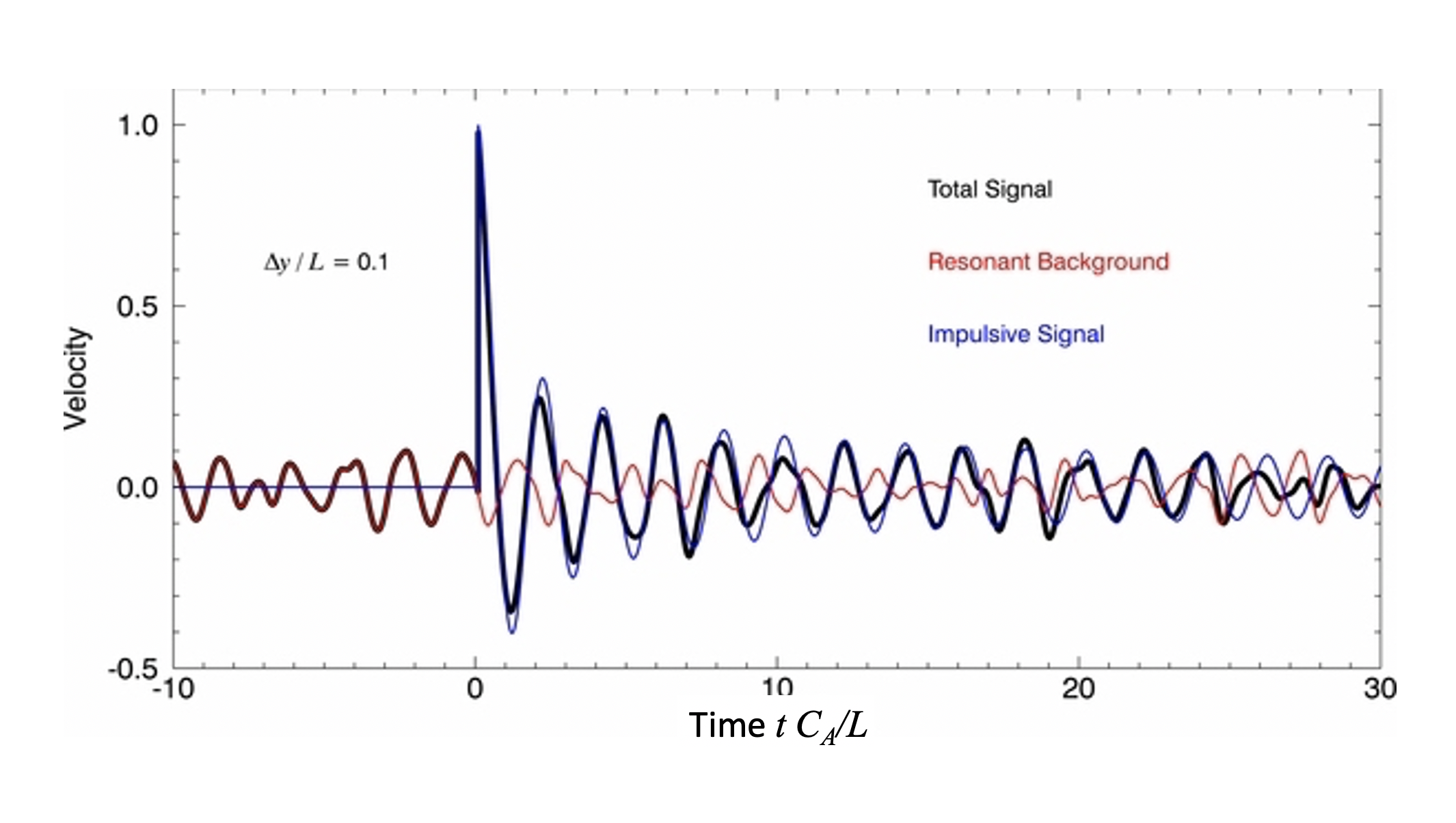}
	\caption{Wave signal as a function of time arising from both wave sources. The red curve is the background signal generated by a distributed stochastic source. The strength of the stochastic source is Gaussian in the wavenumber in the direction perpendicular to the magnetic field. The dominant frequency is the lowest frequency available, corresponding to those waves which propagate parallel to the magnetic field. The blue curve is the signal arising from a single impulsive event that occurred rather close to the observation point along the arcade, $\Delta y = 0.1L$, with $L$ as the length of each field line. The initial pulse corresponds to waves that have propagated straight down the waveguide, while the latter oscillations are an interference pattern arising from waves that have taken a variety of paths down the waveguide. The black curve shows the total wave signal. We have chosen the relative size of the impulsive and background sources such that the background has an amplitude of 10\% of the initial pulse height of the impulsive signal. (From \citet{2014ApJ...784..103H}.}
	\label{HJ_2014}
 \end{figure}
 
An alternative model was proposed in \citet[][see also Fig.~\ref{HJ_2014}]{2014ApJ...784..103H}. There, a bright loop or a bundle of loops as observed in EUV imaging data, rather than being considered as a single oscillator, is assumed to be a part of a magnetic arcade. The arcade is treated as a 2D waveguide. In this model, fast waves triggered by two drivers, a short durational and spatially localised pulse plus a continuous, distributed, stochastic source, evolve into a mode that is trapped between the footpoints of the arcade. The waves freely propagate along the axis of the arcade, across the field. The model does not need to invoke any physical damping, such as resonant absorption  and viscous damping. The apparent decaying oscillations generated from the impulsive component would be the results of interference fringes produced by waves that travel along different paths through the waveguide, rather than being a process of local dissipation or mode coupling.
It is not clear whether this model is capable of reproducing observed statistical properties of decayless kink oscillations. 

The damping could also be counteracted by the amplification of the kink oscillation amplitude caused by the effect of the plasma cooling (see Section~\ref{sec:cool}). However, observational confirmation of this effect has not been detected yet. 

\citet{2016ApJ...830L..22A} proposed that the observed decayless oscillations might be not real but simply the results of line-of-sight integrations effects in density variations due to KH vortices, as shown with forward modelling of the EUV intensity. In this context, loops subject to decayless oscillations would have a transverse structure that would be turbulent.

\section{Possible detection of kink oscillations in the microwave and radio bands} 
\label{sec:radio}

The angular resolution of a telescope is  $\vartheta \approx \lambda / D$ where $\lambda$ is the wavelength of the observed electromagnetic waves, and $D$ is the aperture of the telescope, with $\lambda \ll D$. In particular, angular resolution of the Nobeyama Radioheliograph (NoRH, Japan) is 5'' at 34~GHz and 12'' at 17~GHz. The resolution at MHz could be up to several solar radii. The amplitude of decaying kink oscillations rarely exceeds 10~Mm (see Fig~\ref{fig:nech1}) or 14". Thus, the intrinsically relatively poor spatial resolution of observations in the radio band, including microwaves, makes it difficult or does not allow one at all to observe kink oscillations of coronal loops directly, as transverse displacements of the loops. However, radio observations which are directly probing the coronal magnetic field and electron concentration could provide us with important and sometimes unique information. Moreover, the advanced time resolution achievable in radio observations, e.g., 100~ms in the flare mode of NoRH, allows one to study higher spatial harmonics, fine evolution of the amplitude profiles, and other effects which cannot be detected in EUV. 

In the lack of spatial resolution, one of the principal parameters for the identification of kink oscillations in observational radio data is the ratio of the periods of the multi-modal quasi-periodic pulsations (QPP). The identification of the theoretically predicted ratio of the periods of the first two harmonics ($P^{(1)}_\mathrm{kink}/(2P^{(2)}_\mathrm{kink})$ (see Section~\ref{sec:p1_2p2}) could provide us with the decisive evidence of the kink mode modulating the radio emission. In addition, indirect indications to the kink mode can be found taking into account specific mechanisms for the radio emission. {We need to note that higher harmonics of the kink mode should not be confused with the higher harmonics of the radio frequency itself. The former is detected in the modulation of the intensity and polarisation signals at certain radio frequencies or a broadband radio signal. Oscillation periods of higher kink harmonics are of the same order of magnitude as the fundamental kink mode, i.e., in the range of seconds or minutes. The latter effect is often seen in Type II and III solar radio bursts as doubling the instantaneous radio frequency of the burst. It is associated with the nonlinearity of the radio emission mechanism. }

In particular, microwave emission from a magnetically structured plasma of an active region with the magnetic field $B$ is often associated with the gyrosynchrotron emission produced by non-thermal electrons gyrating around magnetic field lines and interacting with the background plasma \citep[e.g.,][]{2012Ge&Ae..52..883K}. The emission is sensitive to the power-law distribution $f(E)$ of the non-thermal electrons over the energy $E$, i.e., $f(E) \propto E^{-\delta}$ with the spectral index $\delta$ is typically in the range from 2 to 7. For the frequency $f$ greater than the electron plasma frequencies, the intensity of gyrosynchrotron emission $I_f$ at $f$ could be estimated as  
\begin{equation} \label{Dulk}
I_f \propto (\sin \theta)^{-0.43+0.65\delta} (f/f_B)^{1.22-0.9\delta},
\end{equation}
where $f_B$ is the electron gyrofrequency and $\theta$ is the angle between the field and the line-of-sight \citep[e.g.,][]{1982ApJ...259..350D}.  For typical conditions in the corona, the gyrosynchrotron emission has a maximum in the microwave band. The dependence of the intensity of the gyrosynchrotron emission on the magnetic field $B$ and the line-of-sight angle makes the microwave emission a sensitive tool for the detection of kink oscillations. For example, the dependence on the $\theta$ could lead to appearance of the two first harmonics in the Fourier spectrum. The fundamental mode corresponds to the period of the large-scale transverse kink motion of a coronal loop ($P_\mathrm{kink}$).  When the loop is swinging around the equilibrium position, the angle $\theta$ varies twice per period, causing the second harmonic with the period exactly equal to $P_\mathrm{kink}/2$  \citep{2011A&A...525A.105K}.

Another indication of the association of multi-period QPP with kink oscillations is provided by the departure of the $P^{(1)}_\mathrm{kink}/(2P^{(2)}_\mathrm{kink})$ ratio of the detected periods from unity. Such a departure was found, for example, by \citet{2009A&A...493..259I} 
for the light curves in the microwave and hard X-ray bands. However, the period ratio only is not sufficient for the confident identification of kink oscillations. Additional information was provided by the lack of the QPP  in the soft X-ray emission, which could associated with the periodic modulation of the plasma density, indicating the incompressive nature of the observed oscillations. Thus, the authors concluded that the QPPs were related to harmonics of the almost incompressible kink mode.

\begin{figure}
\centering
    \includegraphics[width=1.0\textwidth]{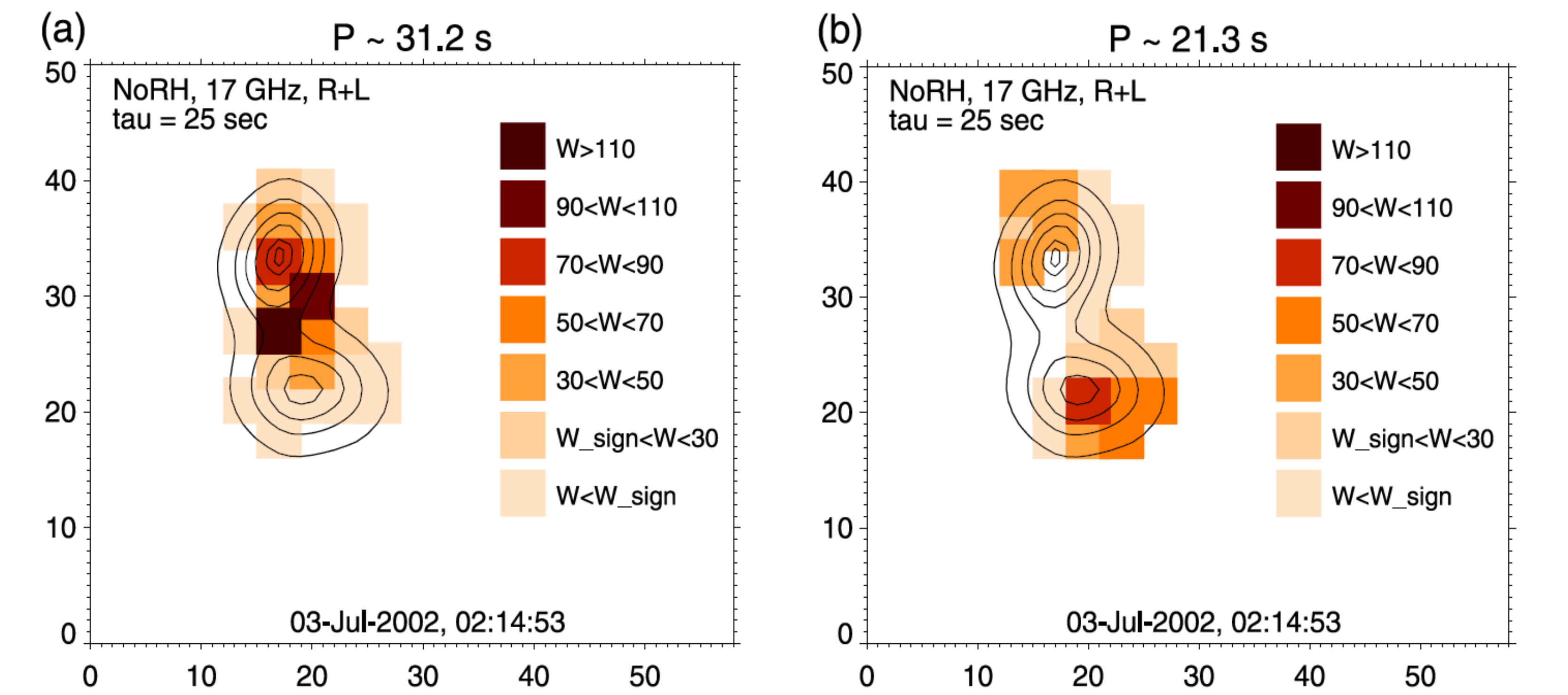}
    \caption{Distribution of the spectral power $W$ over the flare area both for the fundamental kink mode with the mean period $P \approx 31.2$~s (panel a) and the second harmonic of the kink mode with the mean period $P \approx 21.3$~s (panel b). (From \cite{2013SoPh..284..559K}).}
    \label{f:kink_12}
\end{figure}

Another example of a kink oscillation spatially resolved at microwaves was found by \citet{2013SoPh..284..559K}. They analysed microwave maps obtained with NoRH at 17~GHz and found multi-period QPPs.  The authors established that the QPPs with the period $P_1 \approx 30$~s had the maximum amplitude in near the loop top, while the maxima of the amplitude of the QPPs with the period $P_2 \approx 20$~s were localised close to the footpoints (Figure~\ref{f:kink_12}). This information was interpreted as the evidence of the fundamental and second harmonics of kink mode with the ratio $P^{(1)}_\mathrm{kink}/(2P^{(2)}_\mathrm{kink}) \approx 1.46$ corresponds to that both predicted theoretically (Section~\ref{sec:p1_2p2})  and found by \citet{2009A&A...493..259I}, see also \citet{2015A&A...574A..53K} where this period ratio was observed in multi-modal microwave QPPs to be about 1.1. 
In the follow-up study,  \citet{2013PASJ...65S...3K} found additionally that the distance between two sources at the footpoints also varied with the period around 30~s. Moreover, the increase in the microwave emission flux of the spatially integrated signal, was accompanied by a decrease in the distance between the sources. In the assumption of the vertically polarised kink mode modulating the minor radius of the loop \citep[see, e.g.,][]{2011ApJ...736..102A}, small modulation of the magnetic mirror ratio $\sigma = B_\mathrm{max}/B_\mathrm{min}$ could occur. Here $B_\mathrm{max}$ and $B_\mathrm{min}$ are the magnetic field values at the footpoints and loop top, respectively. Such oscillations could displace the sources, and hence can vary the distance between them.

Physical processes associated with solar flares can cover a large range of heights in the solar atmosphere. As the radio emission at longer wavelengths corresponds, roughly, to the higher level of the solar atmosphere, the spectral information allows one to estimate the height of the emission source, compensating the lack of spatial resolution. For example,
\citet{2013A&A...555A..55Z} made diagnostics of the outer corona using radio observations in the 8--32~MHz band by a vertical kink oscillations of the transequatorial loop with the apex at 1.2 of the solar radius.
\citet{2014ApJ...791...44H}  analysed QPP in a broadband flare emission, and found that the period was increasing with the decrease in the wavelength or decrease in frequency, i.e., increase in the height, from 21~s in soft X-rays and 22--23~s at microwaves to 27~s at 100~MHz and 32~s at 50~MHz. The authors linked this effect with a kink oscillation of a loop, triggered by an implosion. As a loop length increases with the height, so does the kink period.

An interesting simultaneous observation of a kink oscillation detected in the microwave and EUV bands in a preflare phase was reported by \cite{2020ApJ...893L..17L}. An oscillatory pattern with a period slowly growing from about 300~s to 500~s was observed at 17~GHz with NoRH. A similar low-amplitude transverse oscillation with a slowly growing period of about 400~s was detected in  a coronal loop in the 171~\AA\ bandpass of AIA. 

\section{Conclusions and outstanding problems}
\label{sec:conc}

Since their discovery more than twenty years ago, kink oscillations of coronal plasma loops remain one of the most intensively studied MHD wave phenomena in the solar atmosphere. There have been a wealth of observational detections of kink oscillations as repetitive transverse displacements of various field plasma nonuniformities of the solar corona, such as loops, plumes, streamers, and jets, in both standing and propagating forms. {Properties of standing kink oscillations are established to be determined by properties of the plasma non-uniformities acting as resonators, and are hence not connected with periodic processes observed in the photosphere and chromosphere.} Two different regimes, the large-amplitude rapidly decaying oscillations and low-amplitude decayless oscillations have been identified in the observational data. Oscillations of the latter type do not appear to be connected with eruption or flare events. Regular detections of kink oscillations in observational data create a ground for the search for statistical relationships between parameters of oscillations and properties of host loops and active regions, which is a power tool for testing theoretical predictions. 

In this review we concentrated mainly on standing kink oscillations and their seismological applications, discussing the tremendous progress reached in this research field in the last decade. Recent breakthroughs summarised here open up several follow-up research steps and important questions which should be addressed by our research community.

1. One of the enigmatic questions of solar physics is the nature of coronal loops. A 3D magnetostatic equilibrium solution describing this ubiquitous plasma structure of the corona has not been found. {It is not clear whether the observed plasma non-uniformities are monolithic or consist of a number of unresolved threads; what the magnetic field geometry in the loop is, and how it could be linked with observed properties of loops, in particular, the apparent constant cross-section;  what the thermal structure is, and whether the thermostatic equilibrium exists and it is sustained.} Kink oscillations which depend on the transverse profile of the plasma {(see Secs.~\ref{sec:2rRAobs} and \ref{sec:pernon})} and the magnetic field twist {(see Sec.~\ref{sec:twisted})}, as well on the variation of the plasma and field along the loop {(see Sec.~\ref{theory1_abi})}, are natural probes for plasma loops, and can shed light on the equilibrium. On the other hand, the continuous movement of the loops, produced by decayless kink oscillations, may suggest that KHI rolls and turbulent flows are an intrinsic feature of a loop. In this case, developing of non-equilibrium models of plasma loops, which would accommodate dynamic processes, becomes of interest. In any case, it is of interest to reproduce a particular observed event of a decaying kink oscillation with the use of truly 3D MHD simulations based on force-free extrapolated magnetic field of the region and inferred density along the loop. 

2. In the (locally) cylindrical geometry, the generation of KHI rolls could be considered as a nonlinear cascade of the wave energy in the azimuthal and possibly radial directions. However, for $m>1$, the ZSER model shows that the wave dispersion increases with the increase in the azimuthal mode number, $m$. The efficiency of nonlinear cascade should decrease with the increase in dispersion. On the other hand, modes with higher radial mode numbers are subject to leakage, which removes wave energy from the waveguide. It would be interesting to investigate the effects of dispersion and leakage on the KHI of kink oscillations.  

3. Re-analysis of the data summarised in the catalogue of \cite{2019ApJS..241...31N}, using the advanced models which account for the combined exponential and Gaussian damping regimes, would allow one to determine the distributions of the characteristic times $\tau_\mathrm{D}$ (for exponential damping), $\tau_\mathrm{g}$ (for Gaussian damping), and $t_\mathrm{switch}$ (for switching between these two damping regimes). It would also be of interest to search for scalings of those times with each other, parameters of the oscillating loops, and the oscillation amplitude. Also, the established dependence of the oscillation quality factor on the initial amplitude suggests that the instantaneous quality factor may evolve with the progression of the oscillation. Such a correlation could be confirmed or disproved observationally. In addition, as it is expected that KHI produced by kink oscillations should develop more in oscillation events of higher amplitude, the catalogue may help to demonstrate that the transverse profiles of the loops are different before and after decaying kink oscillations.   

4. An important outstanding task remains the search for the evidence of resonant absorption of kink oscillations in spectroscopic data, i.e., demonstrations of the increase in the amplitude of incompressive torsional motions during the decay of a kink oscillation. Likewise, direct observational evidence of the development of KHI rolls would be highly important. 

5. Despite the clear observational evidence of the association of decaying kink oscillations with low coronal eruptions, the specific mechanisms for the oscillation excitation remain unrevealed. Why do some loops respond to the mechanical displacement from the equilibrium while the majority of them do not? Numerical modelling of the excitation of kink oscillations by a displacement by {a slowly moving, in comparison with the fast magnetoacoustic speed, agent has not been performed}. Such a displacement of the loop top caused by an eruption should excite not only the fundamental mode, but also higher odd parallel harmonics. However, those harmonics are detected only in a few events. Likewise, if the displacement is asymmetric with respect to the loop top, even harmonics should be produced, which are very rarely detected too. 

6. The interest in the decayless oscillations of coronal loops stems from the unsolved problem of coronal heating. Indeed, observations of persistent oscillations, as those achieved not only by SDO, by Hinode and, perhaps, by CoMP demonstrate that MHD waves are a ubiquitous phenomenon in the solar corona and they could transport energy from the lower to the higher corona and eventually power the acceleration of the solar wind \citep[e.g.][]{2020SSRv..216..140V}. Despite the driver of the decayless oscillations has not been unambiguously identified, it is accepted that the energy source must be of mechanical type, i.e. allowing for the displacement of the loop out of the equilibrium. The resulting global kink mode is then converted to the local Alfv\'en mode via resonant absorption and eventually in the internal energy of the plasma. It remains to establish where this driver is located, if at the loop footpoints (e.g., granulation and supergranulation flows) or higher in the corona (e.g., local flows leading to vortex shedding, coronal rain).
What is the thermal evolution of loops subject to decayless oscillations \citep{2020A&A...638A..89G}?
How to properly relate the decayless oscillations with the dynamics in the chromosphere and the photosphere?
Decayless oscillations are observed as a standing mode in EUV images, whilst CoMP gives the evidence for propagating kink waves along loops. Why this discrepancy between the observations with these two instruments? Are both waves the same phenomenon?

7. Seismological diagnostics of the magnetic field in coronal active regions by decayless kink oscillations opens up promising perspectives for the estimation of free magnetic energy and other parameters, and their evolution before impulsive energy releases \citep{2019ApJ...884L..40A, 2020ApJ...894L..23M}. Knowledge of those parameters is important for forecasting of eruptive and flaring phenomena.  Additional information is provided by the simultaneous detection of several harmonics \citep{2018ApJ...854L...5D}. Thus, development and application of seismological techniques utilising the magnetic field diagnostic potential of decayless kink oscillations is a priority task. 

8. Coronal loops with sheared field-aligned flows, e.g., siphon flows or the flows associated with evaporation or condensation, may be subject to negative energy effects. These effects include instabilities with  thresholds well lower than KHI thresholds. An important regime of negative energy waves is explosive instability which can be responsible for sudden destabilisation of active regions, triggering eruptions and flares. 

Obviously, this list is incomplete and not exclusive. Future observational discoveries anticipated with the cohort of new instruments such as the Extreme Ultraviolet Imager and Spectral Imaging of the Coronal Environment on the recently launched Solar Orbiter spacecraft, and the white-light near-limb observations from the coronagraph ASPIICS of the PROBA-3 mission, and theoretical breakthroughs will add more items. The study of kink oscillations of coronal loops {is of permanent interest}.  

\begin{acknowledgements}
We gratefully acknowledge ISSI-BJ for supporting the workshop on \lq\lq Oscillatory Processes in Solar and Stellar Coronae\rq\rq, during which this review was initiated. 
V.M.N. acknowledges support by the STFC grant ST/T000252/1, the BK21 plus program
through the National Research Foundation funded by the Ministry of Education of Korea, and
the Russian Foundation for Basic Research Grant No. 17-52-80064.
P.A. acknowledges funding from his STFC Ernest Rutherford Fellowship (No. ST/R004285/2). Numerical computations were carried out on Cray XC50 at the Center for Computational Astrophysics, NAOJ.
D.Y.K. thanks STFC for the grant ST/T000252/1, and budgetary funding of Basic Research program II.16.
E.G.K. acknowledges the grant of the Russian Foundation for Basic Research No. 18-02-00856.
D.L. is supported by the NSFC under grant 11973092.
N.M. acknowledges funding through the Newton International Fellowship of the Royal Society (No. R1\textbackslash182293), and the FWO-Vlaanderen (No. 12T6521N).
G.N. acknowledges the support of the CGAUSS project at the University of G\"ottingen by the German Aerospace Centre (DLR) under grant 50OL1901 and the Rita Levi Montalcini 2017 fellowship funded by the Italian Ministry of Education, University and Research.
A.K.S. acknowledges UKIERI (Indo-UK) Research Grant for the support of his scientific research.
D.Y. is supported by the National Natural Science Foundation of China (NSFC, 11803005, 11911530690, 41731067) and the Shenzhen Technology Project (JCYJ20180306172239618).
I.V.Z. is supported by the budgetary funding of Basic Research program ``PLASMA''.
\end{acknowledgements}


\bibliographystyle{spbasic}

 \end{document}